\documentclass[12pt, letterpaper]{article}
\usepackage[margin=1in]{geometry}
\usepackage[utf8]{inputenc}
\usepackage[nodoi]{apacite}
\usepackage{amsmath}
\usepackage{natbib}
\usepackage{amsfonts}
\usepackage[hidelinks]{hyperref}
\usepackage{cleveref}
\usepackage{authblk}
\usepackage{graphicx}
\usepackage{subcaption}
\usepackage{multirow}
\graphicspath{ {./figures/} }
\DeclareGraphicsExtensions{.pdf}
\makeatletter
\def\blfootnote{\xdef\@thefnmark{*}\@footnotetext}
\makeatother

\title{Integrating Deep Neural Networks with Full-waveform Inversion: Reparametrization, Regularization, and Uncertainty Quantification}

\author[1*]{Weiqiang Zhu}
\author[2*]{Kailai Xu}
\author[2,3]{Eric Darve}
\author[1]{Biondo Biondi}
\author[1]{Gregory C. Beroza}
\affil[1]{\small Department of Geophysics, Stanford University, Stanford, CA, 94305}
\affil[2]{\small Institute for Computational and Mathematical Engineering, Stanford University, Stanford, CA, 94305}
\affil[3]{\small Mechanical Engineering, Stanford University, Stanford, CA, 94305}
\affil[ ]{\texttt {\{zhuwq, kailaix, darve, biondo, beroza\}@stanford.edu}}
\date{}

\begin{document}
\renewcommand{\APACrefYearMonthDay}[3]{\APACrefYear{#1}}

\maketitle

\blfootnote{The two authors contributed  equally to this paper.}

\section*{Abstract}
Full-waveform inversion (FWI) is an accurate imaging approach for modeling velocity structure by minimizing the misfit between recorded and predicted seismic waveforms. However, the strong non-linearity of FWI resulting from fitting oscillatory waveforms can trap the optimization in local minima. We propose a neural-network-based full waveform inversion method (NNFWI) that integrates deep neural networks with FWI by representing the velocity model with a generative neural network. Neural networks can naturally introduce spatial correlations as regularization to the generated velocity model, which suppresses noise in the gradients and mitigates local minima. The velocity model generated by neural networks is input to the same partial differential equation (PDE) solvers used in conventional FWI. The gradients of both the neural networks and PDEs are calculated using automatic differentiation, which back-propagates gradients through the acoustic PDEs and neural network layers to update the weights of the generative neural network. Experiments on 1D velocity models, the Marmousi model, and the 2004 BP model demonstrate that NNFWI can mitigate local minima, especially for imaging high-contrast features like salt bodies, and significantly improves the inversion in the presence of noise. Adding dropout layers to the neural network model also allows analyzing the uncertainty of the inversion results through Monte Carlo dropout. NNFWI opens a new pathway to combine deep learning and FWI for exploiting both the characteristics of deep neural networks and the high accuracy of PDE solvers. Because NNFWI does not require extra training data and optimization loops, it provides an attractive and straightforward alternative to conventional FWI.

\section{Introduction}
Full-waveform inversion (FWI) is a high resolution inversion method in exploration seismology \citep{tarantola1984inversion, tarantolaInverse2005, virieux2009overview}, commonly used for estimating subsurface velocity structure based on seismic waves recorded at the surface. FWI falls within the class of PDE-constrained optimization problems. It solves the wave equation, in the acoustic or elastic approximation, to predict seismic waves based on the velocity model. Through the PDE solver, FWI determines the optimal velocity model by minimizing the misfit between predicted and observed seismic waveforms. The gradient of the misfit function can be efficiently calculated by the adjoint-state method \citep{plessix2006review}.
Although FWI can achieve high accuracy when the full seismic waveform is matched, the non-linearity of the objective function poses a challenge to the optimization process. The inversion result of FWI suffers from local minima due to cycle skipping in the wave oscillations used to form the objective function, i.e., a $\mathcal{L}_2$-norm loss. This is particularly challenging when either a good initial model is lacking or low frequency content of the waveform data is missing. Noise in real seismic recording can also contaminate the inversion results. Because artifacts originating from local minima interfere with imaging results and can lead to misinterpretation of geological structures, a great deal of research has focused on improving the stability of FWI. One solution is to recover or predict the missing low frequency content, such as through envelope inversion \citep{bozdaug2011misfit, wu2014seismic}, sparse blind deconvolution \citep{zhangSparse2017}, and phase-tracking methods \citep{liFullwaveform2016}. Another solution is to add regularization and preconditioning, such as Laplacian smoothing \citep{bursteddeAlgorithmic2009}, $l_2$-norm penalty \citep{huSimultaneous2009}, $l_1$-norm penalty (total variation) \citep{guittonBlocky2012, esser2018total, kalita2019regularized}, and prior information as constraints \citep{asnaashariRegularized2013}. Many other solutions have also been proposed and tested such as: multiscale inversion \citep{bunks1995multiscale}, wave-equation traveltime inversion \citep{luoWave1991}, tomographic full-waveform inversion \citep{biondi2014simultaneous}, model extension \citep{barnier2018full}, model reduction \citep{barnier2019waveform}, model reparameterization \citep{guitton2012constrained}, and dictionary learning \citep{zhuSparsepromoting2017, liFull2018},. In addition to the strong non-linearity of FWI, uncertainty analysis of FWI results is challenging due both to the high dimensionality of the model space and to the demanding computational cost of solving the wave equation \citep{gebraadBayesian2020}.

In recent studies, the success of deep learning in computer vision, natural language processing, and many other fields has drawn attention to its potential application in FWI \citep{adler2021deep}. One research direction is to build a direct inverse mapping from observations to subsurface structure by training neural networks on paired data of seismic waveforms and velocity models \citep{wuInversionet2018, yangDeeplearning2019, li2019deep, kazei2021mapping}. This approach does not rely on solving the wave equation but instead treats FWI as a data-driven machine learning problem similar to that in image recognition. The accuracy and generalization of this approach, however, cannot be guaranteed without the PDE constraint in FWI. Another research direction is to apply deep learning as an effective signal processing tool to improve the optimization process of conventional FWI. For example, several studies applied neural networks to extrapolate the missing low frequencies and help mitigate the cycle skipping problem \citep{ovcharenko2019deep,sunExtrapolated2020,hu2021progressive}. In addition to these data-driven approaches, another promising direction is to combine neural networks and PDEs to formulate FWI as a physics-constrained machine learning problem. \citet{richardson2018generative} and \citet{mosser2020stochastic} trained a generative adversarial network (GAN) to build an a prior model of subsurface geological structures and optimized a lower-dimensional latent variable to fit observed data. \citet{wuParametric2019} and \citet{wu2020cnn} proposed a CNN-domain FWI, which reparameterizes the velocity model or the gradient field by a convolutional neural network (CNN) and minimizes the loss by updating the neural network weights. \citet{he2021reparameterized} further analyzed the adaptive regularization effect from the convolutional neural network. However, these works relied on pre-training convolutional neural networks on initial velocity models and attributed the regularization effect to the prior information coming from fitting these initial velocity models. In contrast, \citet{ulyanovDeep2018}'s work on deep image prior demonstrated that the CNN architecture without pre-training can be used as a prior with excellent results in inverse problems of computer vision, such as denoising, super-resolution, and inpainting. Their results showed that although a high-capacity neural network can fit both structured objects and unstructured noise, the parametrization of the neural network offers high impedance to noise during optimization and learns much more quickly towards natural-looking images.
Another limitation of the CNN-domain FWI approach is the complex inversion workflow combining two optimization processes of neural network training and full-waveform inversion. \citet{richardson2018seismic} and \citet{zhuGeneral2020} have implemented FWI using deep learning frameworks so that reverse-mode automatic differentiation and various effective optimization tools in deep learning frameworks can be used for FWI. The similarity between neural network training and FWI demonstrated by these works makes it possible to greatly simplify the previous inversion workflows within one unified framework for both deep learning and FWI.

In this study, we propose a method, NNFWI, to integrate deep neural networks with full-waveform inversion. Similar to \citet{wuParametric2019}'s idea, we use deep neural networks to generate a physical velocity model, which is then fed to a PDE solver to simulate seismic waveforms. The training process of NNFWI is similar to that of conventional FWI, but with gradients calculated by automatic differentiation instead of by the adjoint-state method. Thus we can easily optimize the two systems of neural networks and PDEs together. Unlike conventional FWI, which directly estimates the velocity model, NNFWI reparametrizes the velocity model with a generative neural network model and optimizes the neural network's weights. In contrast to previous work that learns prior information from pre-training, NNFWI does not require pre-training neural networks on the initial velocity model. As demonstrated by \citet{ulyanovDeep2018}'s work on the deep image prior, the inductive bias captured by deep convolutional networks is an effective image prior and can be used as regularization for tasks such as denoising and super-resolution. Due to the regularization effect of neural networks, NNFWI mitigates the effects of local minima and is robust with respect to noise in the data. Furthermore, NNFWI can model uncertainty by adding dropout layers in the neural network. Dropout not only prevents over-fitting during training deep neural networks, but can also approximate Bayesian inference to capture model uncertainty without much extra computational cost \citep{gal2016dropout}. NNFWI has the potential to provide uncertainty quantification for FWI, which otherwise remains a challenging problem. NNFWI exploits the unique advantages of deep neural networks and the high accuracy and generalization of PDEs to improve inversion and uncertainty analysis in FWI.

In the following, we first describe the two components of NNFWI including a generative neural network to parametrize the inversion target (i.e., velocity model) and an acoustic PDE to predict accurate waveforms same as that used conventional FWI. We then compare the performance of NNFWI with conventional FWI and FWI with TV regularization on three benchmark models: 1D velocity profiles, the Marmousi model, and the 2004 BP model. Last, we present an uncertainty estimation method using the Monte Carlo dropout technique and analyzed the computational cost of NNFWI. Additional comparison results are provided in the appendix.

\section{Method}
\begin{figure}[!ht]
\centering
\includegraphics[width=0.8\textwidth]{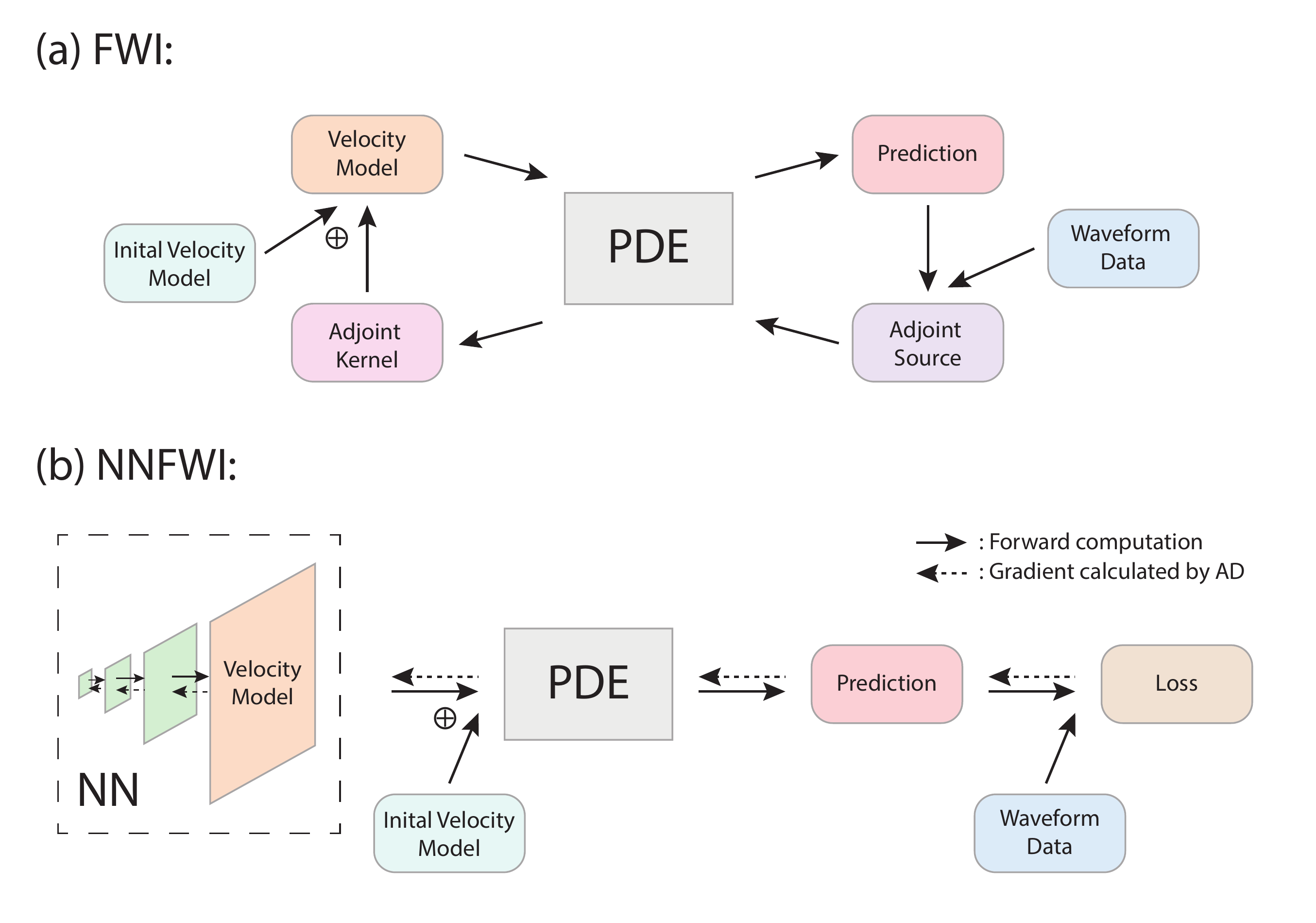}
\caption{(a) Workflow of conventional FWI based on the adjoint-state method. (b) Diagram of NNFWI, which combines neural networks and PDEs of seismic waves. The input velocity to the PDEs is a direct summation of the initial velocity model and the velocity model generated by the neural network. In this way, NNFWI follows the same data format as conventional FWI and can be directly applied to conventional FWI applications. The gradients of both the neural network and PDE are calculated using automatic differentiation in NNFWI using the ADSeismic package \citep{zhuGeneral2020}.}
\label{fig:model}
\end{figure}

FWI aims to minimize the discrepancy between the observed seismic data $u(x, t)$ and the numerical prediction $\hat{u}$, which is the solution to a wave equation (denoted as $F(\hat{u}, m) = 0$), such as the acoustic wave equation:
\begin{equation}
    \frac{\partial^2 \hat{u}}{\partial t^2} = \nabla\cdot(c^2 \nabla \hat{u}) +  f
    \label{eqn:acoustic}
\end{equation}
where $f$ is the source term. The inversion parameter here is the acoustic wave speed $m = c$,
thus the PDE-constrained optimization problem for estimating an unknown field $m$ is given by 
\begin{align*}
    \underset{m}{\mbox{min }} &\  D(\hat{u}(x, t, m), u) =\sum_{i=1}^I  \sum_{j=1}^J \int_0^T || \hat{u}(x_i, x_j, t, m) - u(x_i, x_j, t) ||_2^2 dt\\
    \mbox{s.t.} &\ F(\hat u, m) = 0 
\end{align*}
where $D$ is the misfit function, i.e., $\mathcal{L}_2$-norm loss, 
$\{x_i\}_{i=1}^I$ designates the discrete locations of active sources; and $\{x_j\}_{j=1}^J$ designates the discrete locations where receivers are available. 
The standard FWI derives the gradient $\frac{\partial D}  {\partial m}$ using the adjoint-state method and updates $m$ using a gradient-based optimization method, such as L-BFGS (\Cref{fig:model}a). There are two challenges when this method is used:
\begin{enumerate}
    \item The inverse problem is usually ill-conditioned. One approach is to use a misfit function with a regularization term, such as Tikhonov and total variation regularizations.
    \item Quantifying uncertainties of $m$ is very challenging because the forward simulation, i.e., computing $\hat u$ by solving $F(\hat u, m) = 0$, is very expensive. Conventional methods, such as  Markov Chain Monte Carlo (MCMC), usually require a large number of forward simulations, and therefore are computationally demanding. 
\end{enumerate}

To address these two issues, we propose Neural-Network-based Full Waveform Inversion (NNFWI). In NNFWI, we use a generative neural network to parametrize the velocity model
\begin{equation}\label{equ:mG}
    m = \mathcal{N}(z, w)
\end{equation}
where $\mathcal{N}$ is a generative neural network, $z$ is the latent variable, and $w$ includes the weights of the neural network. \Cref{equ:mG} introduces regularization to the velocity field by representing $m$ with a neural network.
In this work we show the regularization effect of two types of neural networks consisting of fully connected layers or convolutional layers. For neural networks of fully connected layers, the latent variable $z$ is set to be the coordinate $x$, so that $\mathcal{N}(x, w)$ imposes a spatial correlation on $m$, i.e., $\mathcal{N}(x_1, w)$ and $\mathcal{N}(x_2, w)$ are similar if the spatial coordinates $x_1$ and $x_2$ are close. For neural networks of convolutional layers, the local convolutional kernels applied across the entire domain imposes self-similarity and spatial correlation. \citet{ulyanovDeep2018} demonstrated that this deep image prior of convolutional neural network architectures achieved competitive results with handcrafted self-similarity-based and dictionary-based priors, such as the total variation norm, in a variety of image processing problems, such as denoising, super-resolution, and inpainting. Thus, reparametrizing the velocity $m$ with convolutional neural networks naturally introduces regularization to FWI. We set the latent variable $z$ as a fixed random vector and build a generative neural network to generate a velocity model, which applies a series of convolutional layers and upsampling operations to convert the latent vector into a 2D matrix (\Cref{tab:nn}).
The optimization problem of NNFWI can be written as:
\begin{align}
    \underset{w}{\mbox{min}}&\ D(\hat{u}(x, t, \mathcal{N}(z, w)), u(x_i, t)) = \sum_{i=1}^I \sum_{j=1}^J \int_0^T || \hat{u}(x_i, x_j, t, \mathcal{N}(z, w)) - u(x_i, x_j, t) ||_2^2 dt\\ \label{eqn:loss}
    \mbox{s.t.}&\ F(\hat u, \mathcal{N}(z, w)) = 0 
\end{align}
A diagram of the NNFWI approach is shown in \Cref{fig:model}b. The data setting of NNFWI, i.e., the initial velocity model and the waveform data is same as for conventional FWI, which makes NNFWI applicable to all conventional FWI problems. In NNFWI, we directly combine the output from the generative neural networks with the initial velocity model as the input for the PDE, so the generative neural networks are trained to predict the updates over the initial model. Because adjoint-state methods and reverse-mode automatic differentiation are mathematically equivalent, we can calculate the gradients of both the generative neural network and the PDEs by automatic differentiation, which simplifies the optimization of both neural networks and PDEs \citep{zhuGeneral2020,xu2020adcme}. NNFWI is implemented based on ADCME\footnote{\url{https://github.com/kailaix/ADCME.jl}} and ADSeismic\footnote{\url{https://github.com/kailaix/ADSeismic.jl}}, which simulate both acoustic and elastic waveform equations using the finite-difference time-domain (FDTD) method and automatically calculate gradients using the automatic differentiation method based on the deep learning framework, Tensorflow \citep{abadi2016tensorflow}. We use the Adam algorithm \citep{kingma2014adam} for the optimization of NNFWI. For comparison, we also conduct conventional FWI using the L-BFGS algorithm \citep{zhu1997algorithm}.

\begin{table}
\centering
\caption{The architectures of the generative neural networks in NNFWI. The fully connected neural network (FC) model is used for the 1D FWI case and the convolutional neural network (CNN) model is used for the Marmousi model and the 2004 BP models. We apply scaling factors to the output of neural networks depending on the physical parameters and units. Note that the parameters and layers in these architectures can be modified for different applications. }
\label{tab:nn}
\resizebox{\textwidth}{!}{%
\begin{tabular}{l|l|l}
\hline
Model                      & NN layer & Architecture                                                                        \\ \hline
\multirow{5}{*}{FC model}  & Input    & Spatial coordinate $x$                                                              \\
                           & Layer 1  & Fully-connected layer (channels = 30) + Tanh                                        \\
                           & Layer 2  & Fully-connected layer (30) + Tanh                                                   \\
                           & Layer 3  & Fully-connected layer (30) + Tanh                                                   \\
                           & Layer 4  & Fully-connected layer (1) + Tanh                                                   \\ \hline
\multirow{6}{*}{CNN model} & Input    & Random latent vector  (8)  \\
                           & Layer 1  & Fully-connected layer (8) + Tanh + Reshape \\
                           & Layer 2  & 2$\times$2 Upsampling + 4$\times$4 Convolutional layer (128) + Leaky Relu (0.1) + Dropout \\
                           & Layer 3  & 2$\times$2 Upsampling + 4$\times$4 Convolutional layer (64) + Leaky Relu (0.1) + Dropout \\
                           & Layer 4  & 2$\times$2 Upsampling + 4$\times$4 Convolutional layer (32) + Leaky Relu (0.1) + Dropout \\
                           & Layer 5  & 2$\times$2 Upsampling + 4$\times$4 Convolutional layer (16) + Leaky Relu (0.1) + Dropout \\
                           & Layer 6  & 4$\times$4 Convolutional layer (1) + Tanh                                          \\ \hline
\end{tabular}%
}
\end{table}

\section{Results}
In this section, we evaluate the performance of NNFWI by comparing the inversion results between conventional FWI and NNFWI on three cases.

\subsection{1D model}
\begin{figure}
\centering
\includegraphics[width=1.0\linewidth]{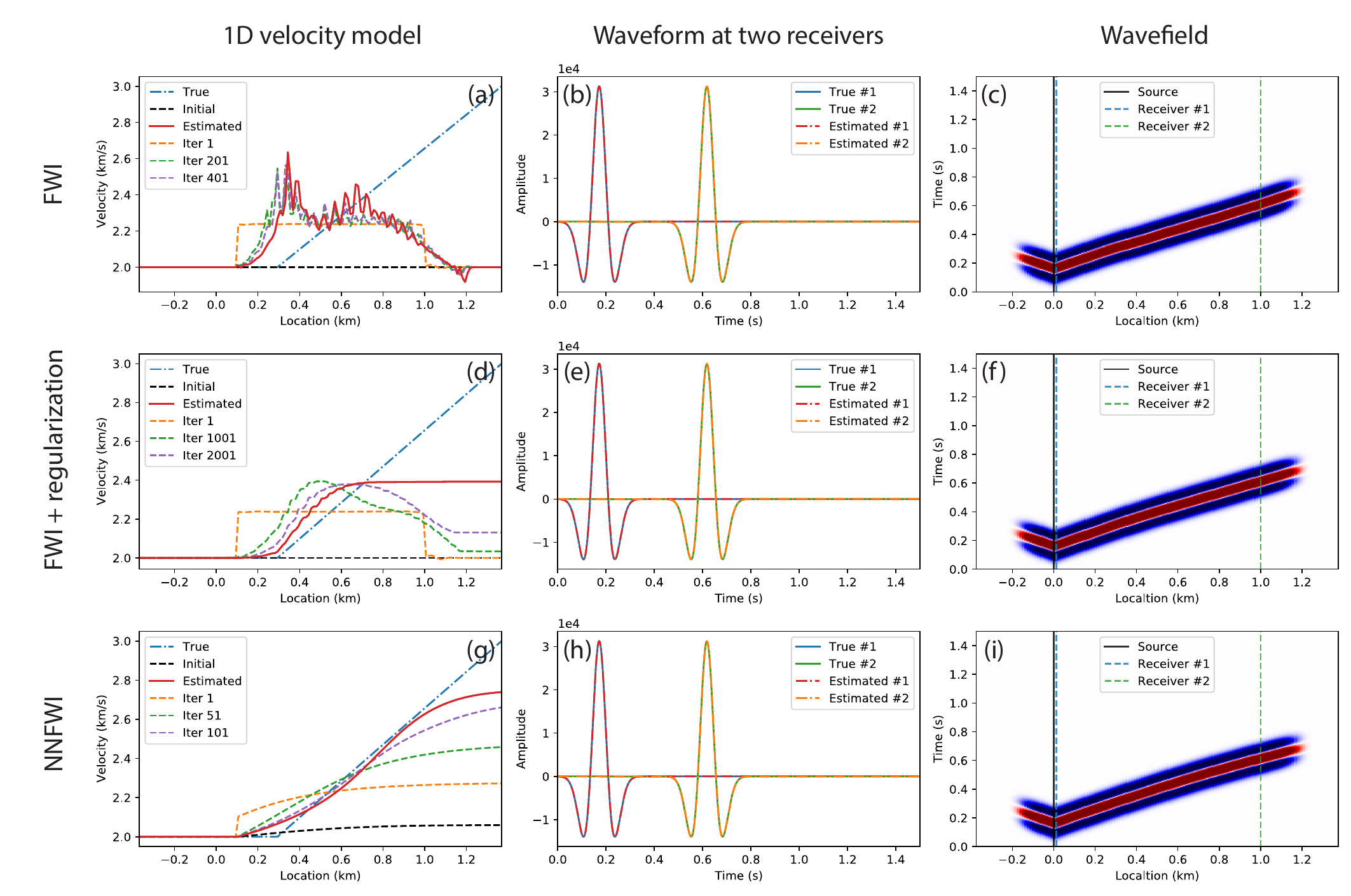}
\caption{Inversion results of an 1D linear velocity profile. The upper panel (a-c) shows results of conventional FWI. The middle panel (d-f) shows results of FWI with Tikhonov regularization. The lower panel (g-i) shows results of NNFWI. The true and estimated velocity profiles are plotted in the left panel (a, d, g). We run 10,000 interactions for the final estimation. The estimated profiles at three selected iterations are also plotted to show the convergence. Note that we only estimate the velocity at $x\ge0.1$ km to avoid extreme gradient values at the source. Because we only placed two receivers at $x=0$ and $x=1.0$, the updates of the first few iterations of FWI mainly occurred in the area between the two receivers, causing a box-like perturbation in the first iteration of FWI in panel (a). NNFWI, however, imposes a spatial regularization to update the whole region, as shown in panel (g). The fitted waveforms are plotted in the middle panel (b, e, h). The whole wavefields are plotted in the right panel (c, f, i).}
\label{figure:1d-linear}
\end{figure}

\begin{figure}
\centering
\includegraphics[width=1.0\linewidth]{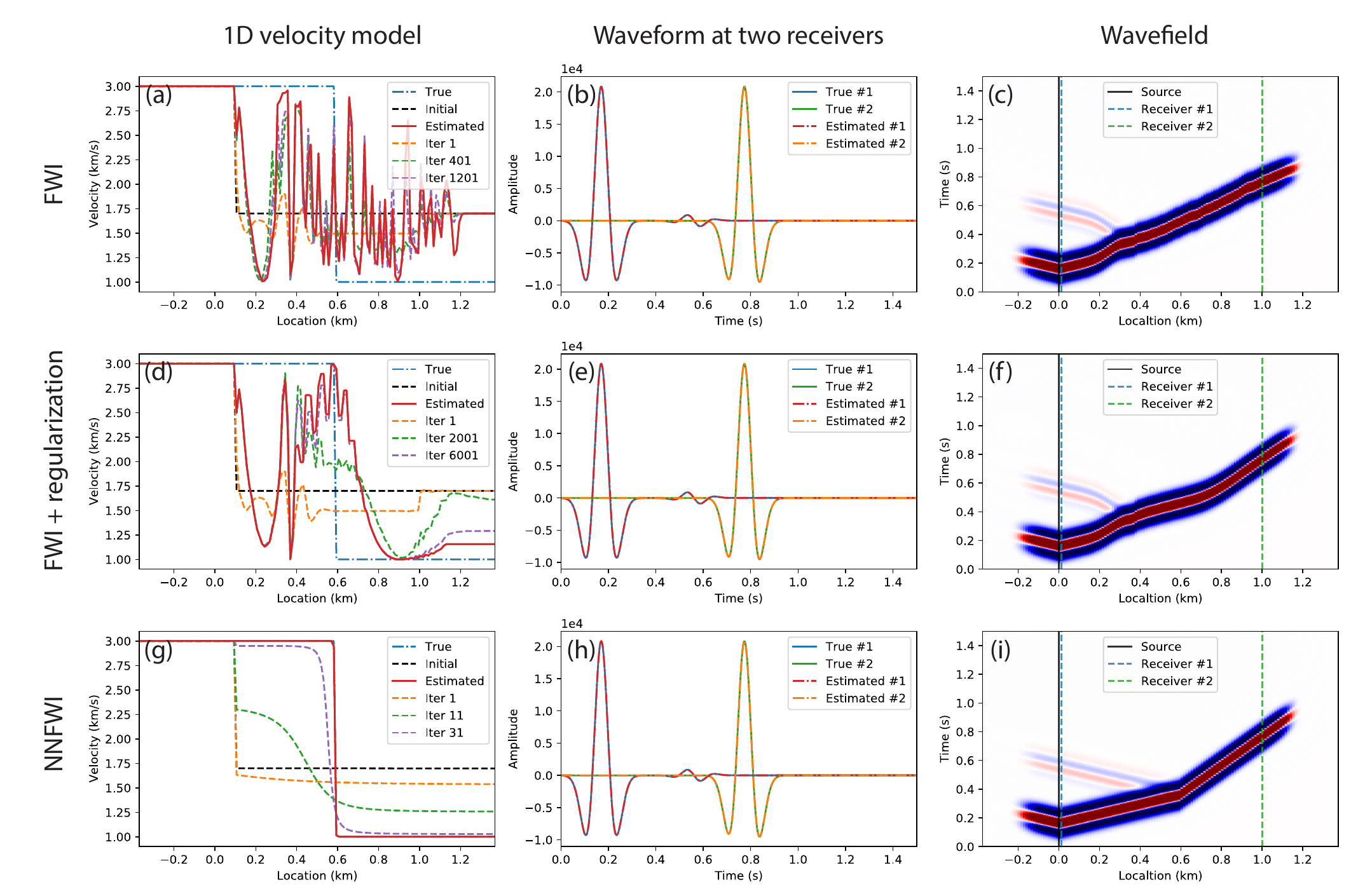}
\caption{Inversion results of an 1D step-change velocity profile. The panels are plotted in the same way as \Cref{figure:1d-linear}. Note that the initial step-change at $x=0.1$ km occurs because we only estimate the velocity at $x\ge0.1$ km and fix the true velocity value at $x<0.1$ km to avoid source effects. The waveforms at two receivers are well fit for all three methods, but we observe significant differences in the wavefields (c, f, i) between the two receivers due to the incorrect velocity models.}
\label{figure:1d-step}
\end{figure}

To demonstrate the regularization effect of neural networks, we design two simple 1D velocity models: one with a linear velocity profile (\Cref{figure:1d-linear}) and another with a step-change profile (\Cref{figure:1d-step}). We implement a 1D acoustic wave equation to carry out forward simulation and inversion for these examples. We inject a Ricker wavelet at $x=0$ km and record the received waveforms (e.g. \Cref{figure:1d-linear}b) at both sides of the domain ($x=0$ km and $x=1.0$ km). The whole wavefield is also plotted in the last columns of \Cref{figure:1d-linear,figure:1d-step}, but only the waveforms at the two receivers are used for FWI as shown in the middle columns. Due to the limited receivers, the velocity model can not be well constrained by conventional FWI (\Cref{figure:1d-linear,figure:1d-step} (a)), even though the two received waveforms can be matched (\Cref{figure:1d-linear,figure:1d-step} (b)). Adding regularization is an effective approach to constrain the inversion results. We add Tikhonov regularization to the loss function, which produces a much smoother velocity profile (\Cref{figure:1d-linear,figure:1d-step} (d)). Note that the weight of regularization is ad-hoc. The inversion results could be improved by searching for a better weight or by adopting a complex weighting strategy, such as imposing strong initial regularization and relaxing it with further iterations. Instead of adding regularization to the loss function, we use a fully connected neural network (the FC model in \Cref{tab:nn}) to reparametrize the velocity model. The neural network learns a continuous function $\mathcal{N}(x, w)$ to represent either the linear or step-change profiles $m(x)$, which implicitly imposes a spatial correlation as regularization. The results demonstrate that the neural network first learns a smooth profile due to the regularization effect and then gradually adapts to complex structures such as the step-change.

\subsection{Marmousi model} \label{sec:marmousi}
\begin{figure}[!ht]
\centering
\begin{subfigure}{0.48\textwidth}
    \centering
    \includegraphics[width=\textwidth]{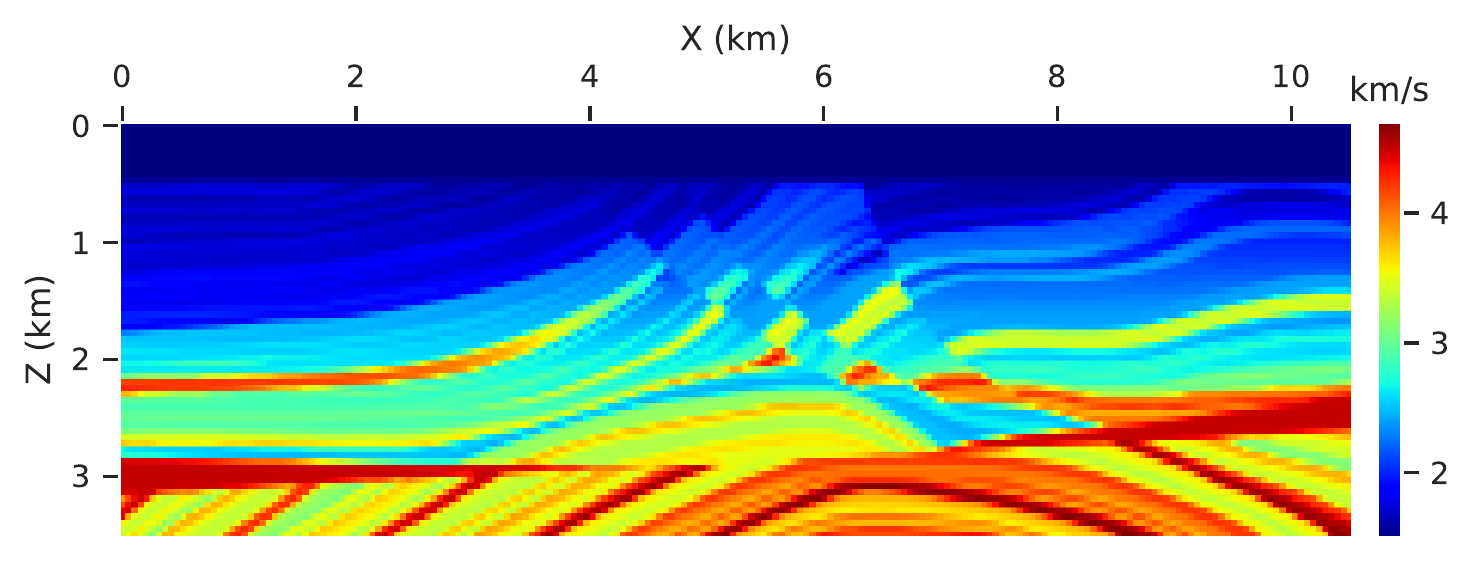}
    \caption{}
\end{subfigure}
\begin{subfigure}{0.48\textwidth}
    \centering
    \includegraphics[width=\textwidth]{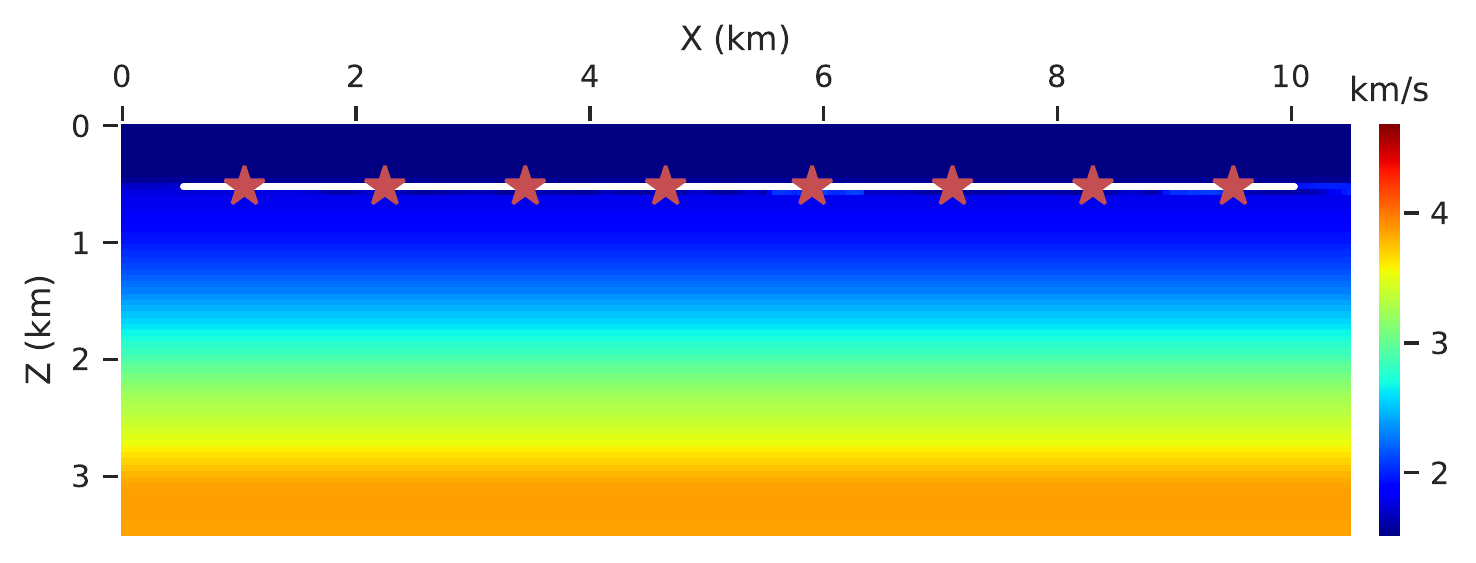}
    \caption{}
\end{subfigure}
\caption{The Marmousi velocity model: (a) the ground-truth model for generating synthetic data; (b) the 1D smoothed initial model for inversion. The source locations (red stars) and the receiver depth (white line) are plotted in (b).}
\label{fig:marmousi}
\end{figure}

The Marmousi model \citep{versteeg1994marmousi} is a benchmark velocity model commonly used for evaluating FWI algorithms. The true velocity model and the 1D initial model used in this case are shown in \Cref{fig:marmousi}. We generated synthetic seismic waveform data using 8 active sources and a sequence of receivers on the surface marked by red stars and a white line respectively in \Cref{fig:marmousi}b. The source spacing is 1.2 km and the receiver spacing is 52 m. The source time function is a Ricker wavelet with a peak frequency of 2.5 Hz. We conducted three experiments by adding, respectively: no random noise, random noise with $\sigma = 0.5\sigma_0$, and random noise with $\sigma = \sigma_0$ to the synthetic data, where $\sigma$ is the standard deviation of Gaussian white noise and $\sigma_0$ is the standard deviation of the synthetic seismic recordings of the entire data set. The same initial model and synthetic data were used for all experiments for both conventional FWI and NNFWI. The number of discretized grid-points of the velocity model in conventional FWI is 13,736, while the number of parameters of the generative neural network in NNFWI is 192,832. The larger number of parameters of NNFWI shows that we can over-parametrize the velocity model with the weights of a generative neural network. However, because of the deep image prior of convolutional neural networks, the inversion results of NNFWI are still robust. The free parameters of NNFWI can be adjusted by changing the architectures and hyper-parameters of the generative neural network. 

\begin{figure}[!ht]
\centering
\begin{subfigure}{0.48\textwidth}
    \centering
    \includegraphics[width=\textwidth]{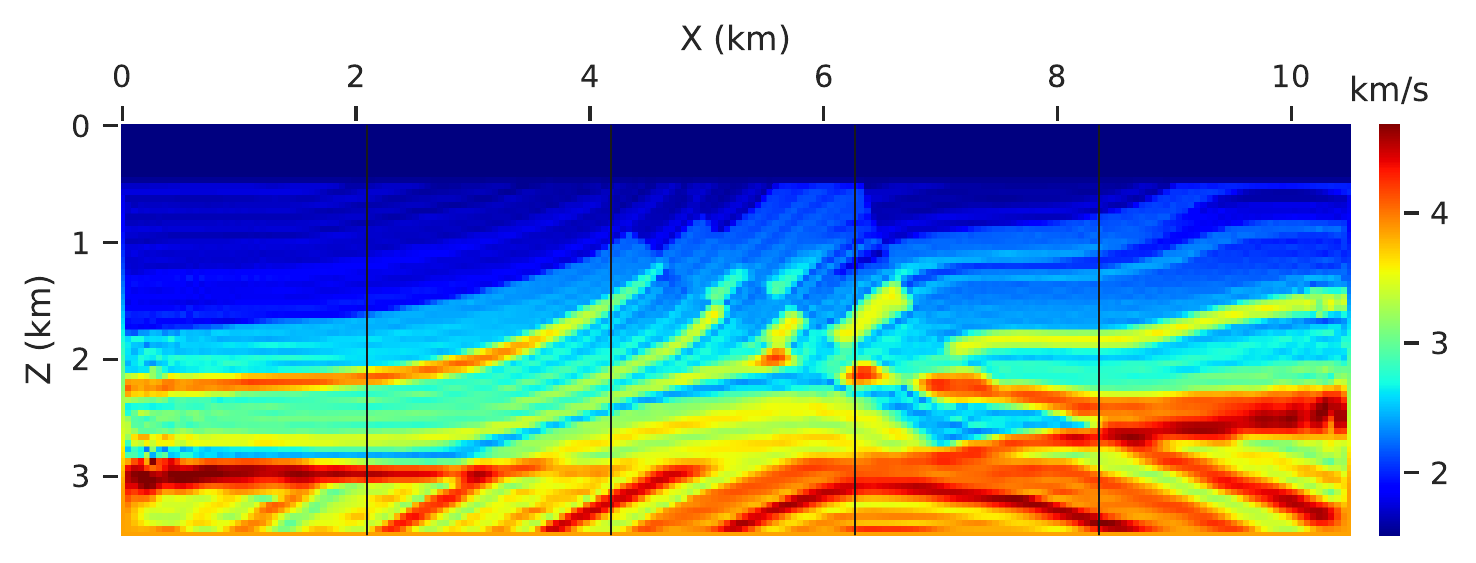}
    \caption{}
\end{subfigure}
\begin{subfigure}{0.48\textwidth}
    \centering
    \includegraphics[width=\textwidth]{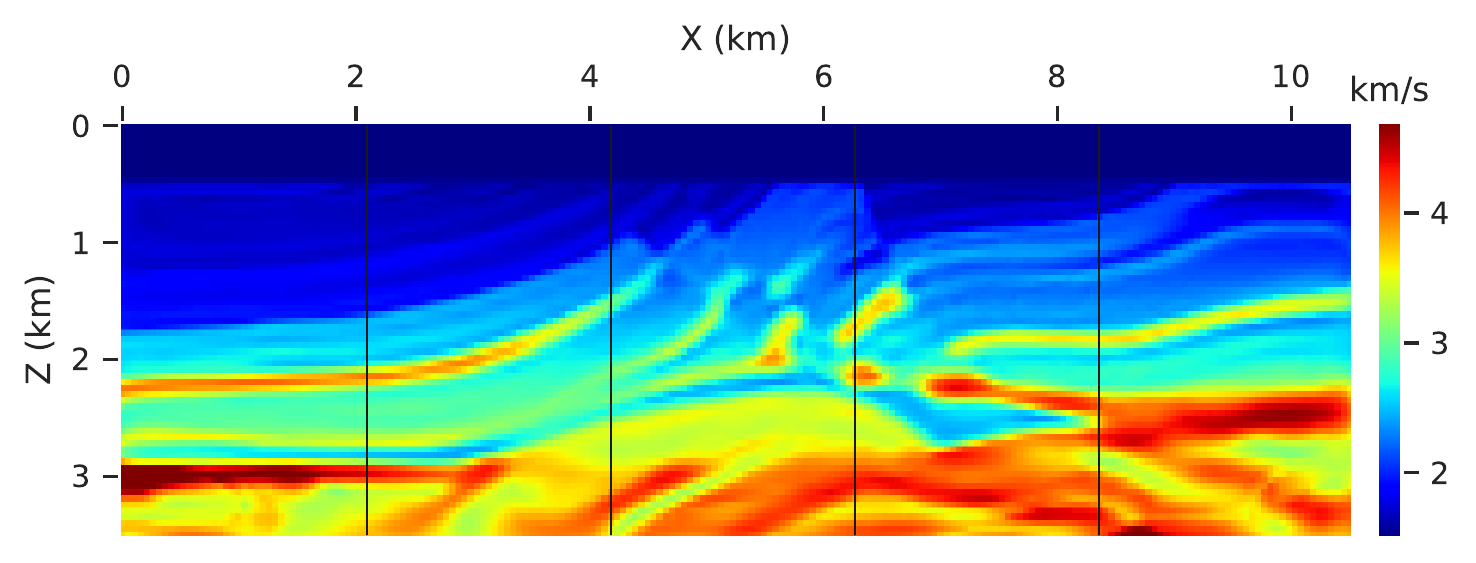}
    \caption{}
\end{subfigure}
\begin{subfigure}{0.48\textwidth}
    \centering
    \includegraphics[width=\textwidth]{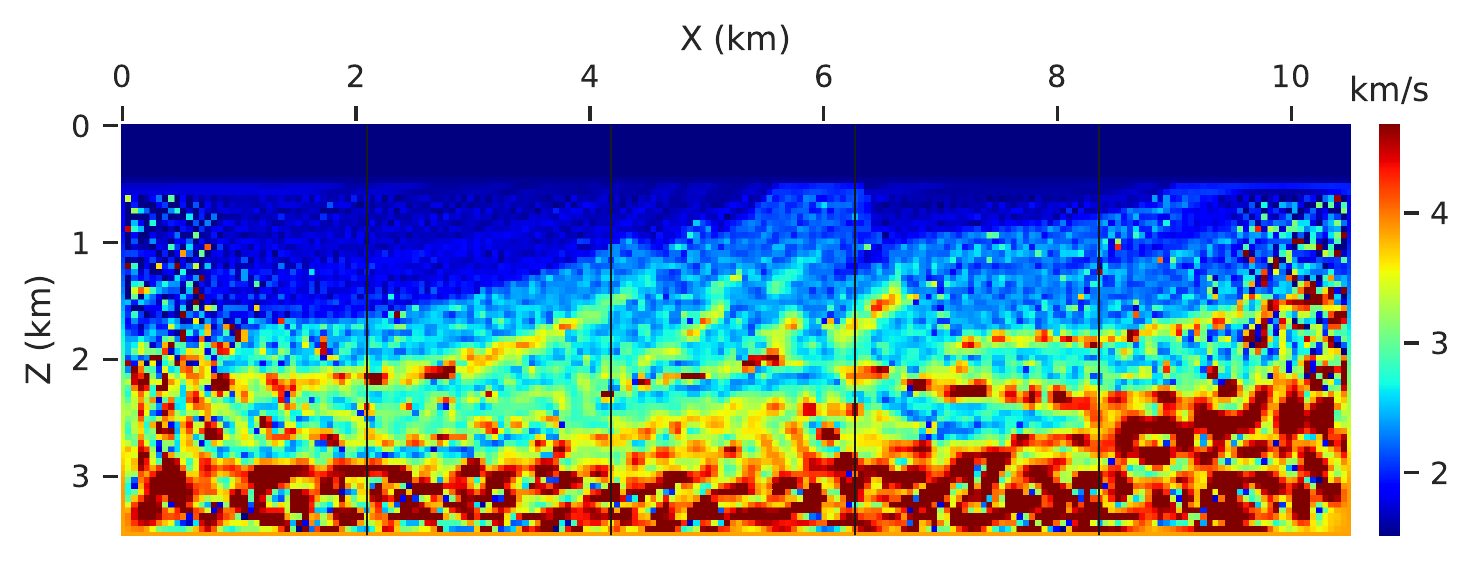}
    \caption{}
\end{subfigure}
\begin{subfigure}{0.48\textwidth}
    \centering
    \includegraphics[width=\textwidth]{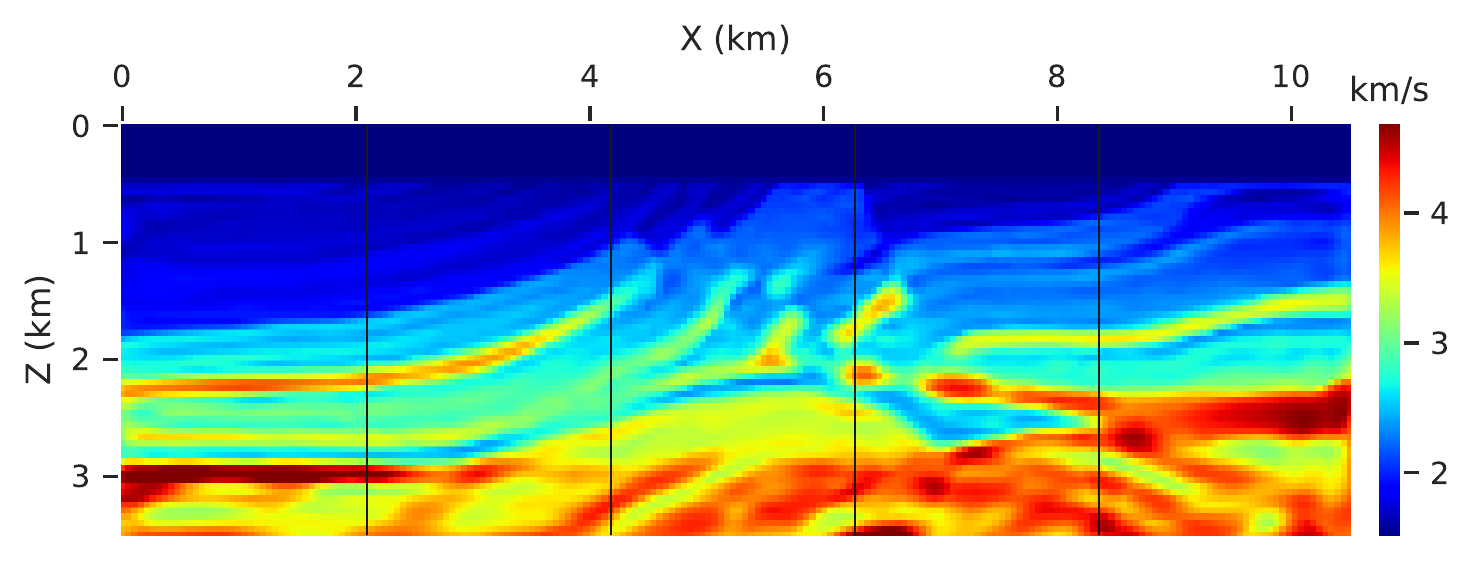}
    \caption{}
\end{subfigure}
\begin{subfigure}{0.48\textwidth}
    \centering
    \includegraphics[width=\textwidth]{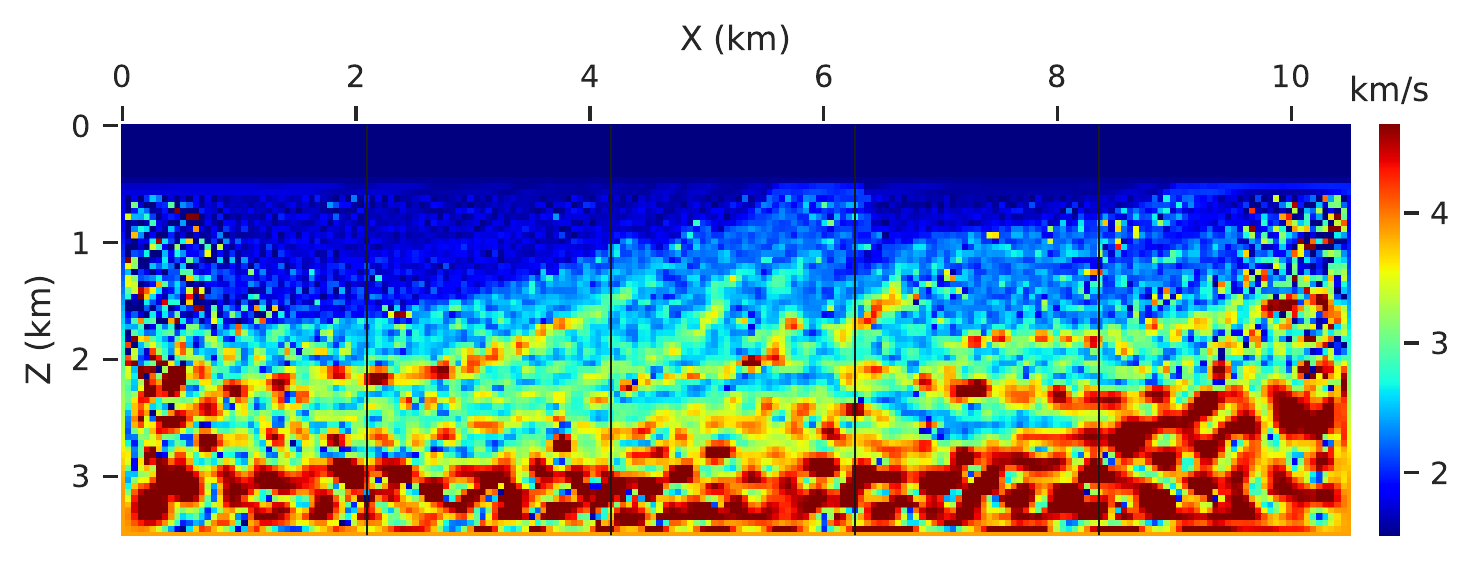}
    \caption{}
\end{subfigure}
\begin{subfigure}{0.48\textwidth}
    \centering
    \includegraphics[width=\textwidth]{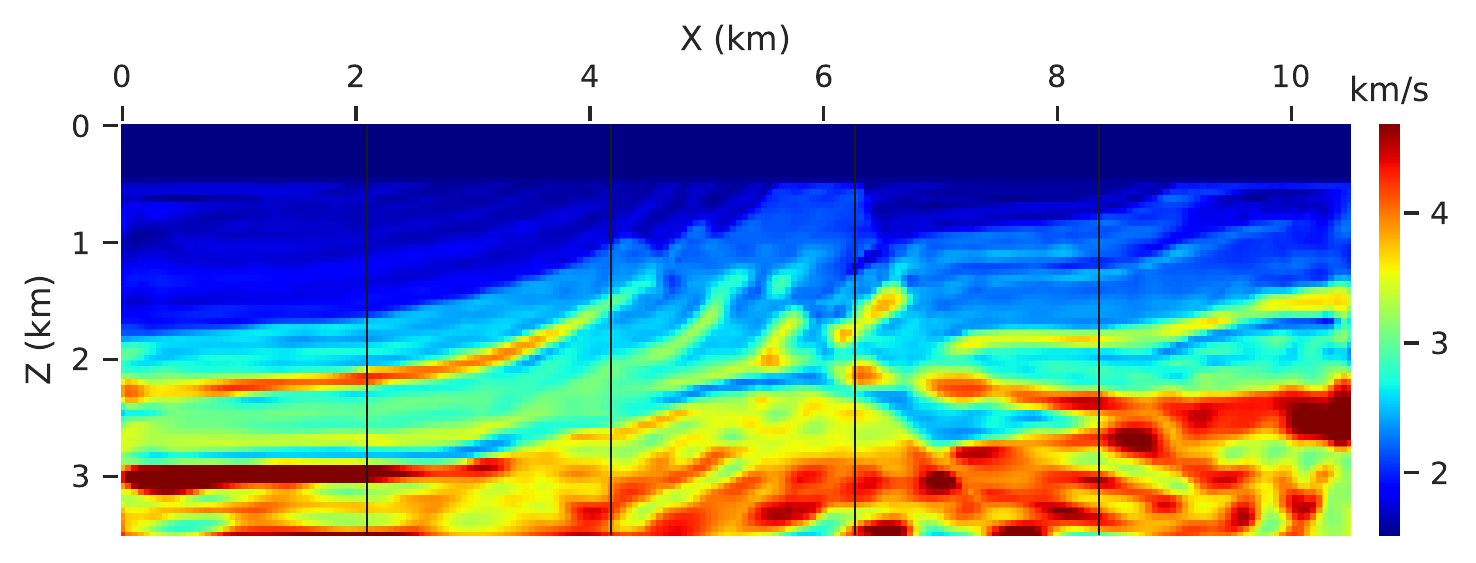}
    \caption{}
\end{subfigure}
\caption{Inversion results of conventional FWI and NNFWI on the Marmousi benchmark model. The left panel: conventional FWI results based on data with (a) no random noise, (c) random noise with $\sigma = 0.5\sigma_0$, and (e) random noise with $\sigma = \sigma_0$. The right panel: NNFWI results based on data with (b) no random noise, (d) random noise with $\sigma = 0.5\sigma_0$, and (f) random noise with $\sigma = \sigma_0$. Here $\sigma_0$ is the standard deviation of the synthetic seismic data. The velocity profiles along the four black vertical lines are shown in \Cref{fig:profile_marmousi}.} 
\label{fig:result_marmousi}
\end{figure}

\begin{figure}[!ht]
\centering
\begin{subfigure}{0.48\textwidth}
    \centering
    \includegraphics[width=\textwidth]{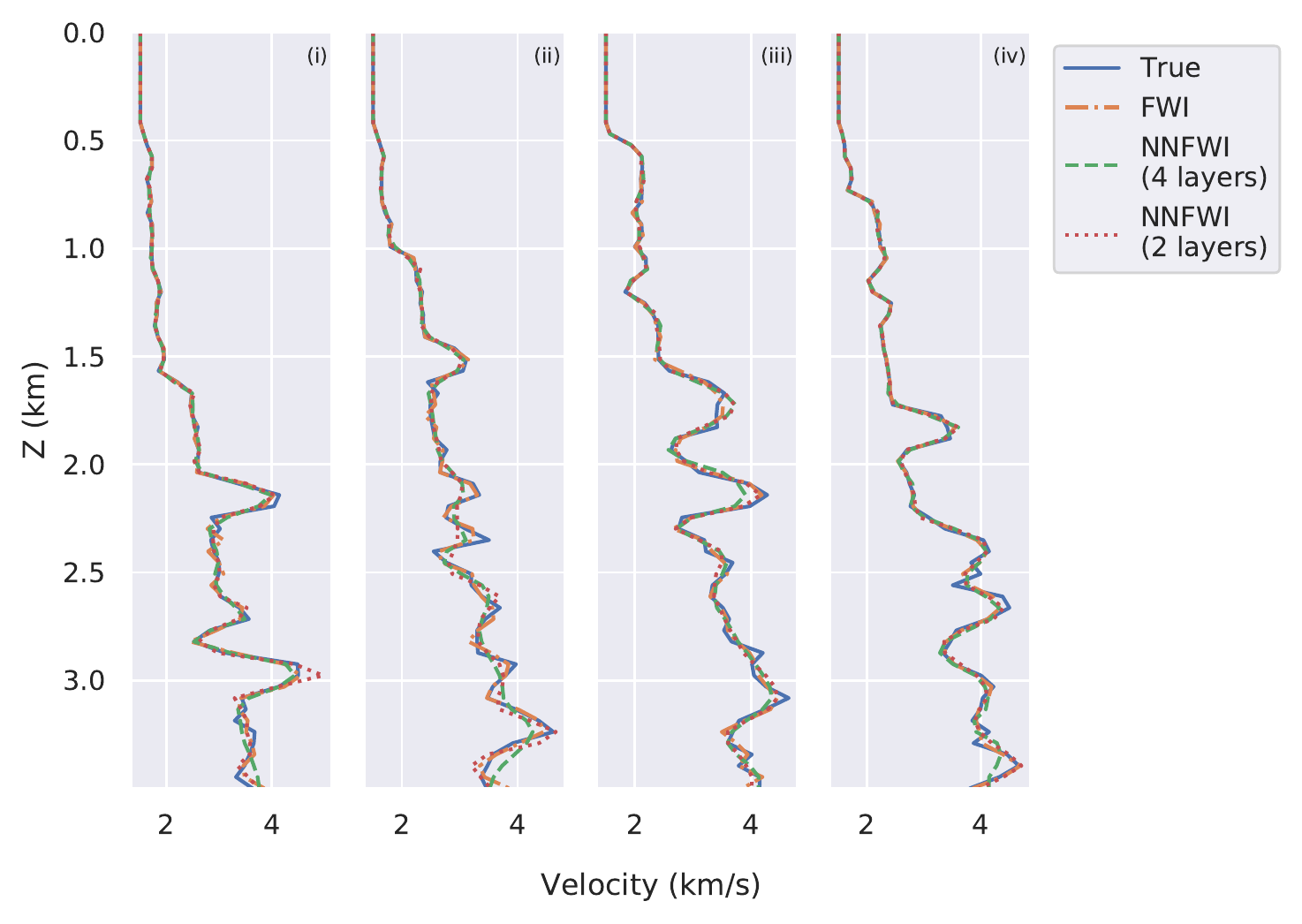}
    \caption{}
\end{subfigure}
\begin{subfigure}{0.48\textwidth}
    \centering
    \includegraphics[width=\textwidth]{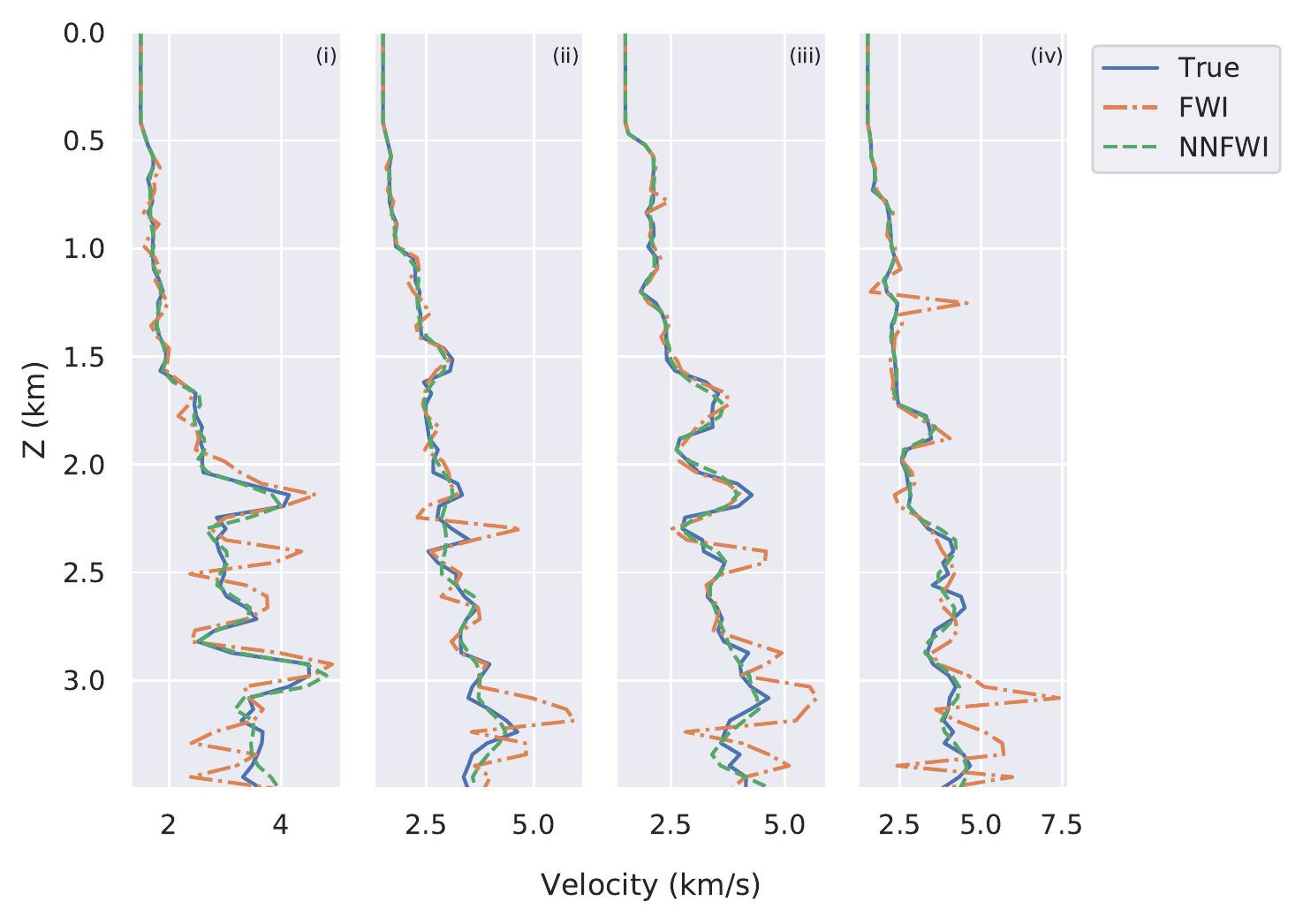}
    \caption{}
\end{subfigure}
\caption{Estimated velocity profiles at four locations of the Marmousi benchmark model: (a) no random noise; (b) random noise ($\sigma = 0.5\sigma_0$). The locations of these four profiles are marked by black vertical lines in \Cref{fig:result_marmousi}.} 
\label{fig:profile_marmousi}
\end{figure}

\begin{figure}[!ht]
\centering
\begin{subfigure}{0.45\textwidth}
    \centering
    \includegraphics[width=\textwidth]{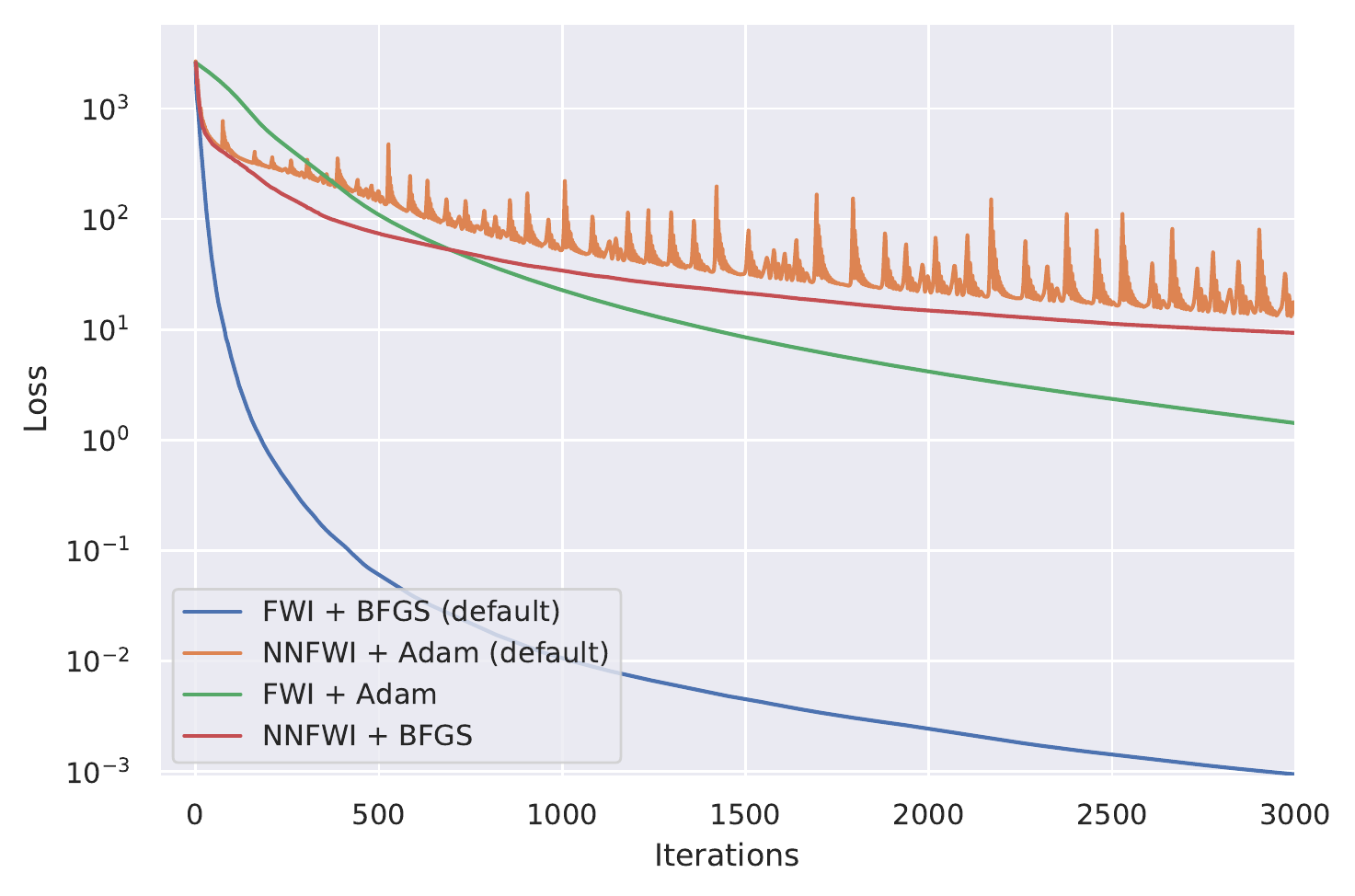}
    \caption{}
\end{subfigure}
\begin{subfigure}{0.45\textwidth}
    \centering
    \includegraphics[width=\textwidth]{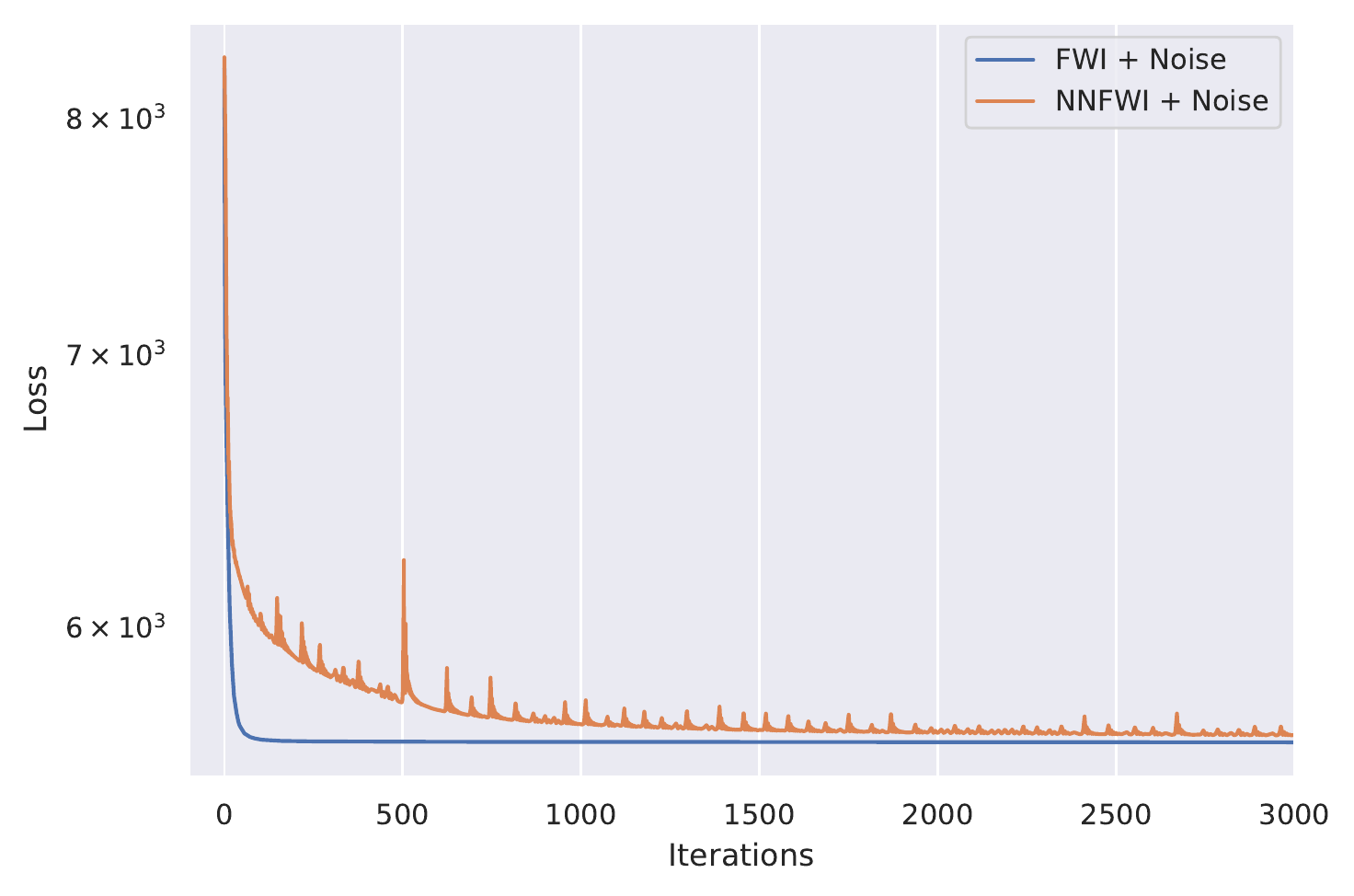}
    \caption{}
\end{subfigure}
\begin{subfigure}{0.45\textwidth}
    \centering
    \includegraphics[width=\textwidth]{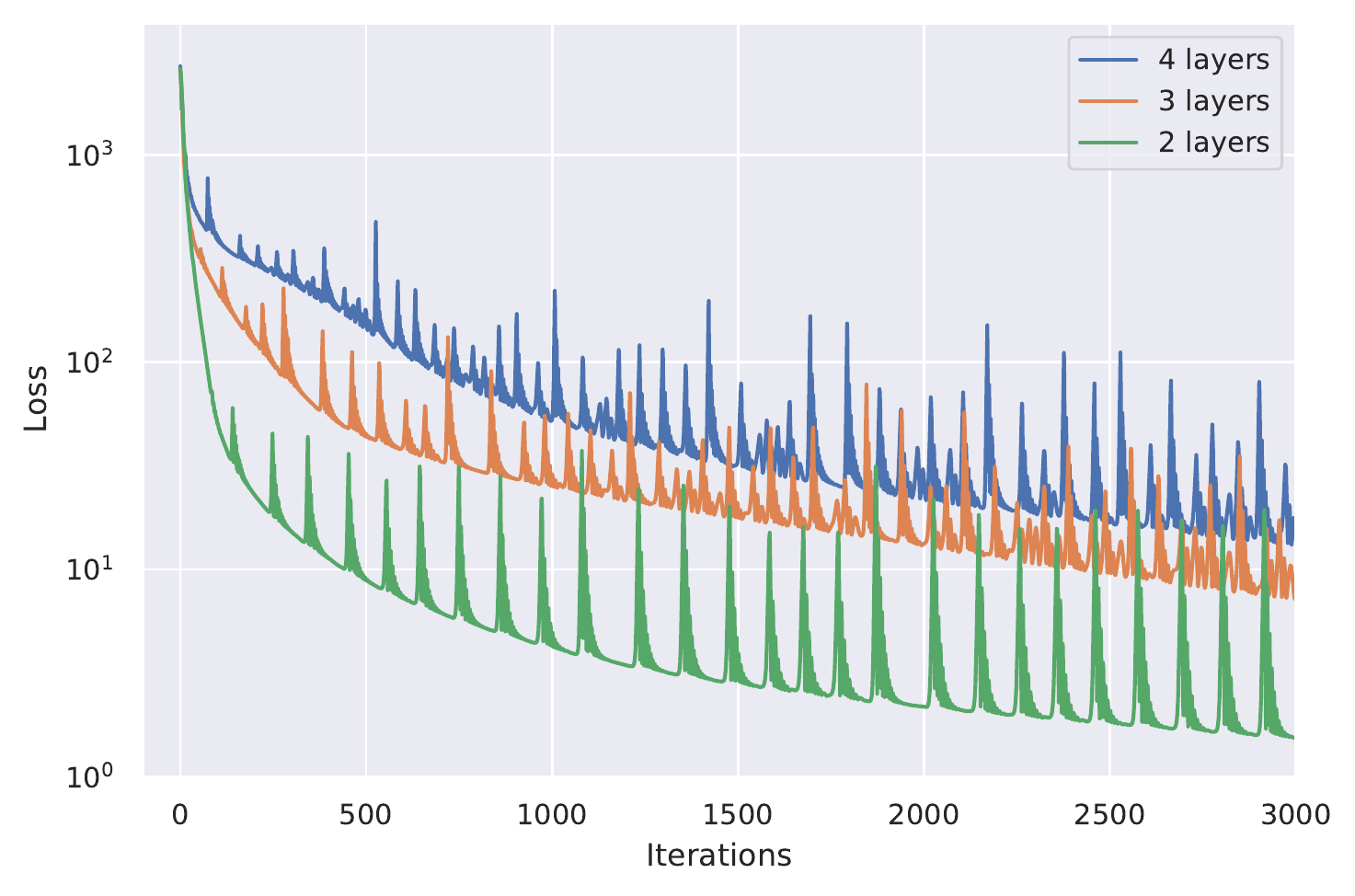}
    \caption{}
\end{subfigure}
\begin{subfigure}{0.45\textwidth}
    \centering
    \includegraphics[width=\textwidth]{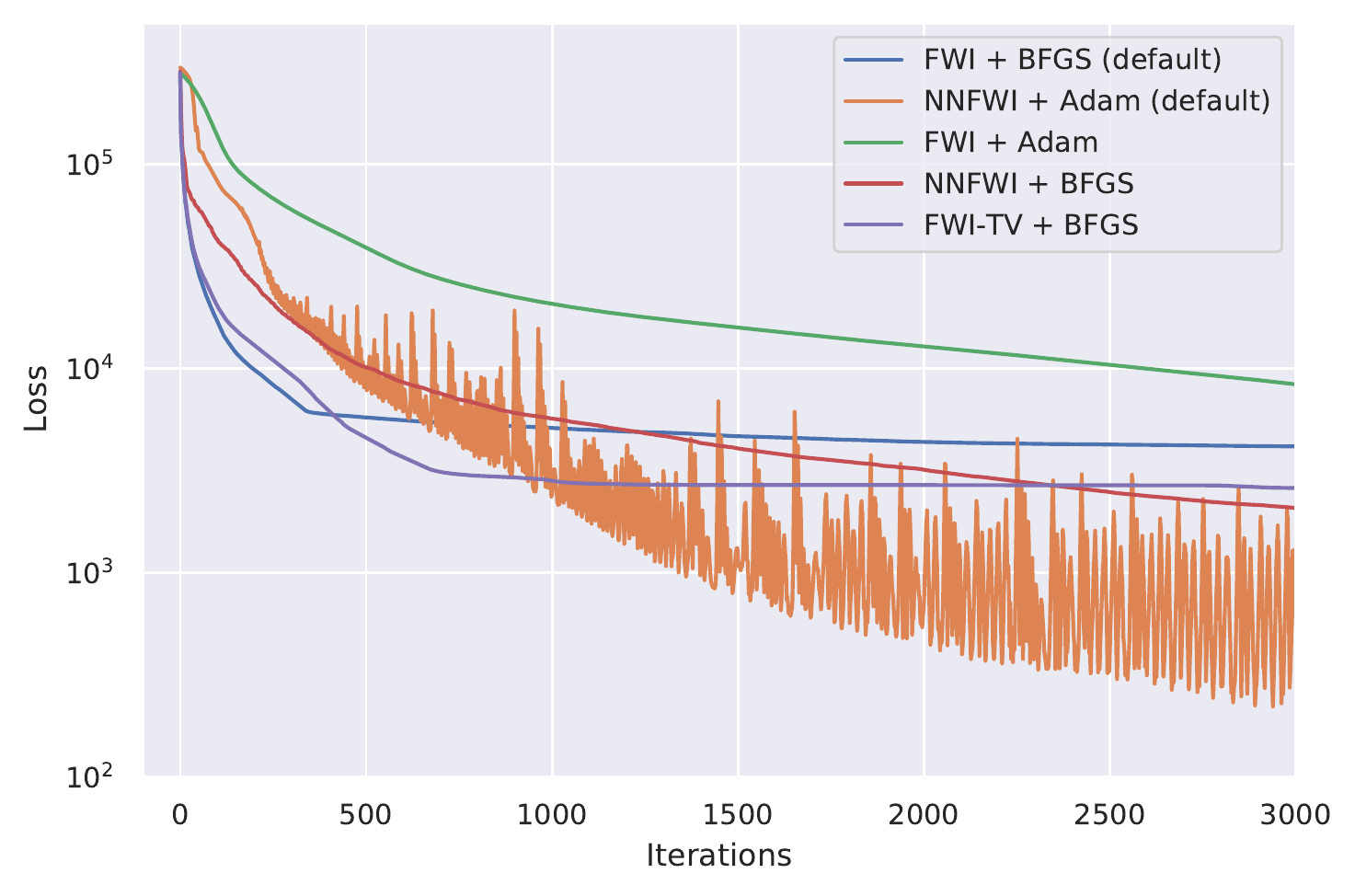}
    \caption{}
\end{subfigure}
\caption{Loss functions: (a) inversions with no random noise of the Marmousi model; (b) inversions with random noise ($\sigma = 0.5\sigma_0$) of the Marmousi model; (c) inversions of NNFWI with different number of convolutional layers; (d) inversion with no random noise of the 2004 BP model (explained in \Cref{sec:bp2004}). (c) is also based on the Marmousi model and the same loss functions based on the 2004 BP model is shown in \Cref{fig:loss_BP_layers}. We use the L-BFGS method for conventional FWI and the Adam method for NNFWI as the default settings in this study.}
\label{fig:loss_marmousi}
\end{figure}

\Cref{fig:result_marmousi} shows the inversion results at different noise levels. We can observe the results of conventional FWI significantly deteriorate with the increase of noise resulting in strong spurious fluctuations in the model. In contrast, the results of NNFWI show much less deterioration. The extracted velocity profiles along depth in \Cref{fig:profile_marmousi} show that NNFWI can estimate accurate velocity profiles with both clean and noisy data. We further analyzed the inversion quality through three metrics: MSE (mean square error), PSNR (peak signal-to-noise ratio), and SSIM (structural similarity index measure) \citep{hore2010image} in \Cref{tab:marmousi_noise}. The metric scores confirm the improved inversion results of NNFWI on noisy data compared with conventional FWI. This comparison demonstrates the regularization effect from the generative neural network with a deep image prior \citep{ulyanovDeep2018}. The inversion results of NNFWI become more smooth and robust compared with those of conventional FWI in the presence of noise. \Cref{fig:loss_marmousi}a and b show the change of loss functions of the two methods. Because of the regularization effect of NNFWI, its loss (labeled as ``NNFWI + Adam") is higher than that of conventional FWI (labeled as ``FWI + BFGS") (\Cref{fig:loss_marmousi}a) and the resolution is also a bit lower than conventional FWI (\Cref{fig:result_marmousi}a and b) for data with no random noise. However, for data with 0.5$\sigma_0$ random noise, both methods converge to a similar loss (\Cref{fig:loss_marmousi}b), but NNFWI gives a much better recovery of the true model (\Cref{fig:result_BP}c and d).
The different intermediate inversion results of conventional FWI and NNFWI at 30\%, 10\% and 3\% of the initial loss can be found in \Cref{fig:updates_marmousi}.

The default architecture of the generative neural network is explained in \Cref{tab:nn}. In this work we focus on analyzing the effects from re-parametrization instead of specific architectures, so we used a basic architecture with fully-connected layers, convolutional layers, and activation functions of tanh and leaky rectified linear unit (ReLU) \citep{maas2013rectifier}. Tuning the network hyper-parameters and using more advanced architectures such as ResNet \citep{he2016deep} may further improve the inversion results. We use 4 upsampling layers of linear interpolation to scale up the latent vector to the size of the velocity model in the experiments above. Meanwhile, we also analyzed architectures using 2 and 3 upsampling layers. The shallow architectures have faster convergence rates (\Cref{fig:loss_marmousi}(c)) and slightly weaker regularization effect (\Cref{tab:marmousi_layers}).
In addition to inversions using Adam for NNFWI and L-BFGS for conventional FWI, we have also tested inversions using L-BFGS for NNFWI and Adam for conventional FWI. The loss functions are also plotted in \Cref{fig:loss_marmousi}a and the metrics scores are listed in \Cref{tab:bfgs_adam}. Both L-BFGS and Adam methods can work well for conventional FWI and NNFWI (\Cref{fig:result_bfgs_adam}). L-BFGS converges faster than Adam for conventional FWI. This is because Adam is a stochastic gradient descent method with a fixed learning rate strategy, while L-BFGS is a quasi-Newton method with an adaptive learning rate determined by line search. However, Adam proves to be more effective for NNFWI due to the optimization in a high dimensional parameter space of deep neural networks. Thus, we choose the L-BFGS method for conventional FWI and the Adam method for NNFWI as the default setting in this study.

\subsection{2004 BP model} \label{sec:bp2004}
\begin{figure}[!ht]
\centering
\begin{subfigure}{0.48\textwidth}
    \centering
    \includegraphics[width=\textwidth]{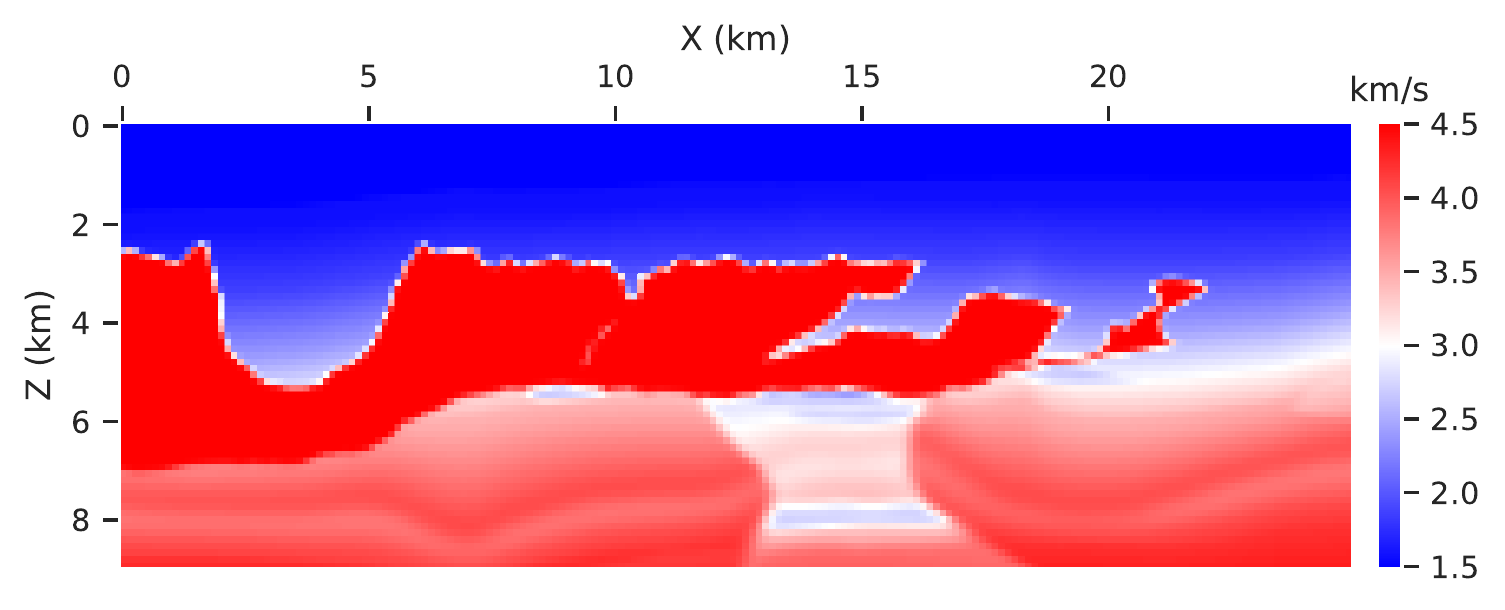}
    \caption{}
\end{subfigure}
\begin{subfigure}{0.48\textwidth}
    \centering
    \includegraphics[width=\textwidth]{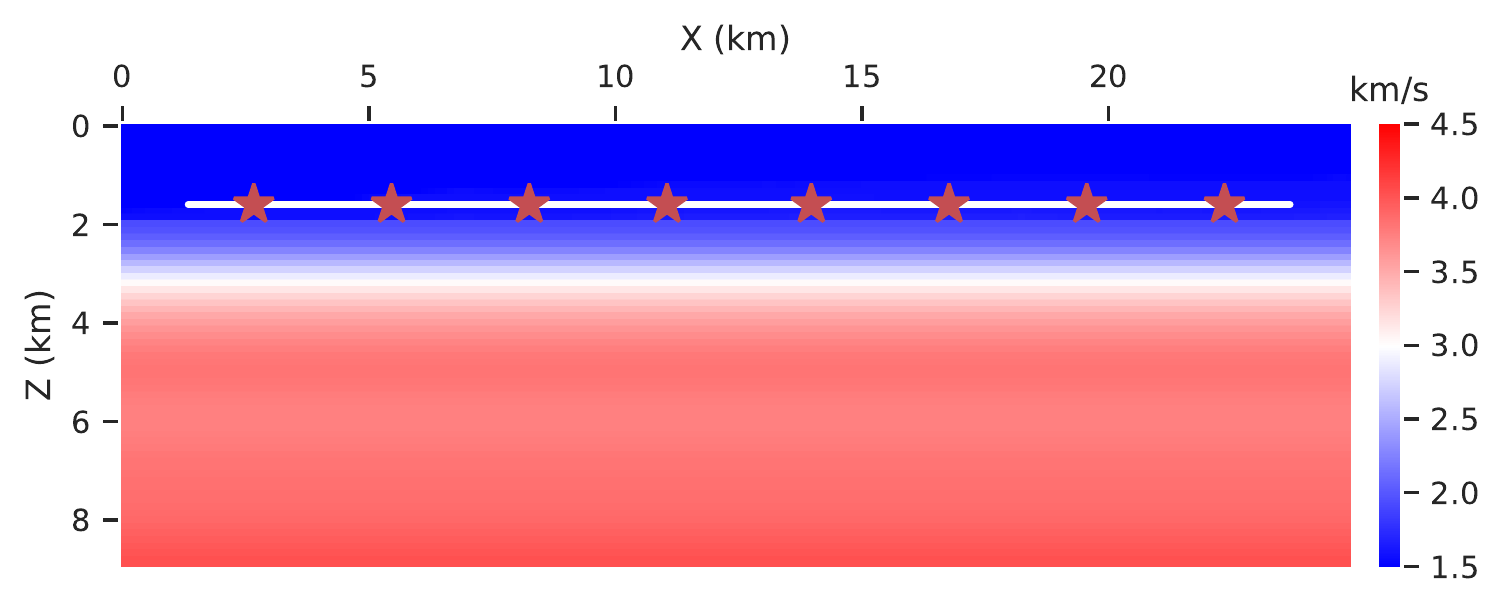}
    \caption{}
\end{subfigure}
\caption{The 2004 BP benchmark model: (a) the ground-truth model for generating synthetic data; (b) the 1D smoothed initial model for inversion. The source locations (red stars) and the receiver depth (while line) are plotted in (b).}
\label{fig:BP}
\end{figure}

\begin{figure}[!ht]
\centering
\begin{subfigure}{0.48\textwidth}
    \centering
    \includegraphics[width=\textwidth]{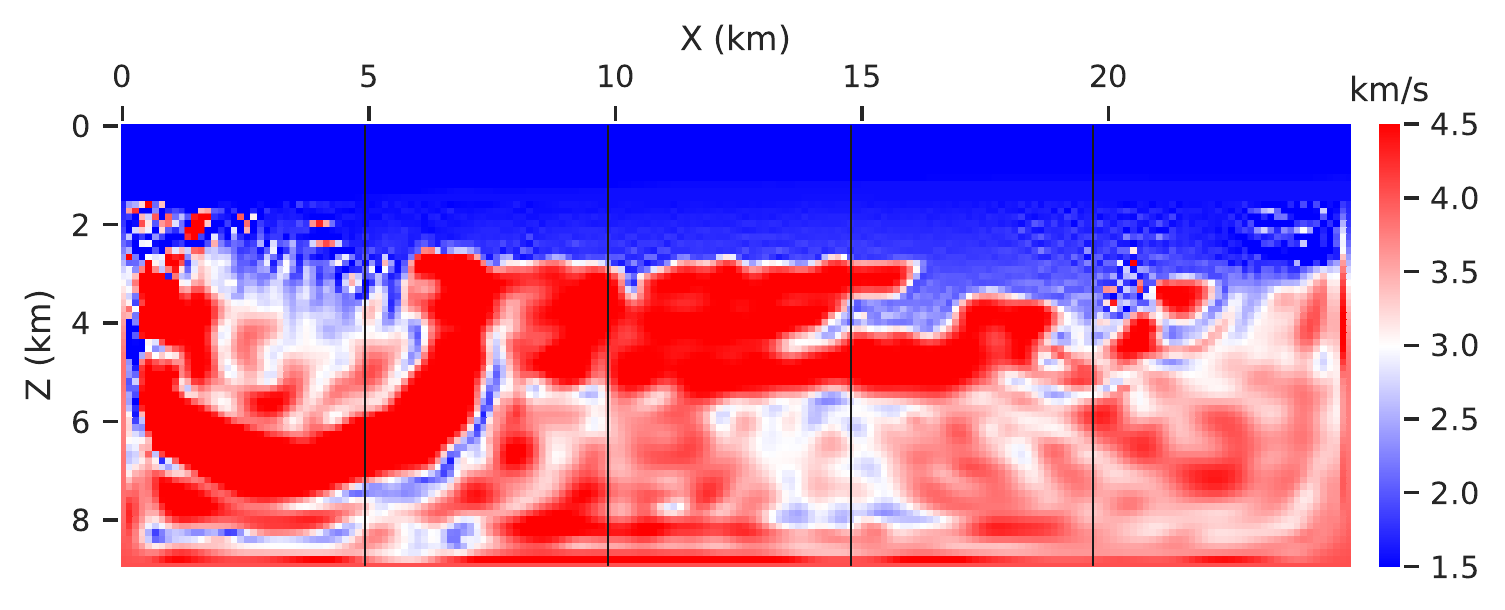}
    \caption{}
\end{subfigure}
\begin{subfigure}{0.48\textwidth}
    \centering
    \includegraphics[width=\textwidth]{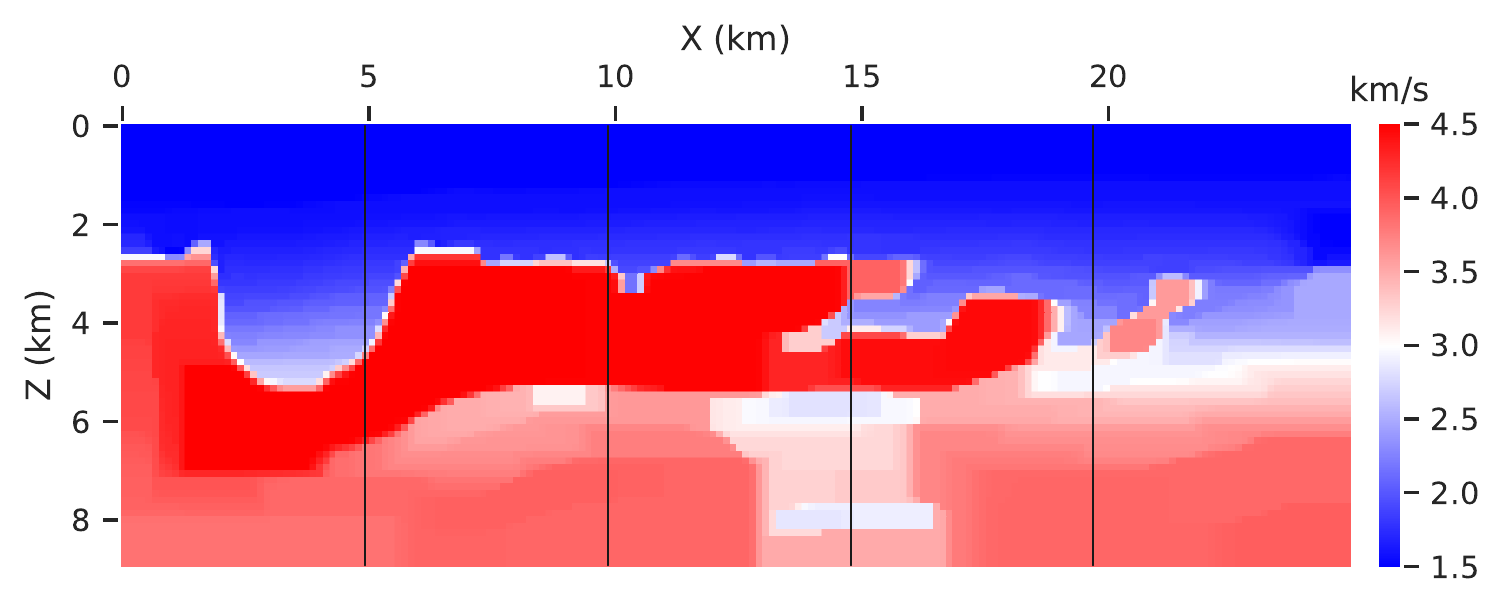}
    \caption{}
\end{subfigure}
\begin{subfigure}{0.48\textwidth}
    \centering
    \includegraphics[width=\textwidth]{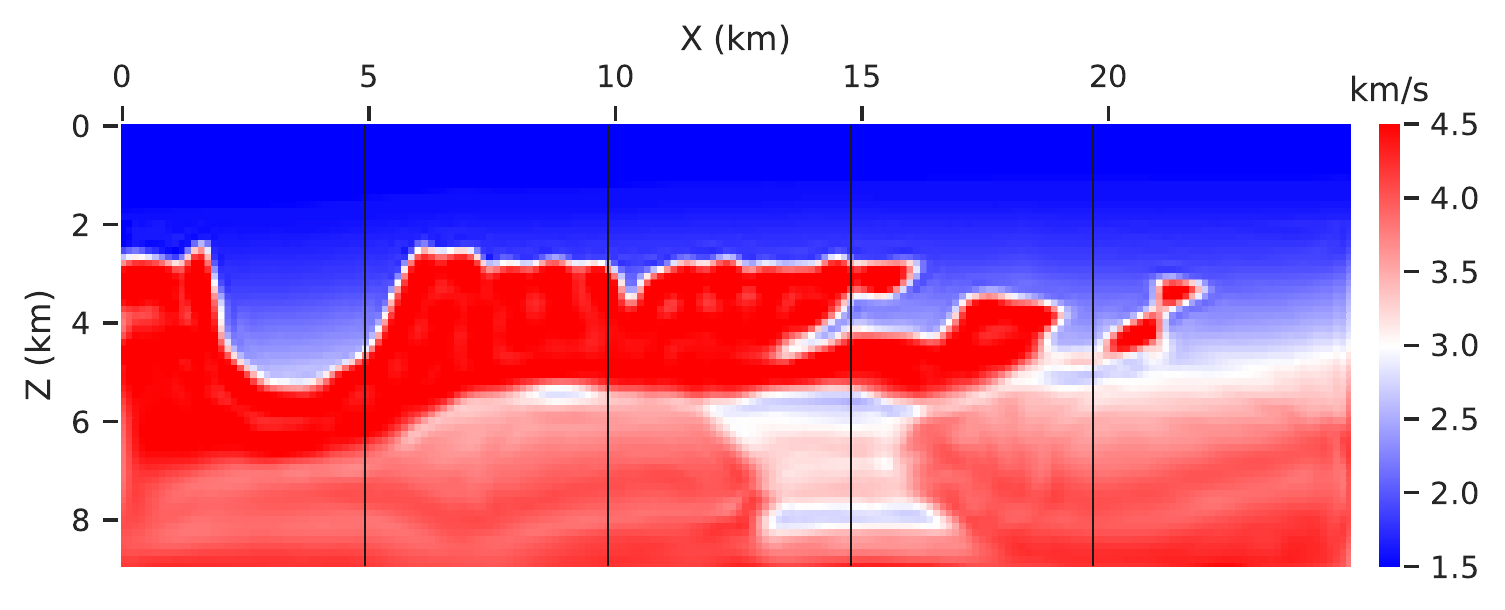}
    \caption{}
\end{subfigure}
\begin{subfigure}{0.48\textwidth}
    \centering
    \includegraphics[width=\textwidth]{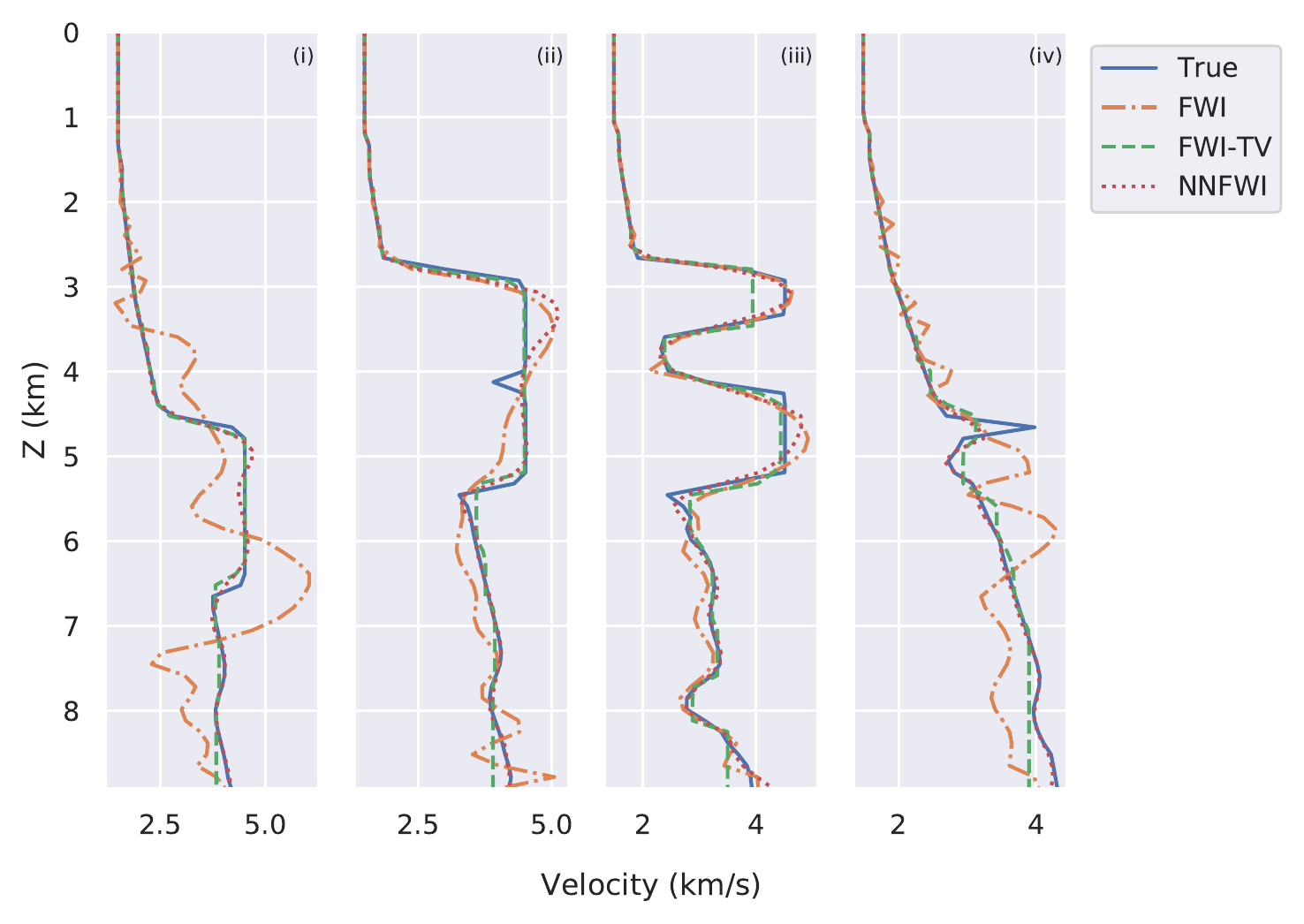}
    \caption{}
\end{subfigure}
\caption{Inversion results of the 2004 BP benchmark model: (a) conventional FWI; (b) FWI with TV (total variation) regularization ($\gamma=10^{-3}$); (c) NNFWI; (d) velocity profiles at four locations that are marked by the black vertical lines in (a, b, c). The other two inversion results with two different TV regularization coefficients of $\gamma=10^{-2}$ and $\gamma=10^{-4}$  are shown in \Cref{fig:BP_TV_2_4}.} 
\label{fig:result_BP}
\end{figure}

\begin{figure}[!ht]
\centering
\begin{subfigure}{0.48\textwidth}
    \centering
    \includegraphics[width=\textwidth]{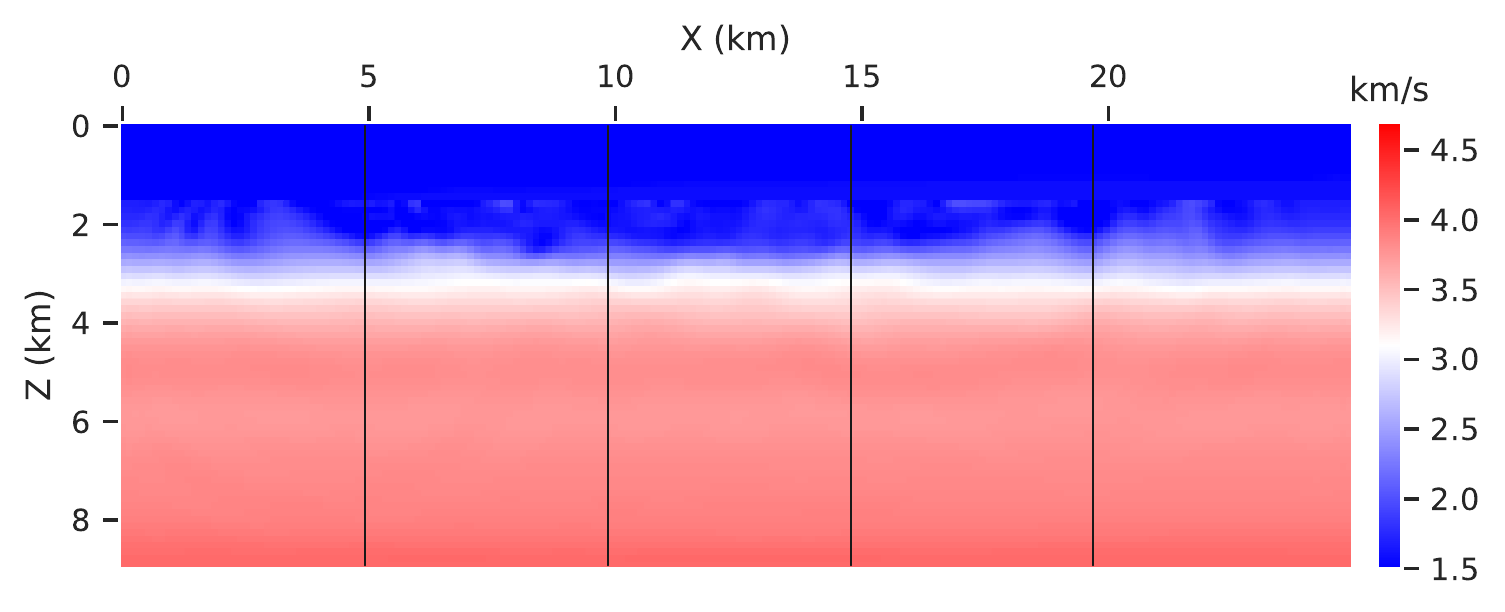}
    \caption{}
\end{subfigure}
\begin{subfigure}{0.48\textwidth}
    \centering
    \includegraphics[width=\textwidth]{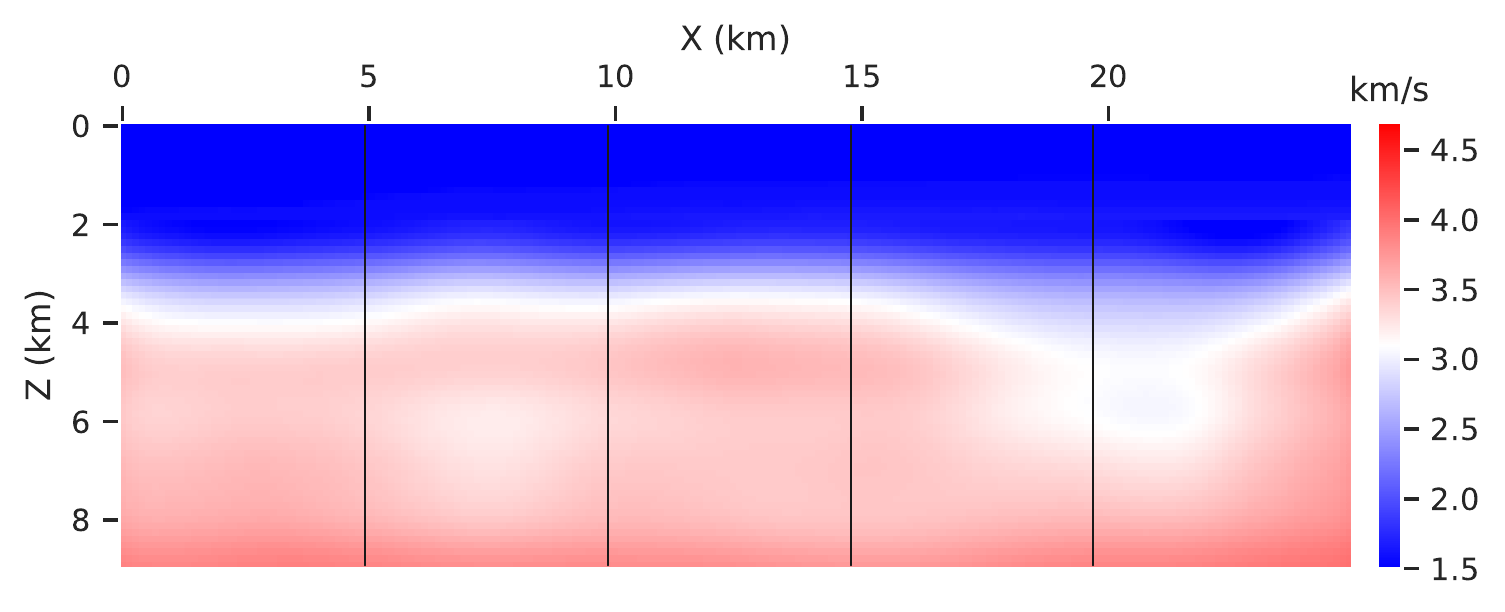}
    \caption{}
\end{subfigure}
\begin{subfigure}{0.48\textwidth}
    \centering
    \includegraphics[width=\textwidth]{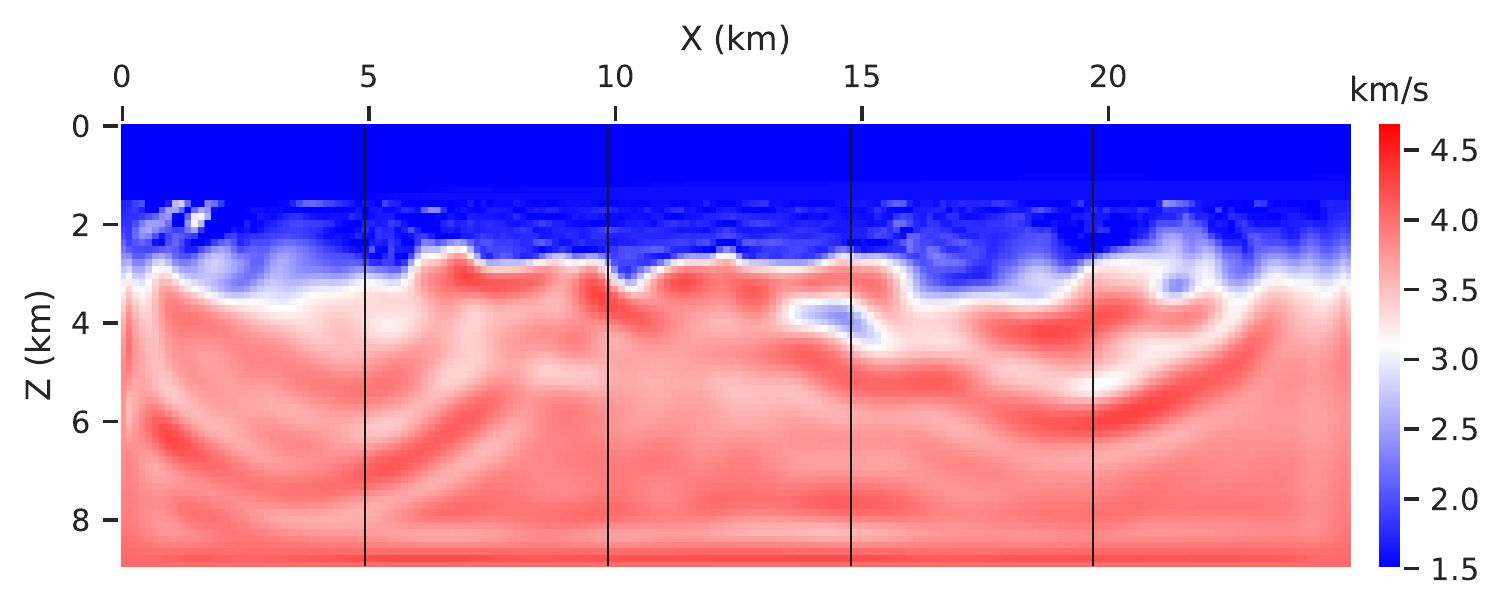}
    \caption{}
\end{subfigure}
\begin{subfigure}{0.48\textwidth}
    \centering
    \includegraphics[width=\textwidth]{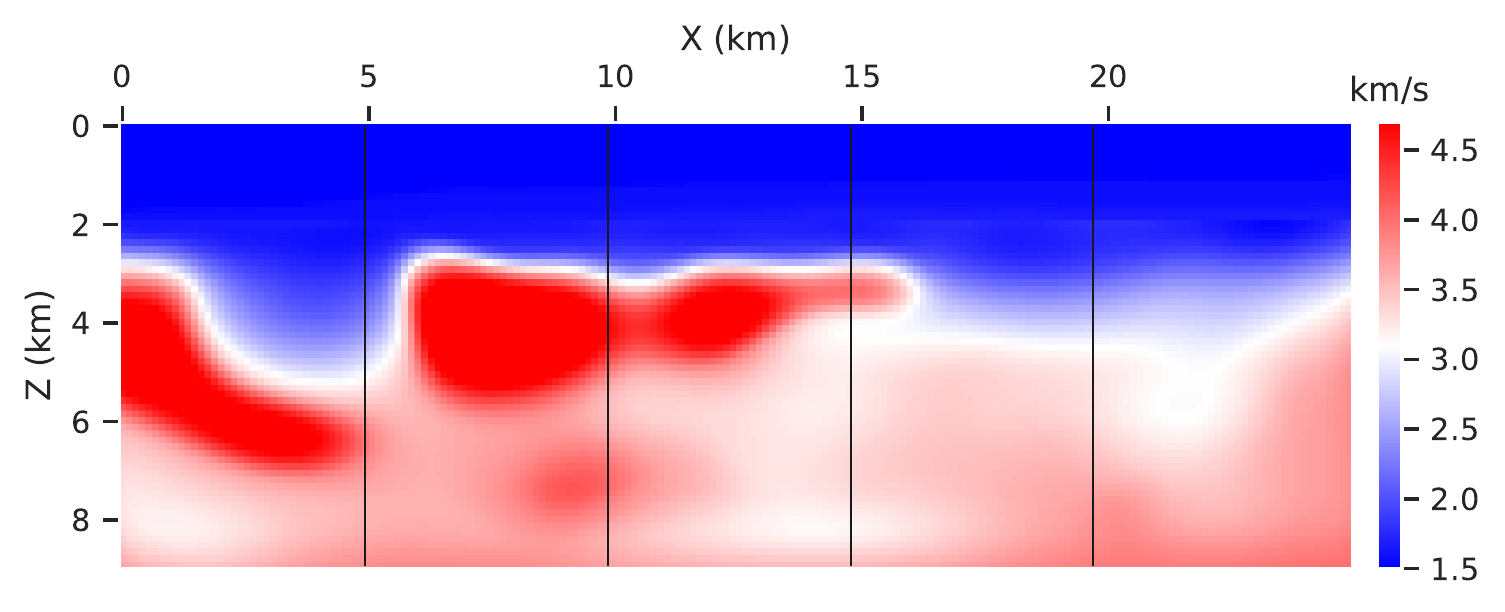}
    \caption{}
\end{subfigure}
\begin{subfigure}{0.48\textwidth}
    \centering
    \includegraphics[width=\textwidth]{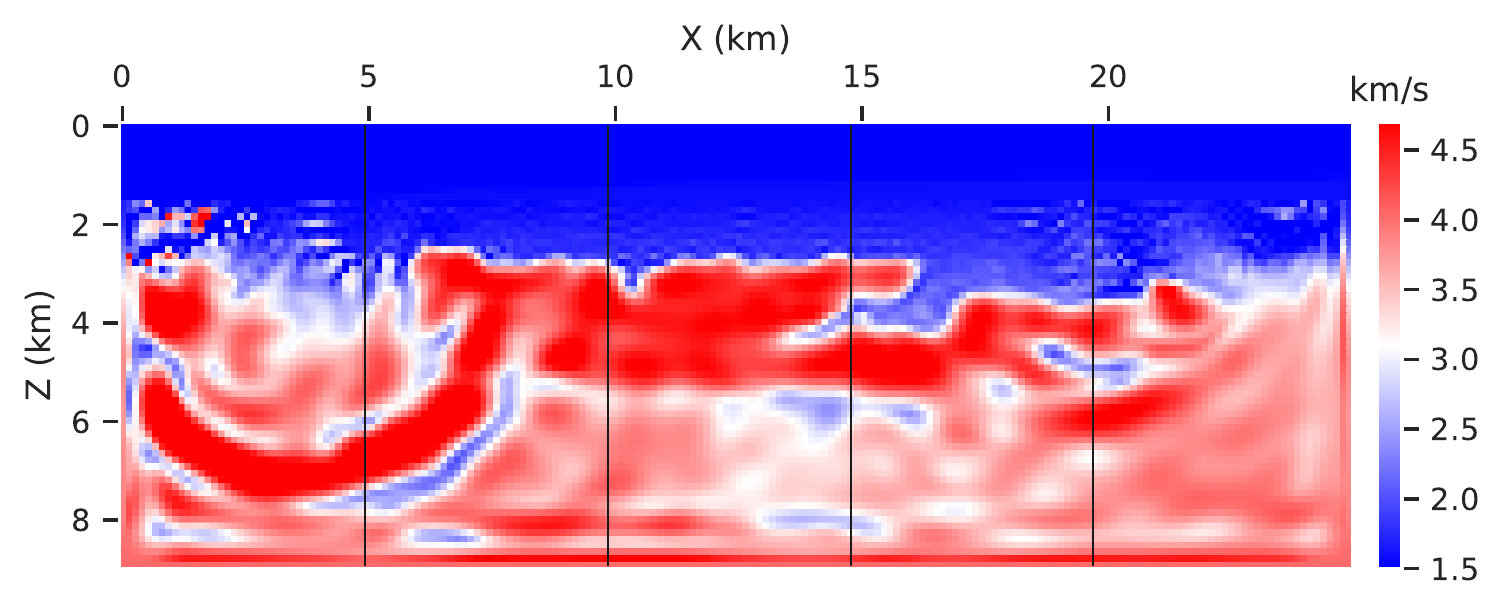}
    \caption{}
\end{subfigure}
\begin{subfigure}{0.48\textwidth}
    \centering
    \includegraphics[width=\textwidth]{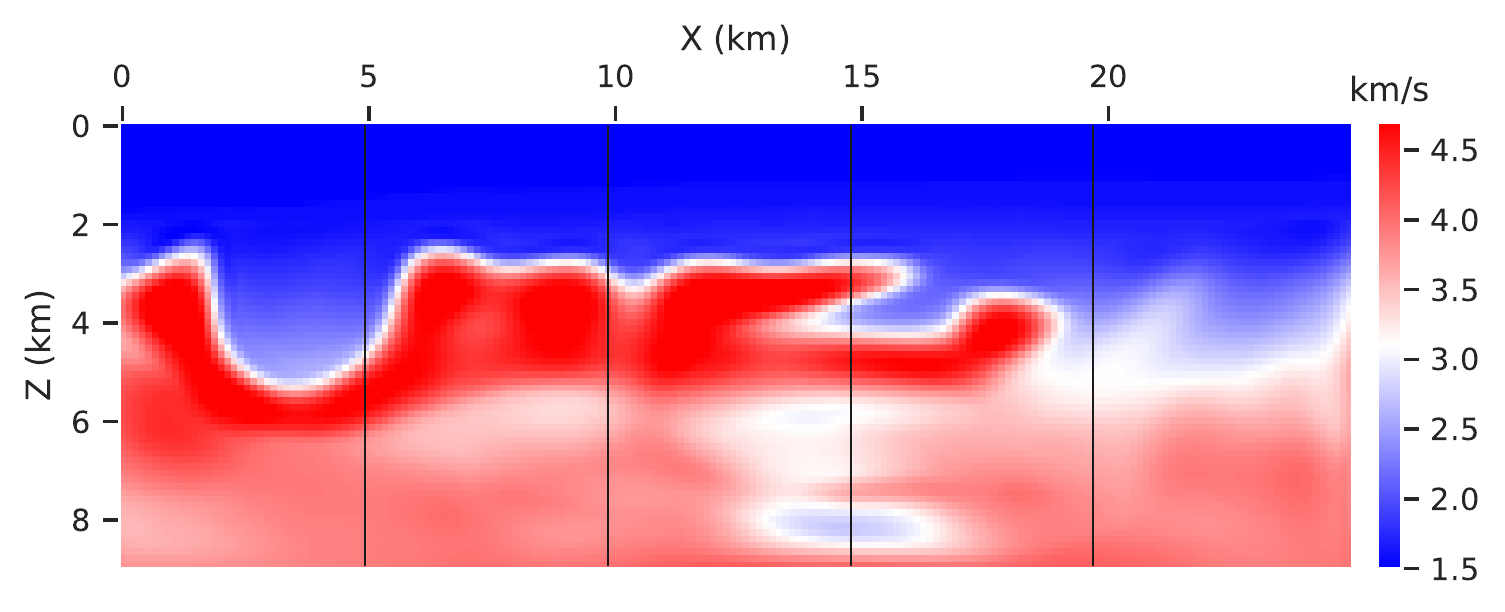}
    \caption{}
\end{subfigure}
\caption{Intermediate inversion results of the 2004 BP model: The left panels (a, c, e) show results of conventional FWI; The right panels (b, d, f) show results of NNFWI. We plot three estimated velocity maps at 30\%, 10\%, and 3\% of the initial loss to show the different optimization processes of conventional FWI and NNFWI. Because of the different convergence speeds (\Cref{fig:loss_marmousi}), the iteration numbers are 10, 50, and 240 for FWI in the left panels and 90, 230, 570 for NNFWI in the right panels. The intermediate inversion results of FWI with TV regularization are plotted in \Cref{fig:updates_BP_TV}. The intermediate inversion results of the Marmousi model are plotted in \Cref{fig:updates_marmousi}.} 
\label{fig:updates_BP}
\end{figure}

The 2004 BP benchmark model \citep{billette20042005} contains complex salt bodies, which are difficult imaging targets for FWI because of cycle-skipping and amplitude discrepancies caused by sharp contrasts and large-scale salt bodies \citep{zhang2018correcting}. We extract the left part of the original 2004 BP benchmark model (\Cref{fig:BP}), which is based on a geological cross section through the Western Gulf of Mexico. We use the default Ricker wavelet with a peak frequency of 1.2 Hz for this case. The source spacing is 2.8 km and the receiver spacing is 133 m. Conventional FWI recovers the shallow central portion of the salt body but fails to resolve the U-shaped structure in the left region and the low velocity layers below the salt body \Cref{fig:result_BP}a). Regularization, such as total variation, is needed to skip local minima and estimate the entire shape of the salt body \citep{esser2018total}. However, a proper regularisation weight needs to be determined by trial and error. We tested three regularization weights ($\gamma$) of total variation and plotted the best inversion result with ($\gamma=10^{-3}$) in \Cref{fig:result_BP}a. The inversion results with a too strong regularization effect ($\gamma=10^{-2}$) and a too weak regularization effect ($\gamma=10^{-4}$) can be found in \Cref{fig:BP_TV_2_4}. In contrast, NNFWI can correctly image the salt body without regularization, especially the left U-shaped target (\Cref{fig:result_BP}c). The loss function in \Cref{fig:loss_marmousi}d shows that the loss of NNFWI is not trapped by local minima and continues to decrease to a lower value than conventional FWI and FWI with total variation (labeled as ``FWI-TV"). The metrics in \Cref{tab:bp2004} show that NNFWI achieves the best inversion quality. Same as the Marmousi case, we also analyzed the performances with 2 and 3 upsampling layers (\Cref{tab:bp2004} and \Cref{fig:loss_BP_layers}) and using both Adam and L-BFGS optimization (\Cref{tab:bfgs_adam} and \Cref{fig:result_bfgs_adam}). We observe the same results that Adam has a better performance for NNFWI and shallower architectures have a slightly weaker regularization effect. To gain insight into the optimization trajectories, we plot the intermediate inversion results when the loss function is reduced to 30\%, 10\%, and 3\% of its initial value in \Cref{fig:updates_BP}. The intermediate inversion results of FWI with total variation can be found in \Cref{fig:updates_BP_TV}. The result of conventional FWI shows that most of the early updates focus on top shallow layers. NNFWI, in contrast, updates over a much broader depth range and recovers a smooth and approximate shape of the salt body. This difference demonstrates the spatial regularization from convolutional neural networks, which applies convolutional filters across the whole domain, so that the updates of convolutional kernels affect the entire generated velocity model.

\subsection{Uncertainty Quantification and Computational Analysis}
To analyze the uncertainty in the inversion results, which is a challenging task for conventional FWI, we conducted another experiment by adding dropout layers to the generative neural network (\Cref{tab:nn}). Dropout was initially proposed to prevent overfitting in neural networks by randomly setting a proportion of neurons and their connections to zero \citep{srivastava2014dropout}. \citet{gal2016dropout} demonstrated that dropout in neural networks can also be interpreted in a Bayesian framework to estimate model uncertainty. The dropout operation with a dropout ratio $q$ is the same as applying a binary sampling vector that follows a Bernoulli distribution with a probability $p=1 - q$ to the weight metrics in a neural network. They proved that a neural network with dropout applied before every weight layer mathematically approximates variational inference for a Gaussian process such that model uncertainty can be efficiently estimated by performing stochastic forward passes through the neural network and collecting statistics of the results. We ran 100 Monte Carlo samplings based on the optimized neural network model to calculate the standard deviation of the sampled velocity models. Because the sampling process only requires inference using the generative neural network, uncertainty estimation is orders of magnitude faster than other sampling-based methods that require solving the PDEs. 

\begin{figure}[!ht]
\centering
\begin{subfigure}{0.48\textwidth}
    \centering
    \includegraphics[width=\textwidth]{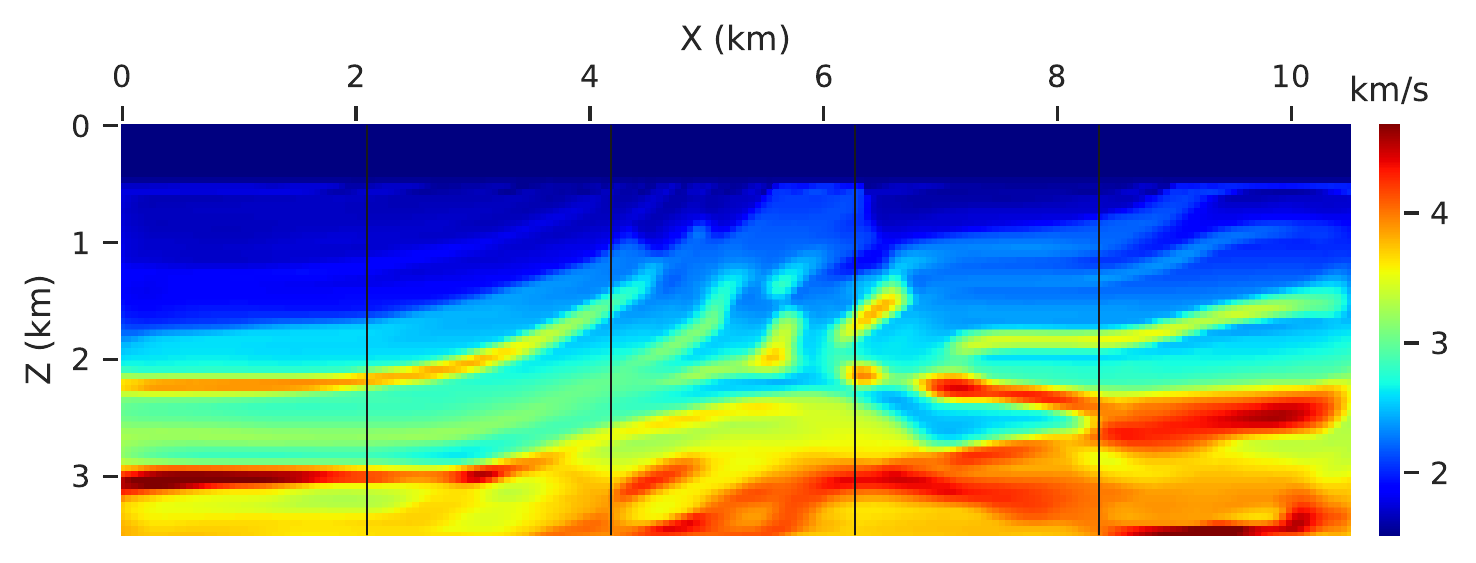}
    \caption{}
\end{subfigure}
\begin{subfigure}{0.48\textwidth}
    \centering
    \includegraphics[width=\textwidth]{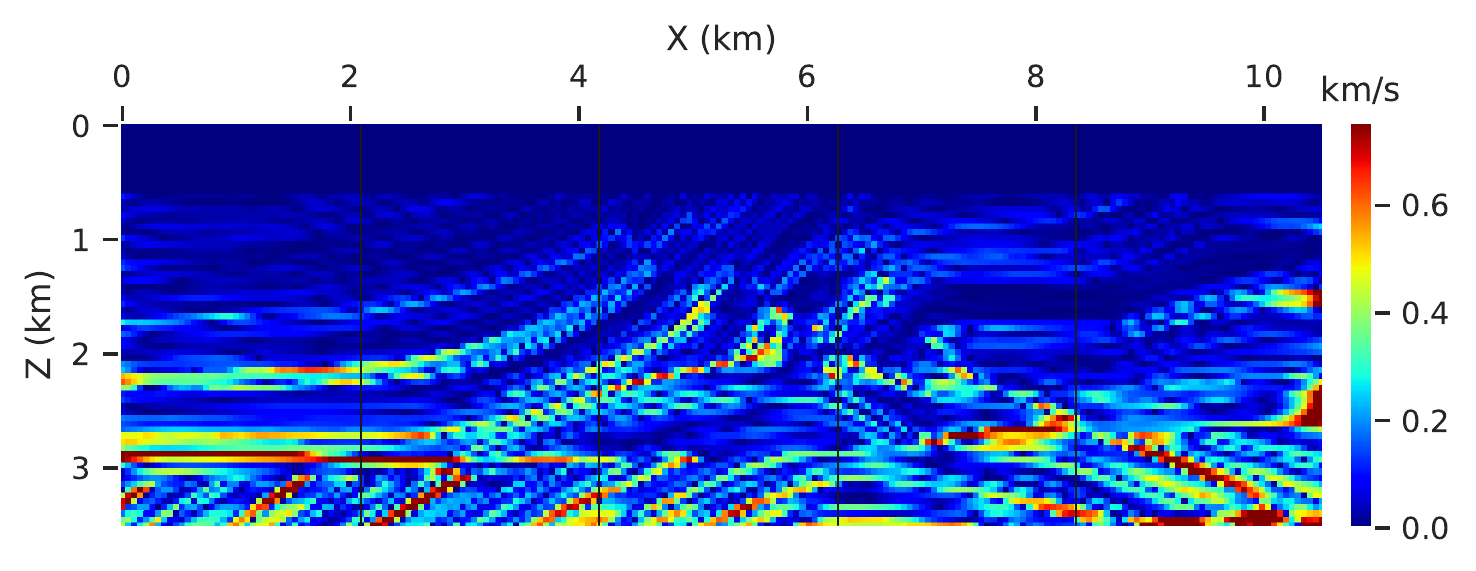}
    \caption{}
\end{subfigure}
\begin{subfigure}{0.48\textwidth}
    \centering
    \includegraphics[width=\textwidth]{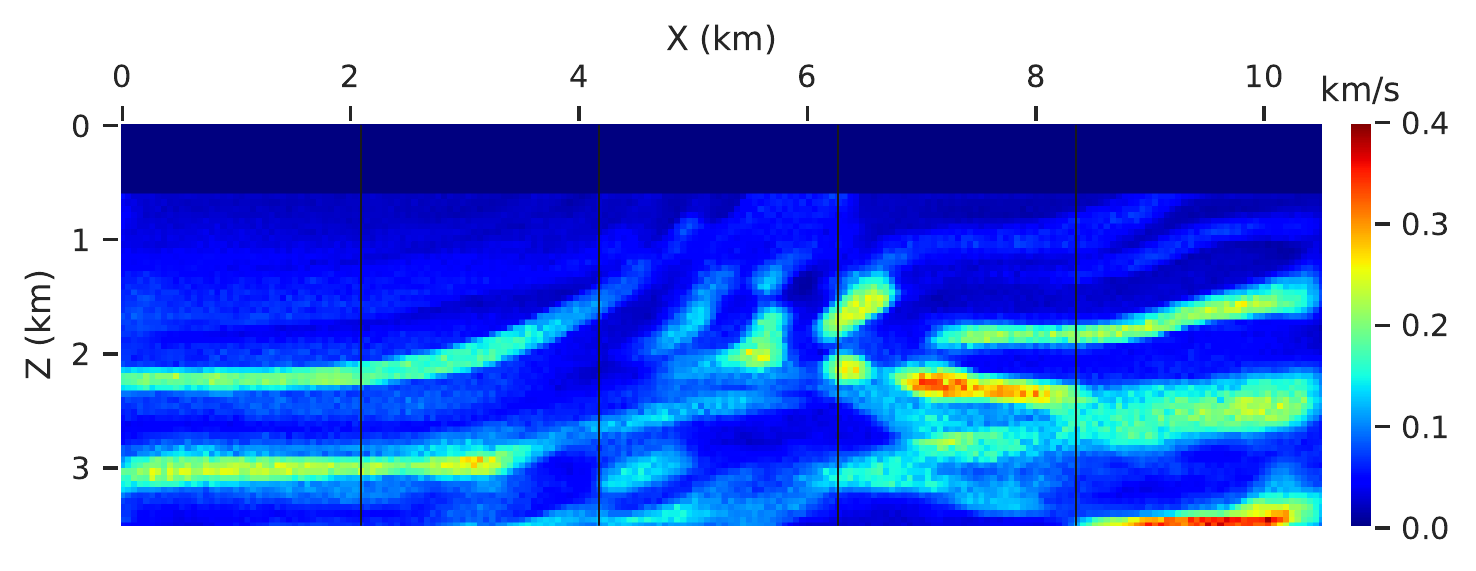}
    \caption{}
\end{subfigure}
\begin{subfigure}{0.48\textwidth}
    \centering
    \includegraphics[width=\textwidth]{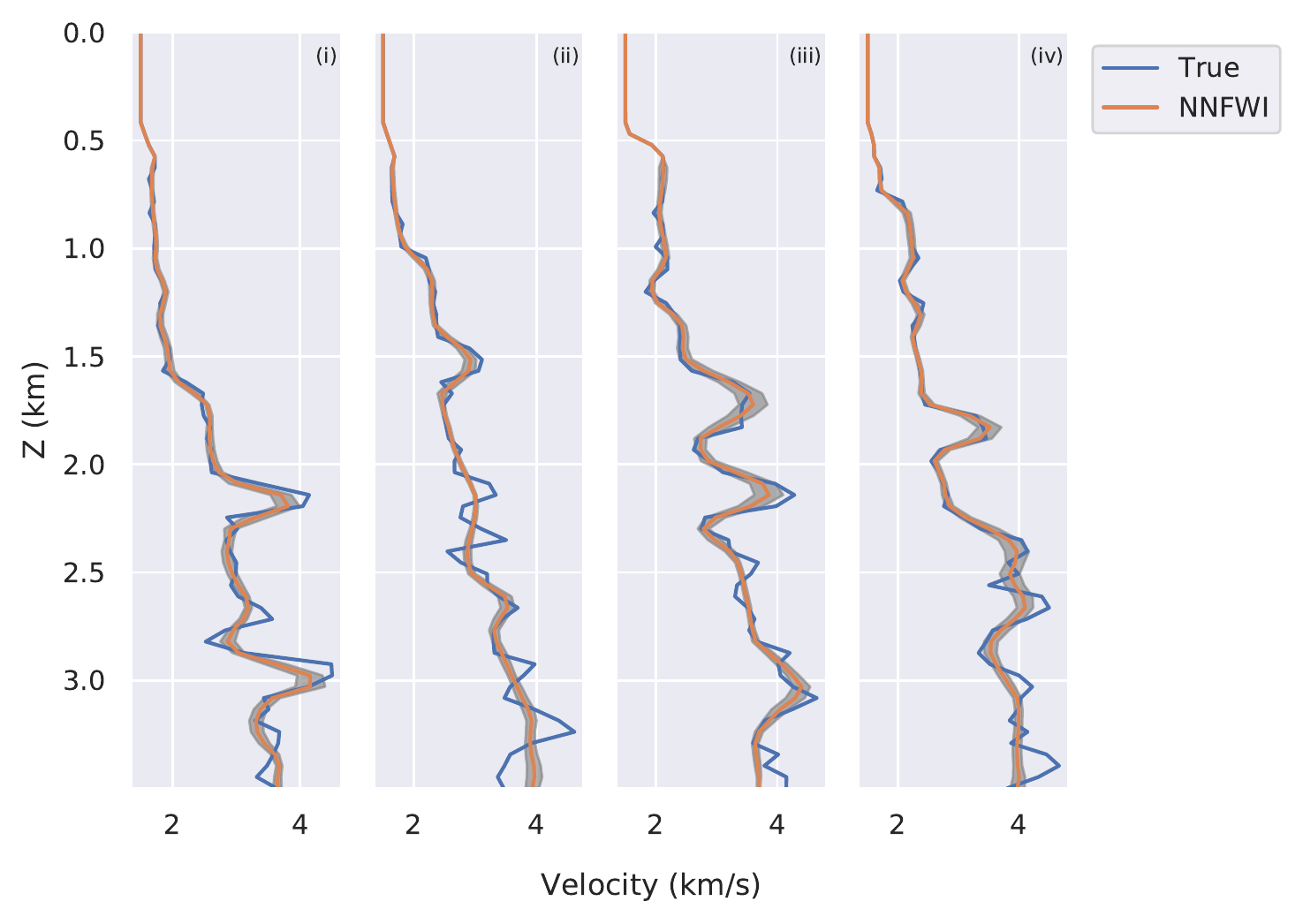}
    \caption{}
\end{subfigure}
\caption{Uncertainty quantification of NNFWI by adding dropout layers during training: (a) inversion result of the Marmousi model; (b) inversion error map; (c) estimated standard deviation calculated through Monte Carlo  sampling with a dropout rate of 0.1; (d) velocity profiles at four locations (marked by black lines in (a, b, c)). The standard deviation ranges are plotted in gray. The results of uncertainty quantification using three convolutional and dropout layers and using a dropout rate of 0.2 are shown in \Cref{fig:marmousi_UQ_3} and \Cref{fig:marmousi_UQ_dp2} respectively.}
\label{fig:marmousi_uq}
\end{figure}

\begin{figure}[!ht]
\centering
\begin{subfigure}{0.48\textwidth}
    \centering
    \includegraphics[width=\textwidth]{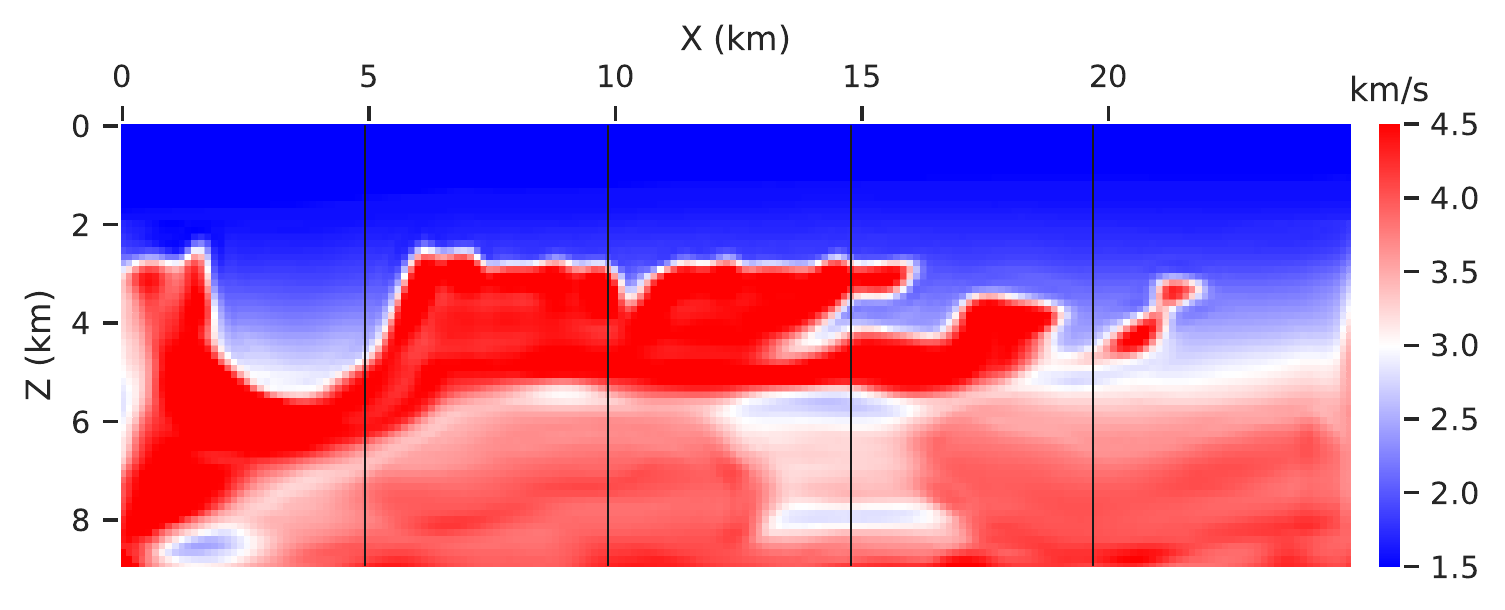}
    \caption{}
\end{subfigure}
\begin{subfigure}{0.48\textwidth}
    \centering
    \includegraphics[width=\textwidth]{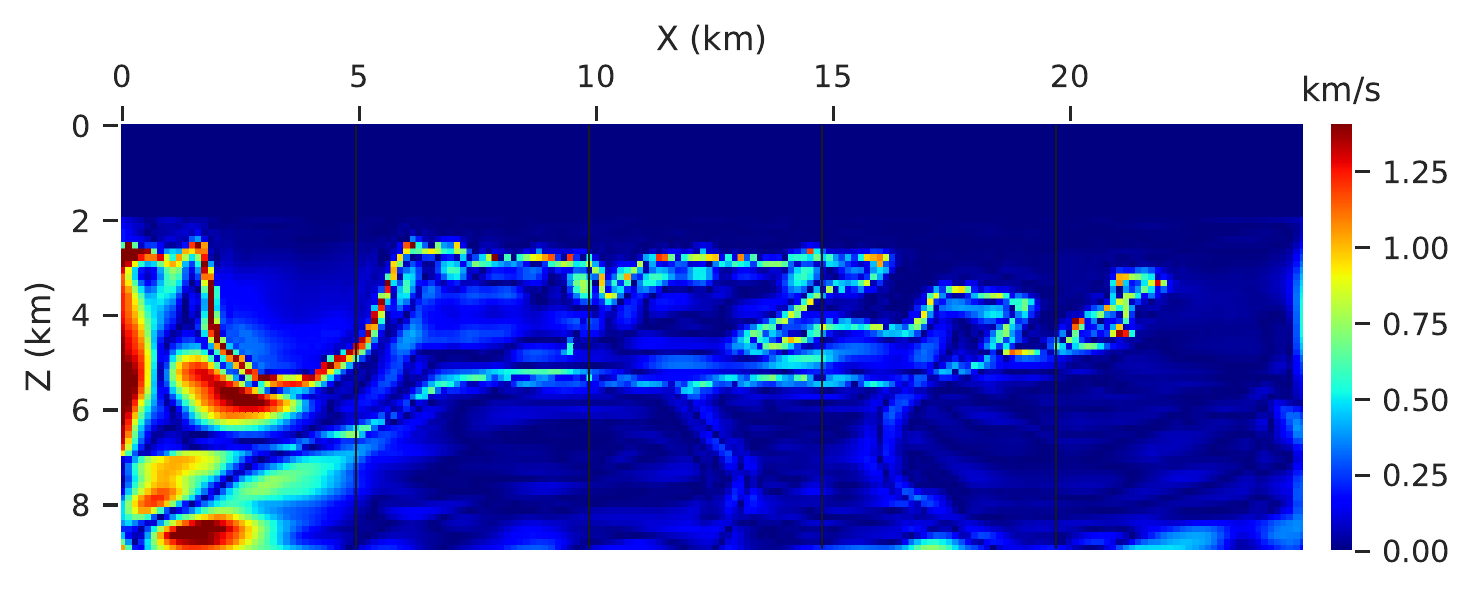}
    \caption{}
\end{subfigure}
\begin{subfigure}{0.48\textwidth}
    \centering
    \includegraphics[width=\textwidth]{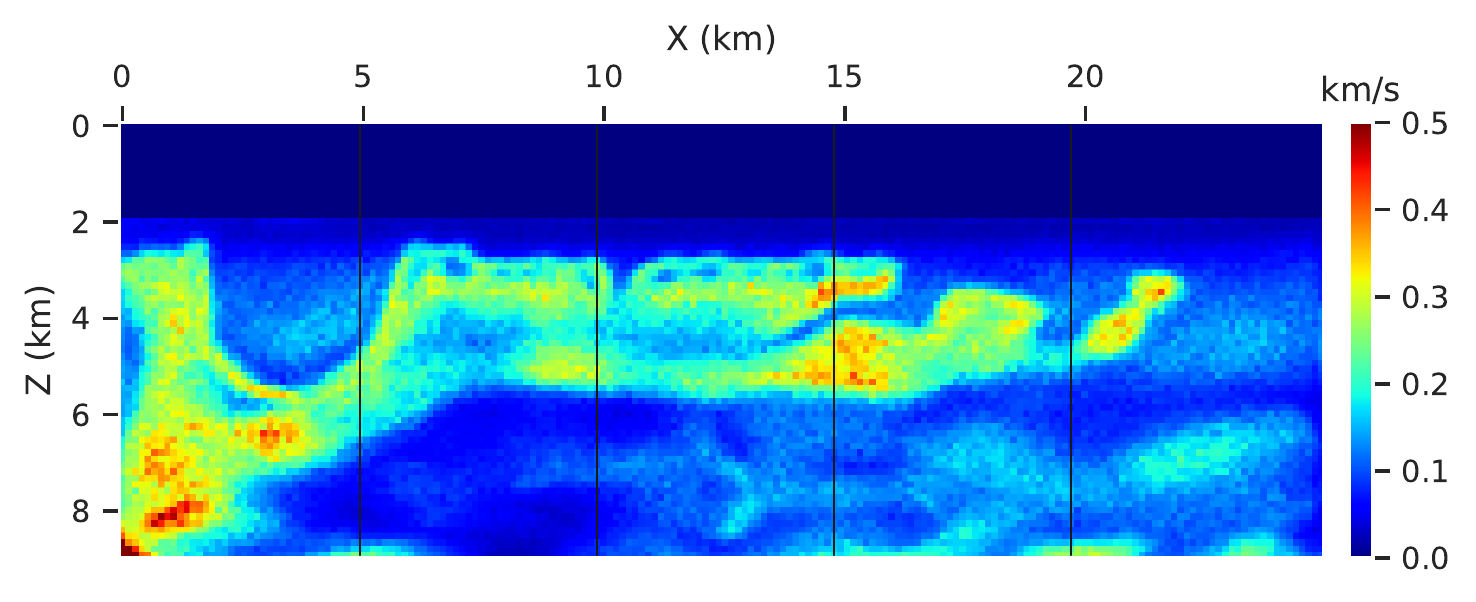}
    \caption{}
\end{subfigure}
\begin{subfigure}{0.48\textwidth}
    \centering
    \includegraphics[width=\textwidth]{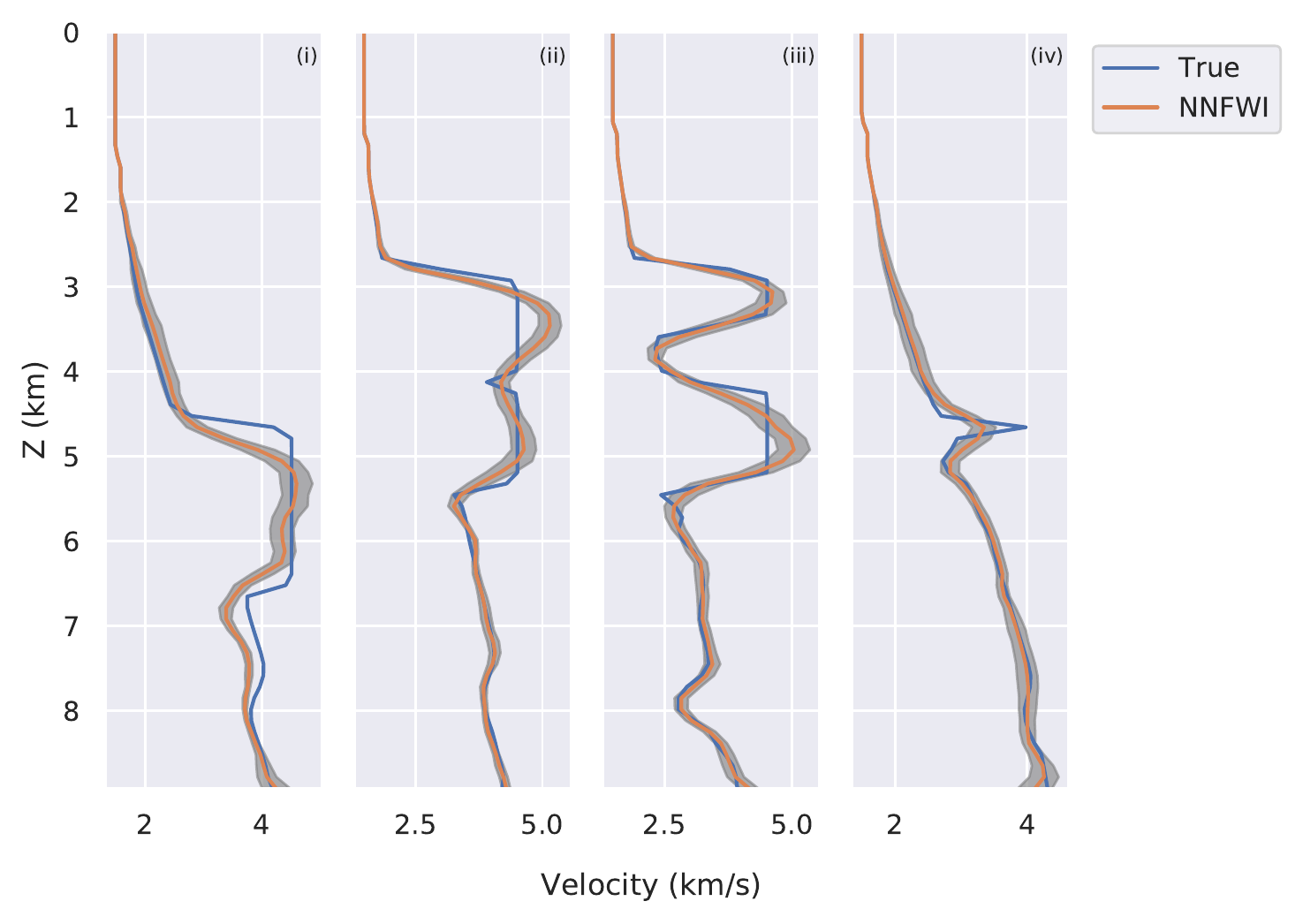}
    \caption{}
\end{subfigure}
\caption{Uncertainty quantification of NNFWI by adding dropout layers during training: (a) inversion result of the Marmousi model; (b) inversion error map; (c) estimated standard deviation through Monte Carlo sampling with a dropout rate of 0.1; (d) velocity profiles at four locations (marked by black lines in (a, b, c)). The results of uncertainty quantification using three convolutional and dropout layers and using a dropout rate of 0.2 are shown in \Cref{fig:BP_UQ_3} and \Cref{fig:BP_UQ_dp2} respectively.}
\label{fig:BP_uq}
\end{figure}

\Cref{fig:marmousi_uq}a and \Cref{fig:BP_uq}a show the estimated velocity models of the Marmousi benchmark model and the 2004 BP model respectively, which are similar to the inversion results in \Cref{fig:result_marmousi}b and \Cref{fig:result_BP}c. \Cref{fig:marmousi_uq}b and \Cref{fig:BP_uq}b show the absolute error between the estimated models and the true models; and \Cref{fig:marmousi_uq}c and \Cref{fig:BP_uq}c show the estimated uncertainty. The estimated velocity profiles and estimated uncertainty along depth are shown in \Cref{fig:marmousi_uq}d and \Cref{fig:BP_uq}d.
The estimated uncertainty by dropout in these two experiments does not exactly reproduce the inversion error but follows a similar trend with the error map. This is not unexpected, since the uncertainty is a statistical estimate, while the model error results from a single computation. 
To analyze if the estimated uncertainty is consistent using different numbers of network layers and dropout ratios, we added one example using 3 upsampling and dropout layers instead of the default 4 layers and another example with a dropout ratio of 20\% instead of the default 10\%. The results of the Marmousi model can be found in \Cref{fig:marmousi_UQ_3} and \Cref{fig:marmousi_UQ_dp2}; and the results of the 2004 BP model can be found in \Cref{fig:BP_UQ_3} and \Cref{fig:BP_UQ_dp2}. The overall estimated uncertainty maps are consistent with each other among these experiments. Meanwhile we can also observe that the estimated standard deviation is slightly smaller for the experiment with a smaller number of upsampling and dropout layers and slightly larger for the experiment with a higher dropout ratio.
Due to the lack of uncertainty benchmarks in FWI, further research is needed to verify the accuracy of uncertainty estimation by dropout used in NNFWI.

Finally, we address the computational demands of NNFWI. Based on the compute times per integration on both CPUs and GPUs in \Cref{tab:time}, NNFWI introduces negligible additional computation compared with conventional FWI. This is because generating the velocity model using neural networks is much less expensive than solving the PDEs. Moreover, the optimization of the neural network and PDEs are done simultaneously with automatic differentiation without additional optimization loops as used in dictionary learning \citep{zhuSparsepromoting2017}. The computational efficiency of neural networks allows the exploration of deeper and more complex neural network architectures in future research.

\begin{table}[]
\centering
\caption{Comparison of FWI and NNFWI under different noise levels based on the Marmousi model.}
\label{tab:marmousi_noise}
\resizebox{0.45\textwidth}{!}{%
\begin{tabular}{llll}
\hline
Experiment                 & MSE & SSIM  & PSNR \\ \hline
FWI (no noise)             & 114 & 0.993 & 49.5 \\
FWI (0.5$\sigma$ noise)    & 568 & 0.851 & 42.6 \\
FWI (1.0$\sigma$ noise)    & 610 & 0.837 & 42.4 \\
NNFWI (no noise)           & 152 & 0.987 & 48.2 \\
NNFWI (0.5$\sigma$ noise)  & 171 & 0.983 & 47.8 \\
NNFWI (1.0$\sigma$ noise)  & 233 & 0.969 & 46.4 \\ \hline
\end{tabular}%
}
\end{table}

\begin{table}[]
\centering
\caption{Comparison of FWI and NNFWI with different numbers of convolutional layers based on the Marmousi model.}
\label{tab:marmousi_layers}
\resizebox{0.45\textwidth}{!}{%
\begin{tabular}{llll}
\hline
Experiment       & MSE & SSIM  & PSNR \\ \hline
FWI              & 114 & 0.993 & 49.5 \\
NNFWI (4 layers) & 152 & 0.987 & 48.2 \\ 
NNFWI (3 layers) & 147 & 0.988 & 48.4 \\ 
NNFWI (2 layers) & 146 & 0.988 & 48.5 \\ \hline
\end{tabular}%
}
\end{table}

\begin{table}[]
\centering
\caption{Comparison among FWI, FWI with total variation regularization, and NNFWI with different numbers of convolutional layers based on the 2004 BP model}
\label{tab:bp2004}
\resizebox{0.45\textwidth}{!}{%
\begin{tabular}{llll}
\hline
Experiment             & MSE          & SSIM           & PSNR          \\ \hline
FWI                    & 622          & 0.866          & 41.7          \\
FWI-TV ($\gamma=10^{-4}$) & 568          & 0.887          & 42.1          \\
FWI-TV ($\gamma=10^{-3}$) & 181          & 0.988          & 47.0          \\
FWI-TV ($\gamma=10^{-2}$) & 494          & 0.901          & 42.7          \\ 
NNFWI (4 layers)       & 134          & 0.993          & 48.3          \\ 
NNFWI (3 layers)       & 209          & 0.984          & 46.4          \\
NNFWI (2 layers)       & 258          & 0.976          & 45.5          \\ \hline
\end{tabular}%
}
\end{table}

\begin{table}[]
\centering
\caption{Comparison of FWI and NNFWI using different optimization algorithms.}
\label{tab:bfgs_adam}
\resizebox{0.54\textwidth}{!}{%
\begin{tabular}{lllll}
\hline
Model & Experiment   & MSE  & SSIM   & PSNR \\ \hline
\multirow{4}{*}{\begin{tabular}[c]{@{}l@{}}Marmousi\\ Model\end{tabular}} & FWI + BFGS & 114 & 0.993 & 49.5 \\
      & FWI + Adam   & 204  & 0.976  & 47.0 \\
      & NNFWI + BFGS & 193  & 0.978  & 47.2
      \\
      & NNFWI + Adam & 152  & 0.987  & 48.2 
      \\ \hline
\multirow{4}{*}{\begin{tabular}[c]{@{}l@{}}2004 BP\\ Model\end{tabular}}  & FWI + BFGS & 622 & 0.866 & 41.7 \\
      & FWI + Adam   & 734  & 0.827  & 41.0 \\
      & NNFWI + BFGS & 225  & 0.981  & 46.1 \\
      & NNFWI + Adam & 134. & 0.993  & 48.3 \\ \hline
\end{tabular}%
}
\end{table}

{\setlength{\tabcolsep}{18pt}
\begin{table}[]
\centering
\caption{The compute time per iteration. The CPU time is averaged over 100 iterations running on a 40-core CPU server with Intel(R) Xeon(R) CPU E5-2698 v4 @ 2.20GHz. The GPU time is averaged over 100 iterations running on a 8 GPU server with Nvidia Tesla V100-SXM2-32GB-LS.}
\label{tab:time}
\begin{tabular}{l|ll}
\hline
Computing time    & CPU (s)  & GPU (s) \\ \hline
Conventional FWI  & 6.25     & 1.91    \\
NNFWI             & 6.33     & 1.92    \\ \hline
\end{tabular}
\end{table}
}

\section{Discussion}

NNFWI combines deep neural networks and PDEs for FWI. Compared with data-driven inversion methods relying on only neural networks for direct inverse prediction, NNFWI can use both the properties of neural networks and the physical information represented by PDEs such that its accuracy is similar to that of conventional FWI. Moreover, direct data-driven inverse prediction relies on training on a large number of data pairs of velocity models and corresponding seismic data, which is not always available, so it may be susceptible to poor generalization when data does not follow a similar distribution to the training data. NNFWI does not need extra datasets for training the neural network. It applies to the same data settings as conventional FWI. Additionally, NNFWI adds little extra computational cost compared with conventional FWI because the computational cost of deep neural networks is much less than PDEs. The built-in GPU acceleration of deep learning frameworks further speeds up the simulation and optimization processes of NNFWI \citep{zhuGeneral2020}.
NNFWI can also be combined with conventional regularization methods in FWI to improve inversion results. For example, we can add the total variation of the generated velocity model $\mathcal{N}(z, w)$ or add the $\mathcal{L}_2$-norm of the weights $w$ of neural networks to the loss function in \Cref{eqn:loss} as regularization.

In conventional FWI, adding additional degrees of freedom by extending velocity into non-physical dimensions shows promise in overcoming local minima and improving the inversion results \citep{symesMigration2008, biondi2014simultaneous, barnier2018full}. The generative neural network in NNFWI can easily over-parametrize physical velocity models. Although theories of deep learning are still developing, over-parametrization is believed to be a key factor in the effective optimization and generalization of deep learning methods \citep{aroraOptimization2018, allen-zhuConvergence2019, caoGeneralization2019}. 
In this study, we have shown that parameterizing the velocity model by a generative neural network can be used for both regularization and uncertainty quantification for FWI. The regularization effect of NNFWI is similar to dictionary learning in FWI. Dictionary-learning-based FWI adds an extra training loop, which is not needed in NNFWI, to learn sparse representations of complex features in the velocity model using many small training patches collected from previous iterations. In NNFWI, the generative neural network serves as a rich feature bank to represent the complex velocity model. Compared with the complex and computationally expensive dictionary learning step, the update of the generative neural network in NNFWI directly uses the gradients calculated by automatic differentiation. The gradients directly back-propagate from the loss through the PDEs and neural network layers to the weights of the generative neural network. Moreover, no extra training loops of the neural networks are needed, making the training workflow of NNFWI as simple as conventional FWI.

In addition to the architecture of the generative neural network used in this study, other neural network architectures can also be used as the generator in NNFWI. A variety of components of NNFWI can be tuned and added, such as the number of neural network layers, the choice of activation function, batch normalization layer, recurrent neural network layer, attention layer, and a group of optimization algorithms in deep learning. We can also replace the acoustic wave equation used in this work with the elastic wave equation as the PDE solver in NNFWI for elastic FWI applications. In other words, NNFWI is not limited to the specific configuration used in this study. Our work provides a general framework for incorporating deep neural networks with PDEs for FWI applications. NNFWI also enables more advanced deep learning modeling in FWI. For example, we can use Bayesian neural networks \citep{kendallWhat2017a}, which model the epistemic uncertainty in the model, as the generator in NNFWI. We can replace the input latent variable $z$, which is a randomized vector in this work, with the waveform observation $y$. In this way, NNFWI becomes similar to an encoder-decoder model \citep{goodfellow2016deep} with the neural network as the encoder and the PDE solver as the decoder. The neural network after training learns a mapping from observations to the velocity model. These extensions are natural directions for future research.

\section{Conclusions}
We have introduced NNFWI, an approach to integrate deep generative neural networks to the PDE-constrained optimization of FWI. NNFWI represents the velocity model of interest using a generative neural network and optimizes the weights of the neural network instead. The gradients of both the neural network and PDEs are calculated using automatic differentiation, which simplifies the optimization of NNFWI without the need of pretraining or extra optimization loops. Our results demonstrate that NNFWI achieves similar accuracy as conventional FWI. More importantly, the deep image prior of generative neural networks automatically filter out noise and introduce a regularization effect, which mitigates cycle-skipping similar as TV regularization and significantly improves the inversion results for noisy seismic data. Additionally, NNFWI provides uncertainty quantification using the dropout technique, which does not require additional computation during training, making NNFWI a much more efficient way to estimate uncertainty in the high dimensional model space of velocity estimation. NNFWI can be directly applied to the same datasets as conventional FWI to improve inversion performance. Extensions to the generative neural work in NNFWI and validating its uncertainty quantification as an absolute error proxy are promising directions for future work. 



\newpage
\clearpage
\section*{Appendix}

\renewcommand\thefigure{A\arabic{figure}}    
\setcounter{figure}{0}  

The appendix includes additional figures to compare inversion results between NNFWI and conventional FWI and to analyze uncertainty quantification of NNFWI. The figures are referenced and explained in the main text. 

\begin{figure}[!ht]
\centering
\begin{subfigure}{0.48\textwidth}
    \centering
    \includegraphics[width=\textwidth]{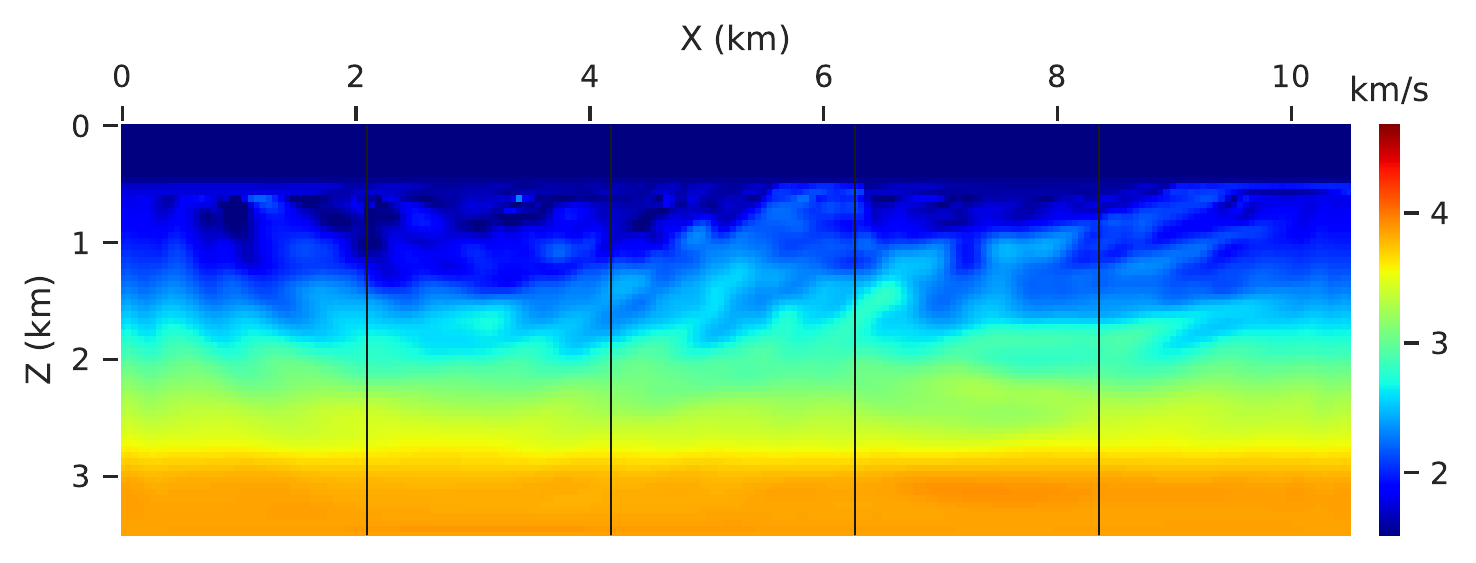}
    \caption{}
\end{subfigure}
\begin{subfigure}{0.48\textwidth}
    \centering
    \includegraphics[width=\textwidth]{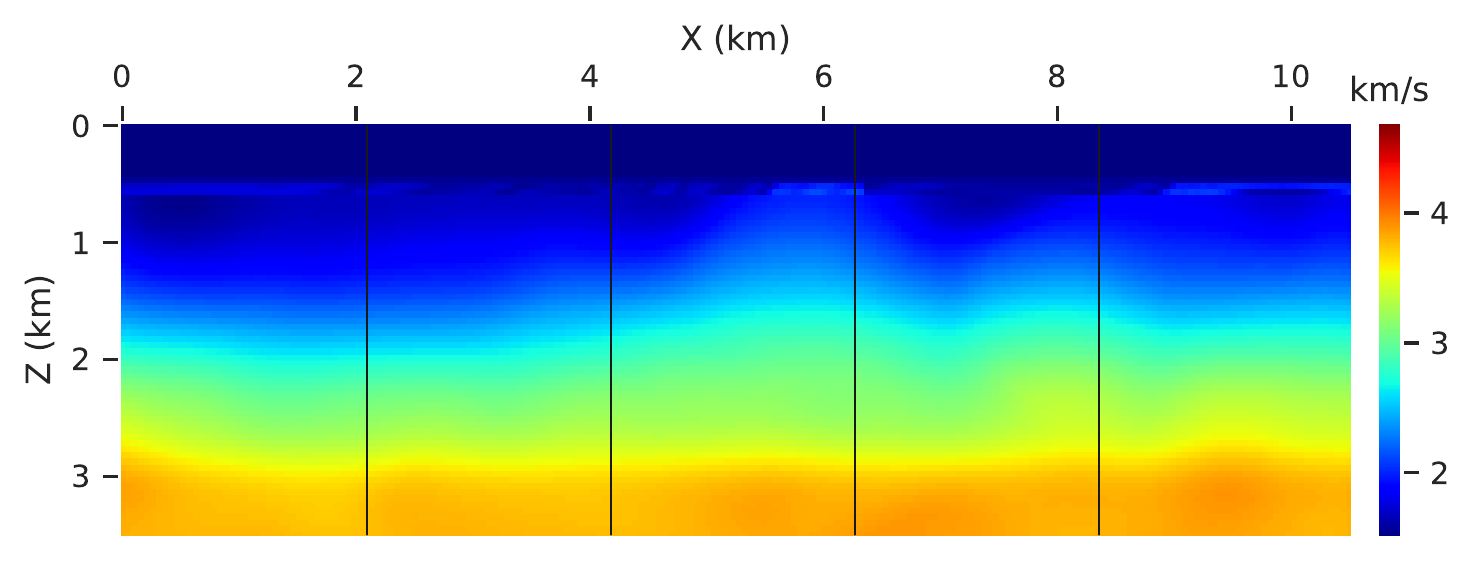}
    \caption{}
\end{subfigure}
\begin{subfigure}{0.48\textwidth}
    \centering
    \includegraphics[width=\textwidth]{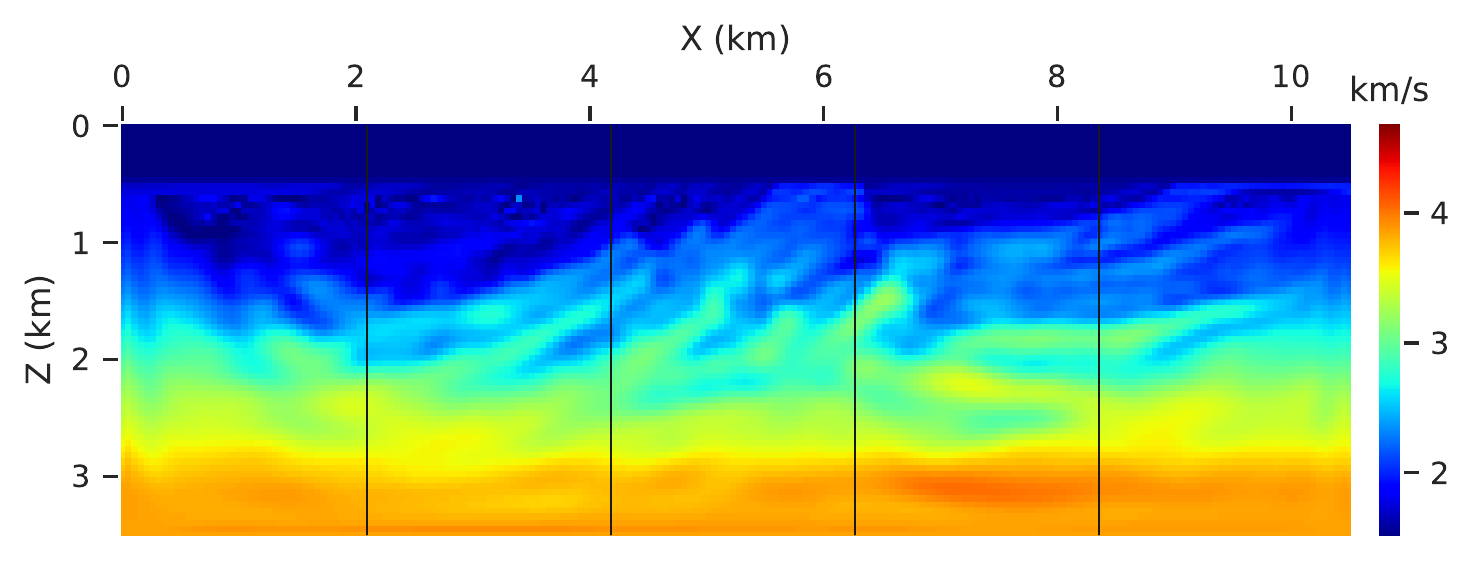}
    \caption{}
\end{subfigure}
\begin{subfigure}{0.48\textwidth}
    \centering
    \includegraphics[width=\textwidth]{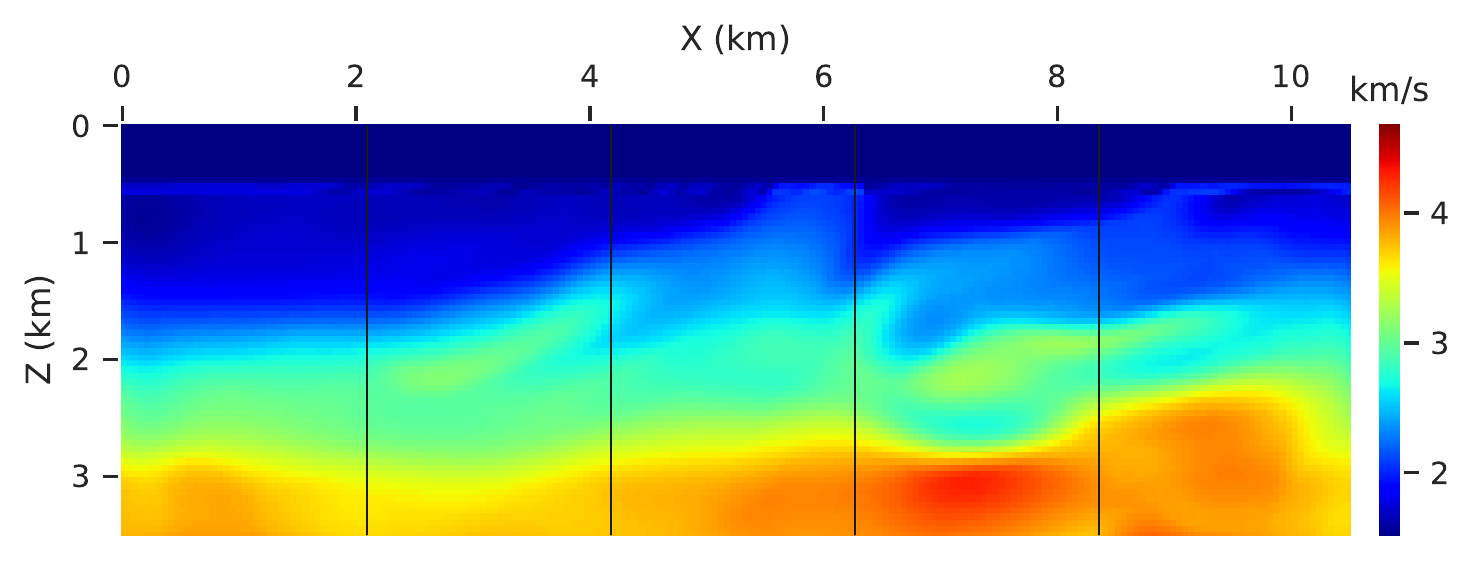}
    \caption{}
\end{subfigure}
\begin{subfigure}{0.48\textwidth}
    \centering
    \includegraphics[width=\textwidth]{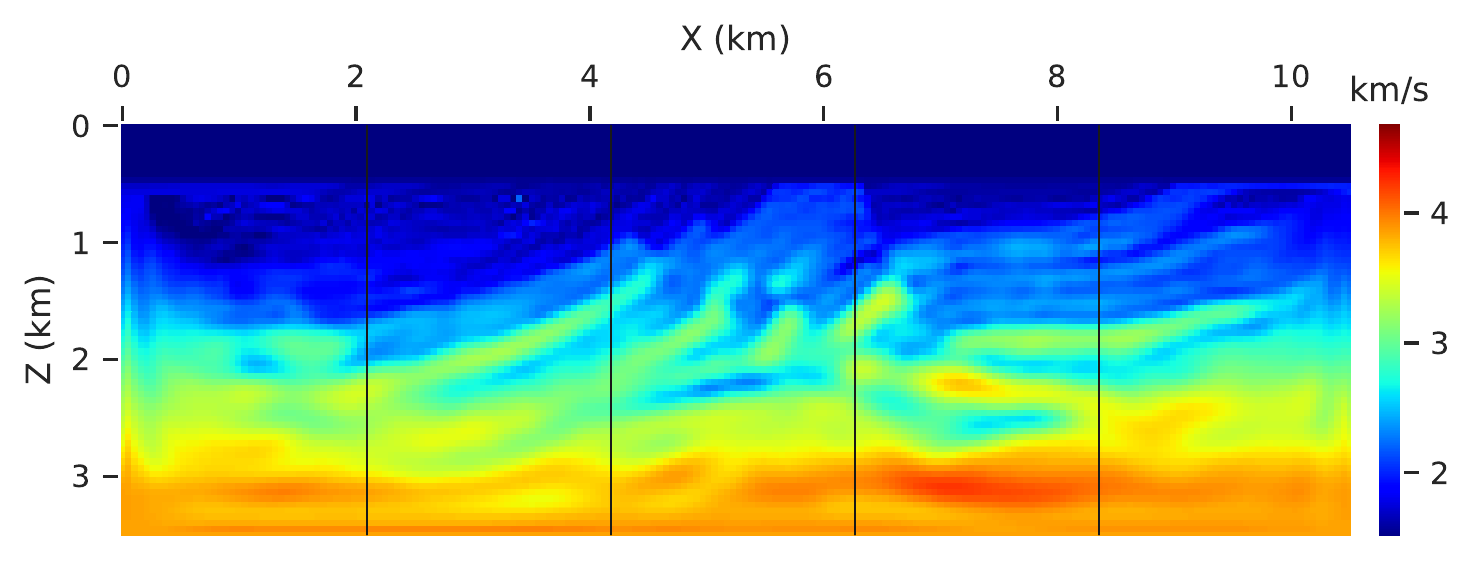}
    \caption{}
\end{subfigure}
\begin{subfigure}{0.48\textwidth}
    \centering
    \includegraphics[width=\textwidth]{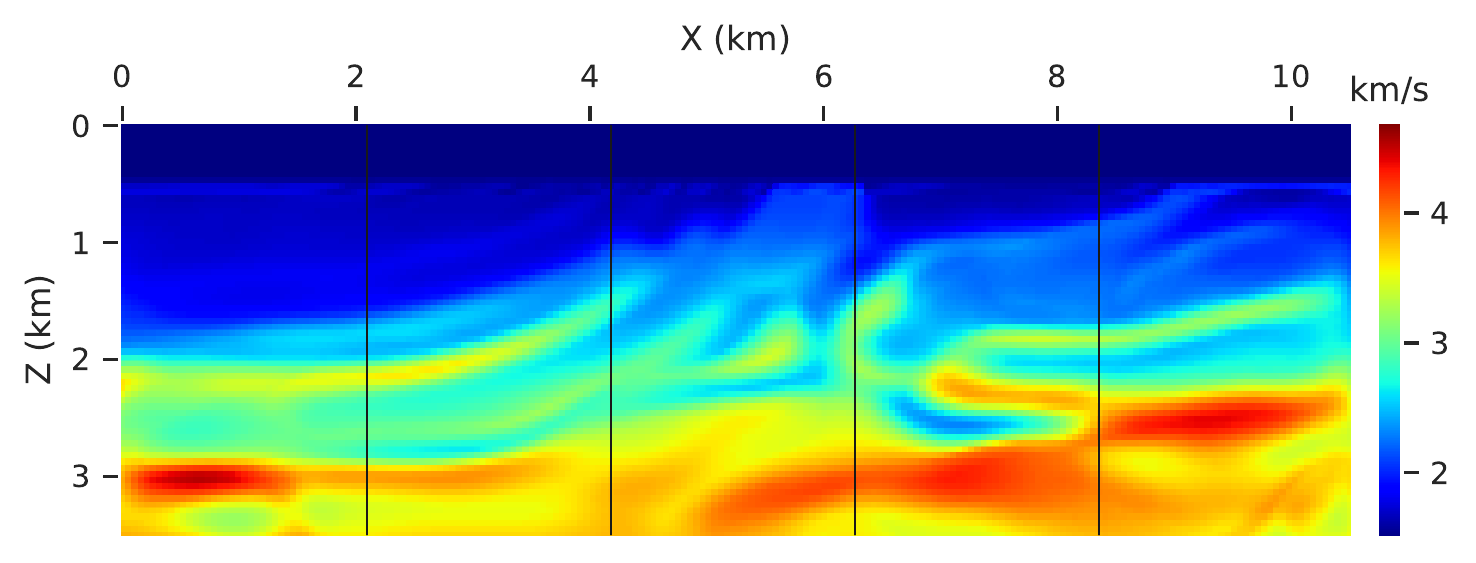}
    \caption{}
\end{subfigure}
\caption{Intermediate inversion results of the Marmousi model: the left panels (a, c, e) show results of conventional FWI; and the right panels (b, d, e) show results of NNFWI. We plot three estimated velocity maps at 0.3, 0.1, and 0.03 of the initial loss to show the different optimization processes of conventional FWI and NNFWI. The iteration numbers are 10, 20, and 30 for FWI in the left panels and 10, 270, 890 for NNFWI in the right panels.} 
\label{fig:updates_marmousi}
\end{figure}

\begin{figure}
    \centering
    \begin{subfigure}{0.48\textwidth}
        \centering
        \includegraphics[width=\textwidth]{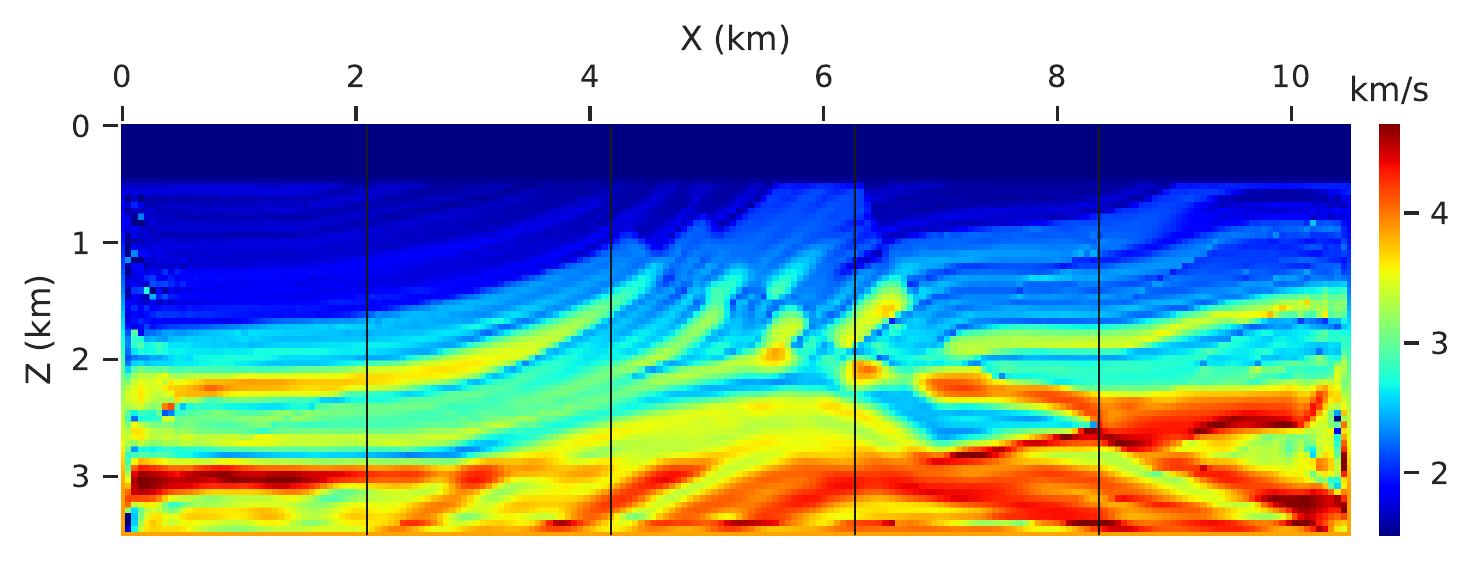}
        \caption{}
    \end{subfigure}
    \begin{subfigure}{0.48\textwidth}
        \centering
        \includegraphics[width=\textwidth]{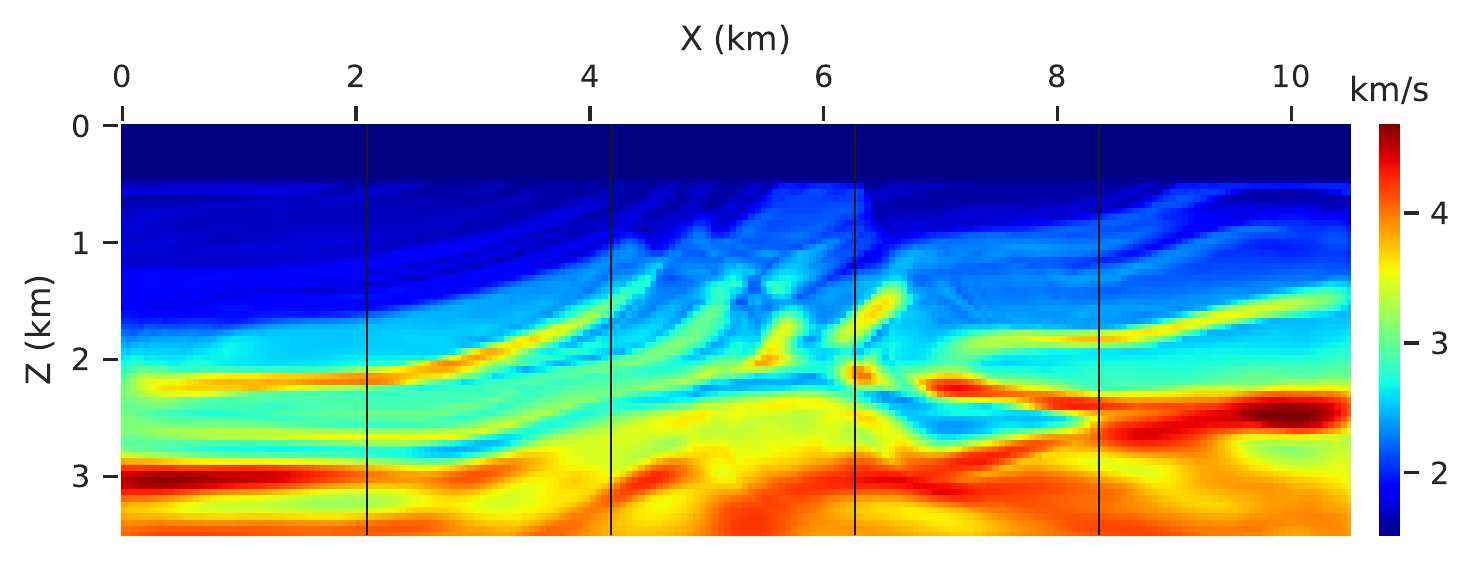}
        \caption{}
    \end{subfigure}
    \begin{subfigure}{0.48\textwidth}
        \centering
        \includegraphics[width=\textwidth]{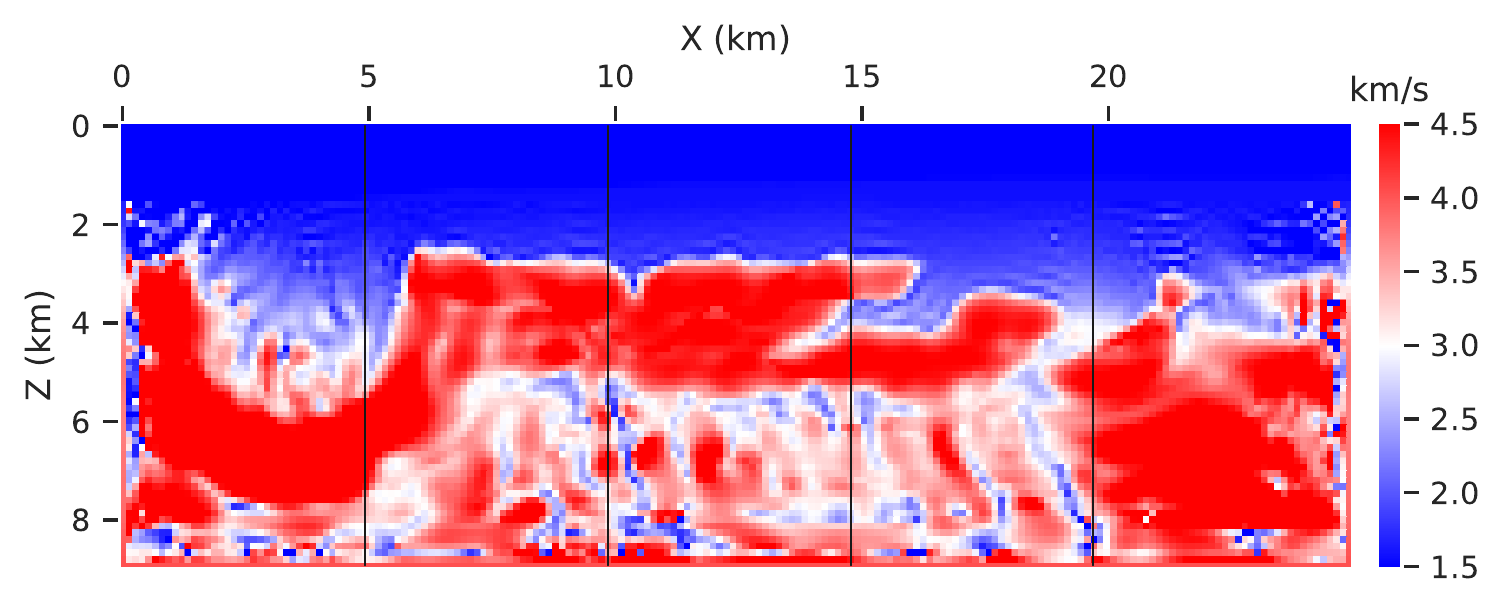}
        \caption{}
    \end{subfigure}
    \begin{subfigure}{0.48\textwidth}
        \centering
        \includegraphics[width=\textwidth]{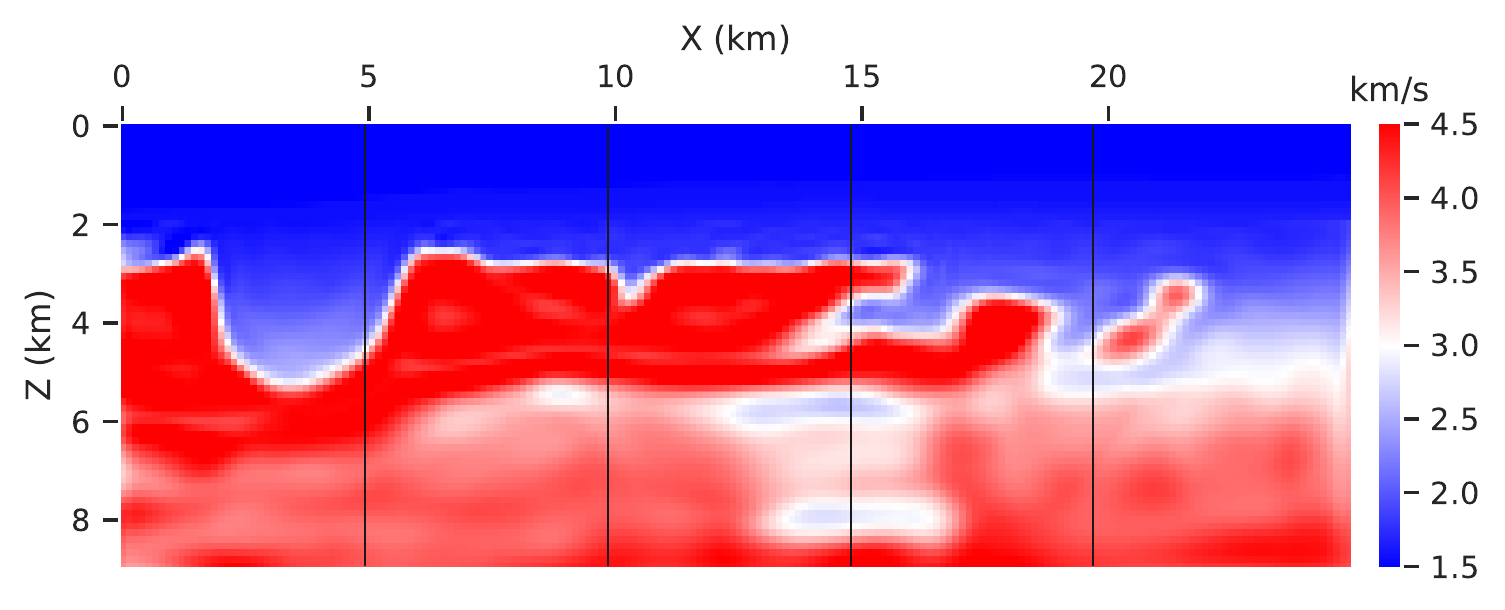}
        \caption{}
    \end{subfigure}
    \caption{Inversion results of (a) conventional FWI with Adam based on the Marmousi model, (b) NNFWI with L-BFGS based on the Marmousi model, (c) conventional FWI with Adam based on the 2004 BP model, and (d) NNFWI with based on the 2004 BP model.}
    \label{fig:result_bfgs_adam}
\end{figure}

\begin{figure}
    \centering
    \begin{subfigure}{0.48\textwidth}
        \centering
        \includegraphics[width=\textwidth]{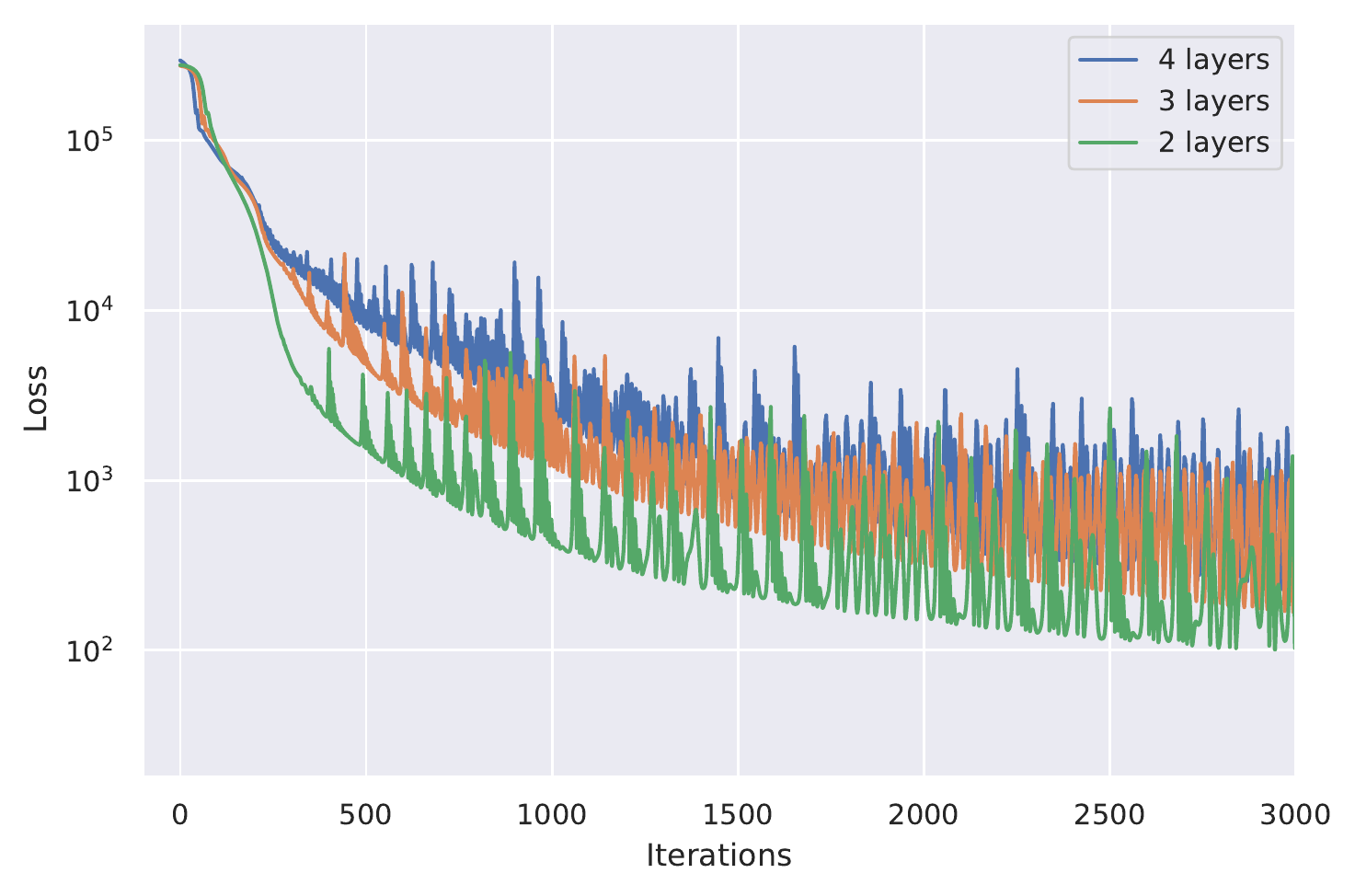}
    \end{subfigure}
    \caption{Loss functions of NNFWI with different convolutional layers based on the 2004 BP model.}
    \label{fig:loss_BP_layers}
\end{figure}

\begin{figure}
    \centering
    \begin{subfigure}{0.48\textwidth}
        \centering
        \includegraphics[width=\textwidth]{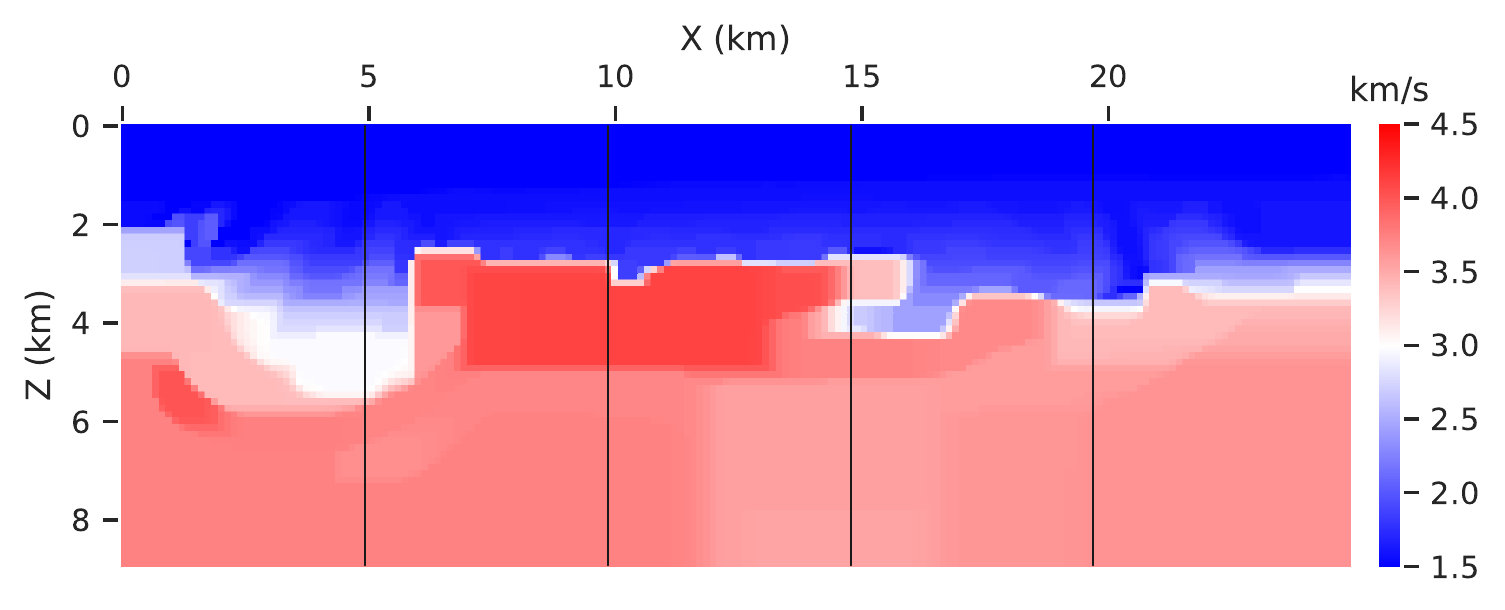}
        \caption{}
    \end{subfigure}
    \begin{subfigure}{0.48\textwidth}
        \centering
        \includegraphics[width=\textwidth]{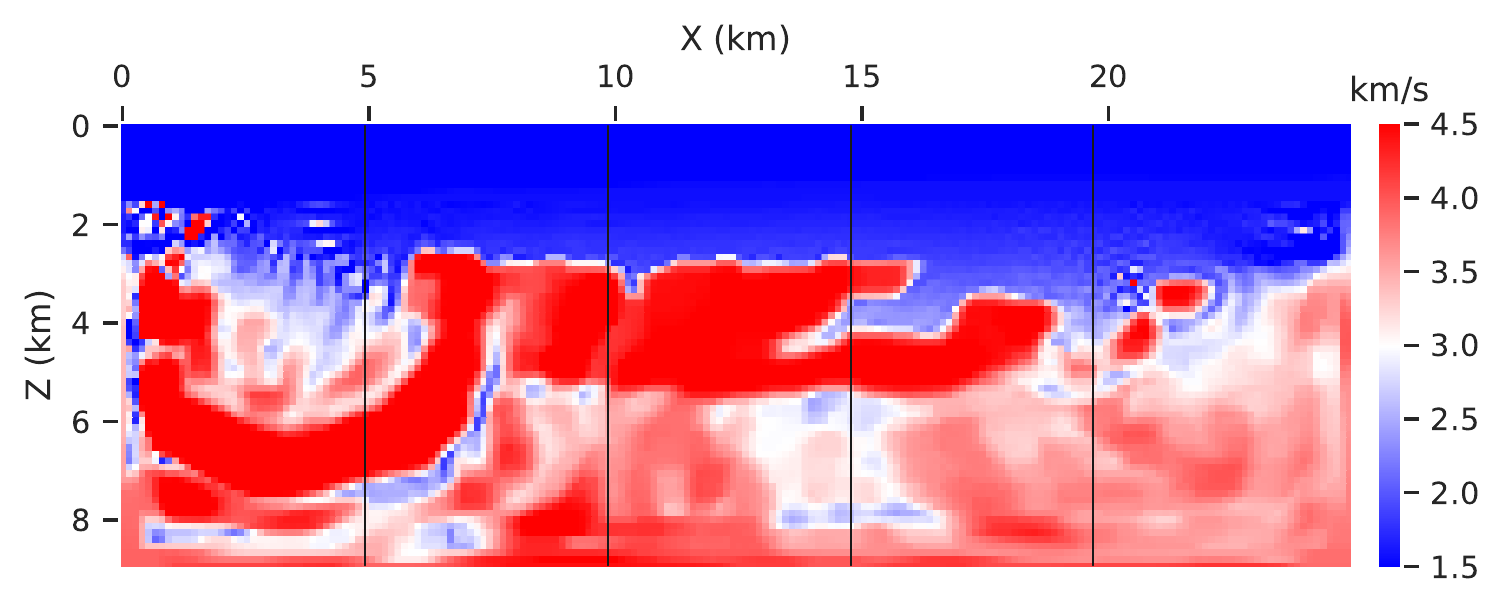}
        \caption{}
    \end{subfigure}
    \caption{Inversion results of conventional FWI with total variation regularization: (a) $\gamma$ = 0.01; and (b) $\gamma$ = 0.0001. The inversion result with $\gamma$ = 0.001 is shown in \Cref{fig:BP}b.}
    \label{fig:BP_TV_2_4}
\end{figure}

\begin{figure}
    \centering
    \begin{subfigure}{0.48\textwidth}
        \centering
        \includegraphics[width=\textwidth]{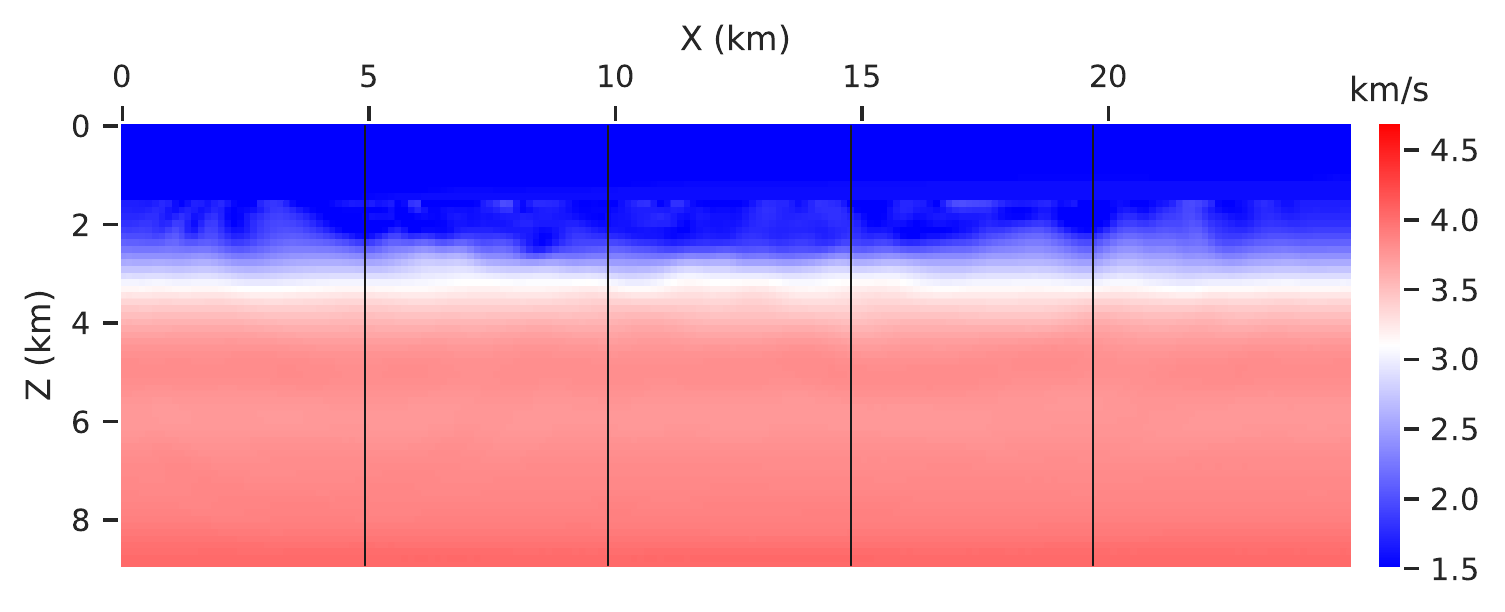}
        \caption{}
    \end{subfigure}
    \begin{subfigure}{0.48\textwidth}
        \centering
        \includegraphics[width=\textwidth]{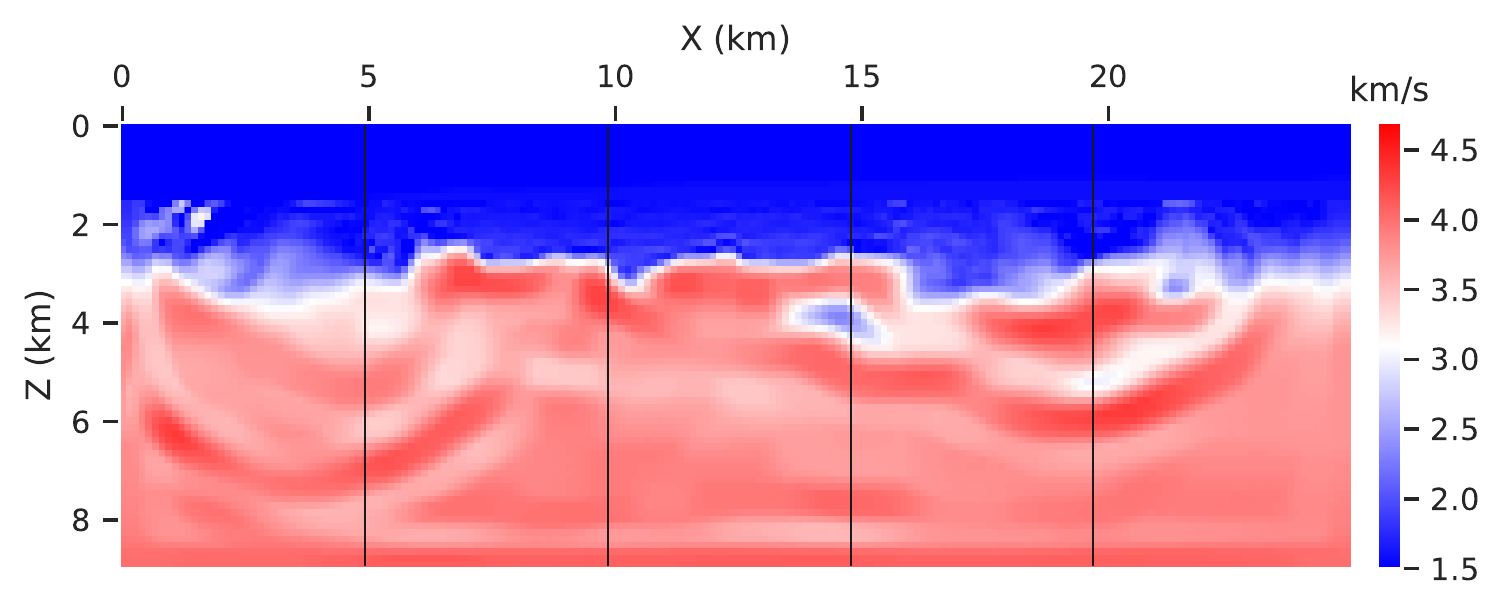}
        \caption{}
    \end{subfigure}
    \begin{subfigure}{0.48\textwidth}
        \centering
        \includegraphics[width=\textwidth]{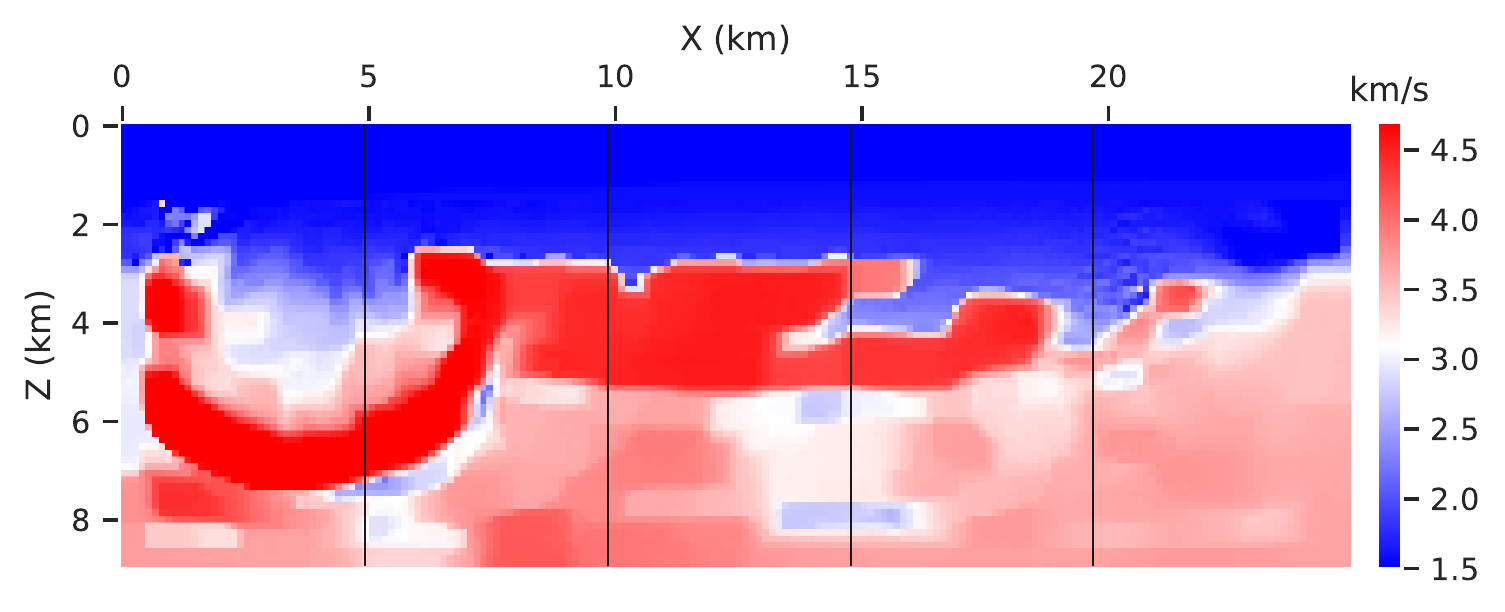}
        \caption{}
    \end{subfigure}
    \caption{Intermediate inversion results of conventional FWI with total variation regularization. We plot three estimated velocity maps at 0.3, 0.1, and 0.03 of the initial loss in (a), (b), and (c) respectively. The iteration numbers are 10, 60, and 330. The results of conventional FWI and NNFWI for comparison are plotted in \Cref{fig:updates_BP}.}
    \label{fig:updates_BP_TV}
\end{figure}

\begin{figure}
    \centering
    \begin{subfigure}{0.48\textwidth}
        \centering
        \includegraphics[width=\textwidth]{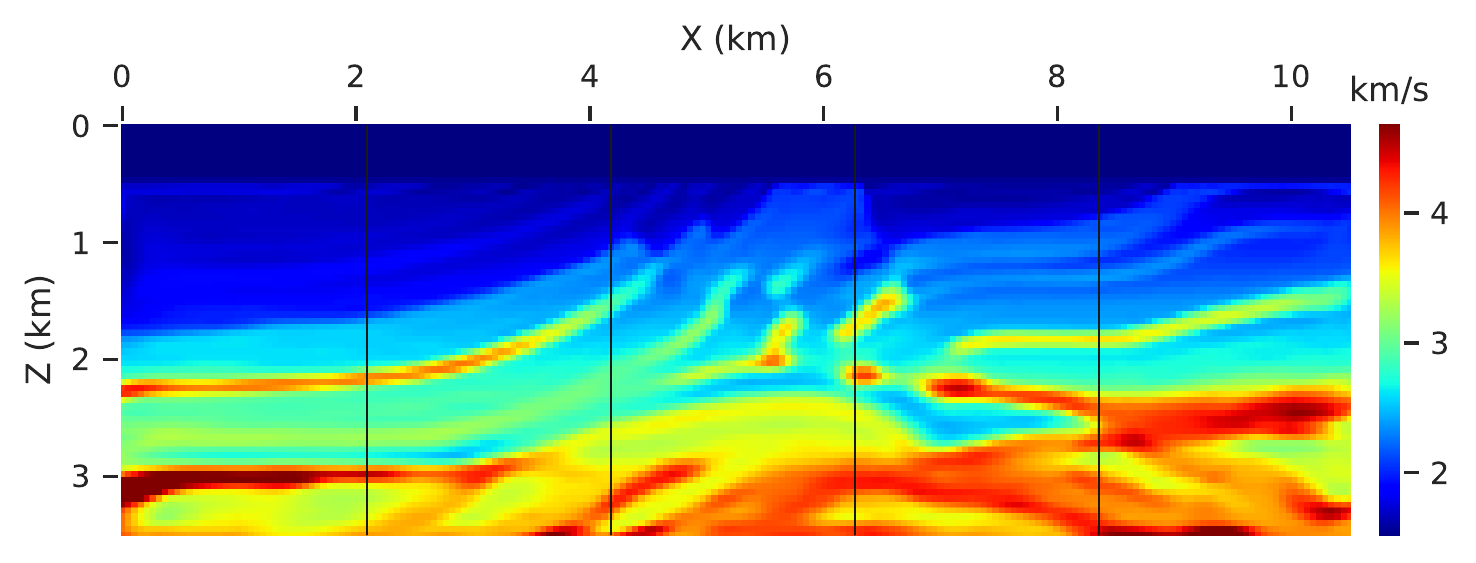}
        \caption{}
    \end{subfigure}
    \begin{subfigure}{0.48\textwidth}
        \centering
        \includegraphics[width=\textwidth]{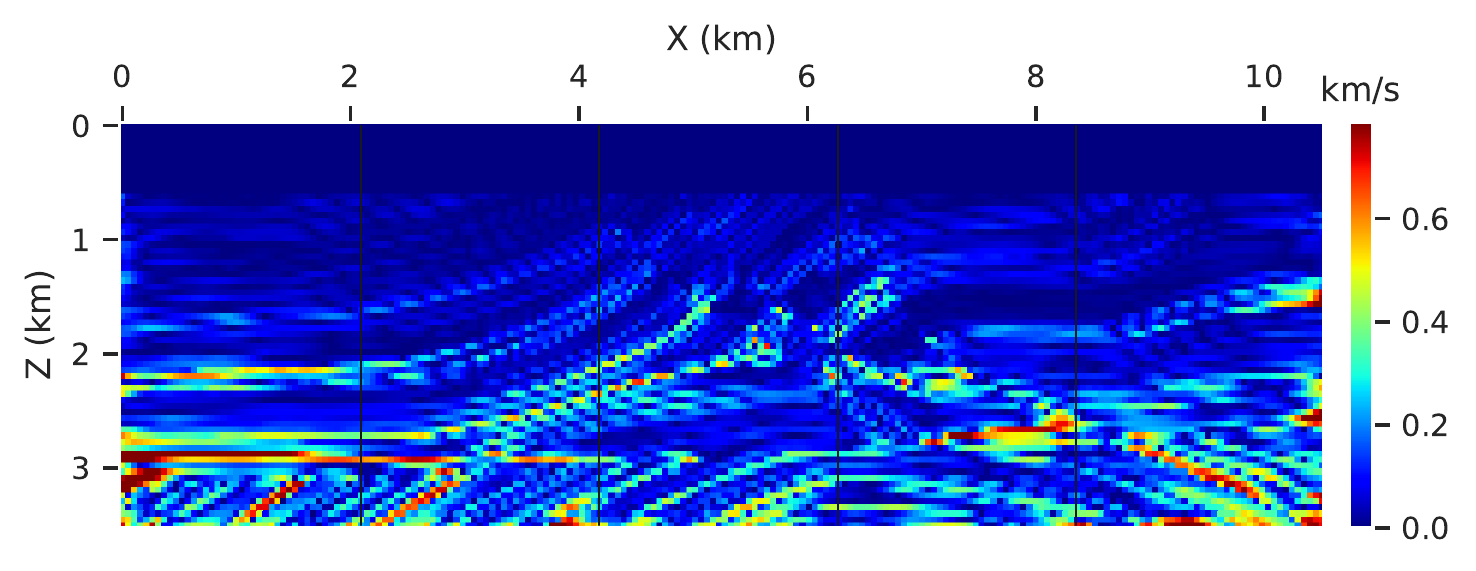}
        \caption{}
    \end{subfigure}
    \begin{subfigure}{0.48\textwidth}
        \centering
        \includegraphics[width=\textwidth]{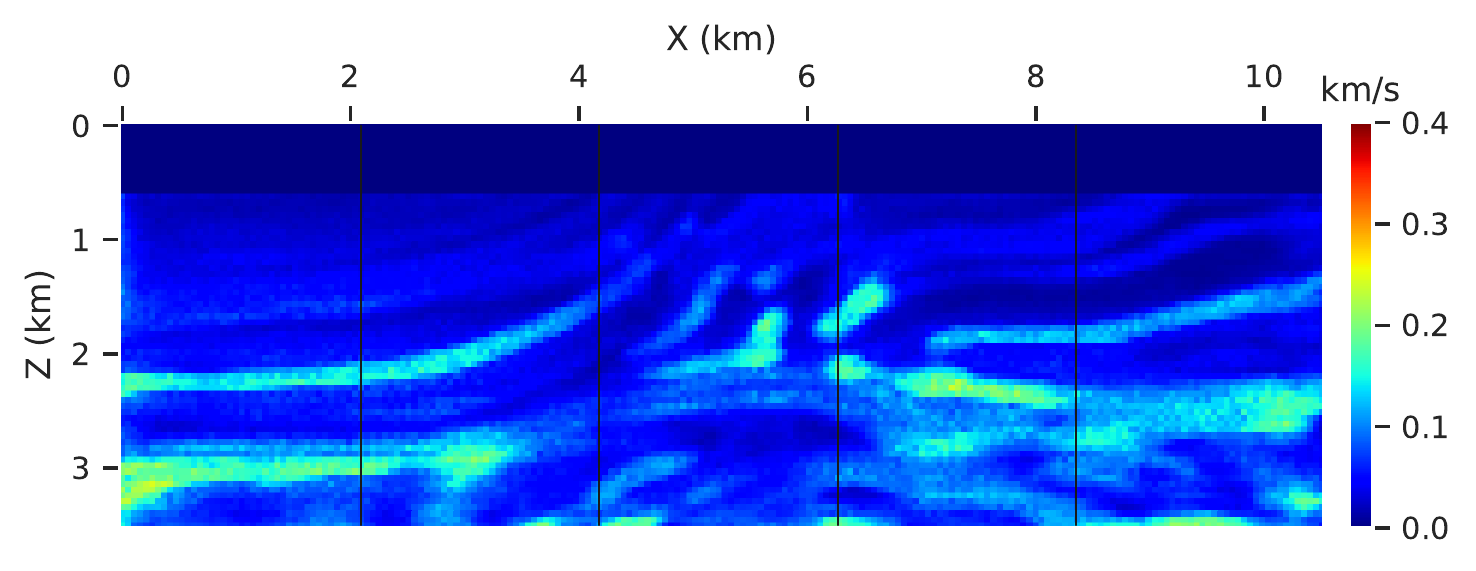}
        \caption{}
    \end{subfigure}
    \begin{subfigure}{0.48\textwidth}
        \centering
        \includegraphics[width=\textwidth]{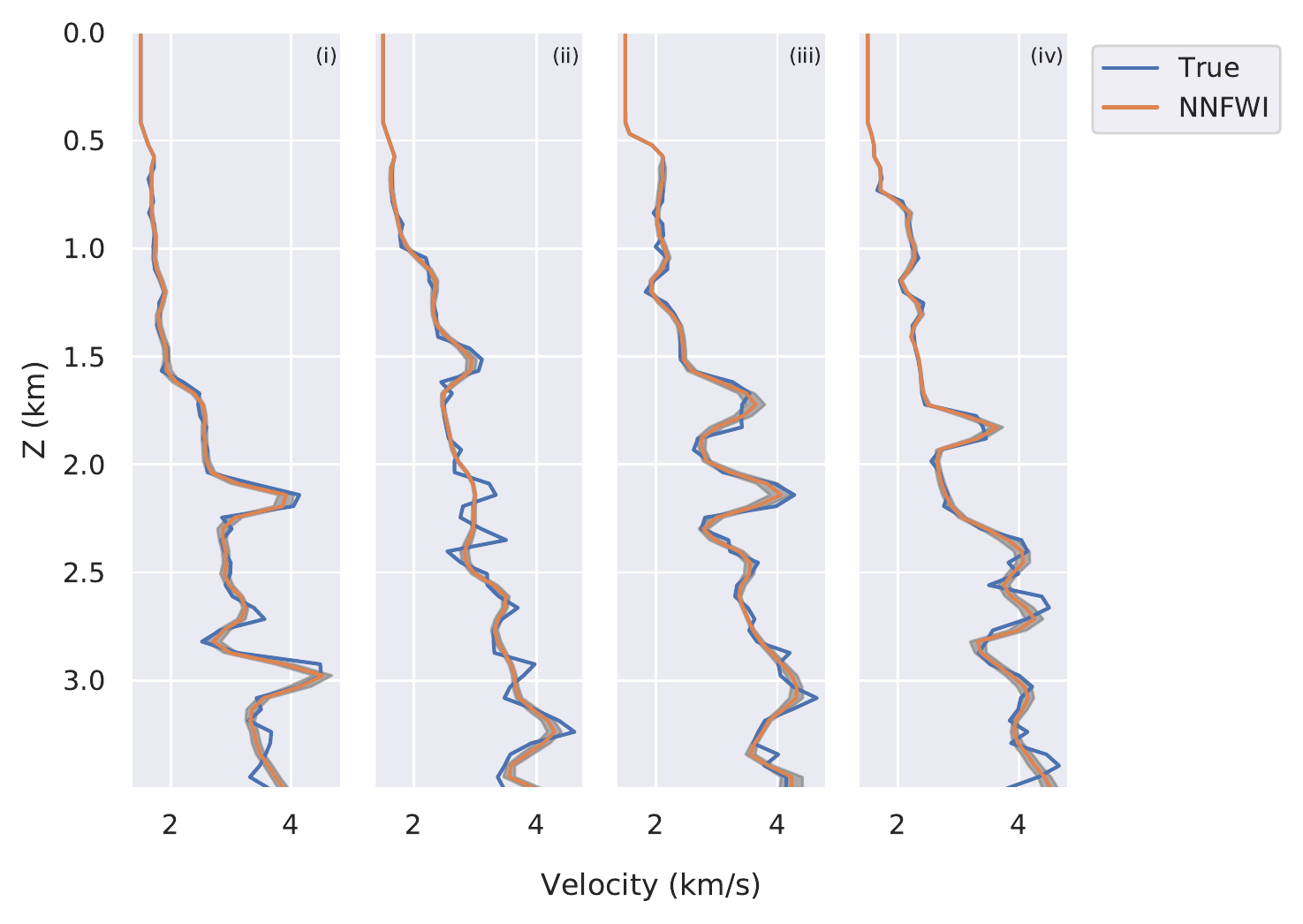}
        \caption{}
    \end{subfigure}
    \caption{Uncertainty quantification of NNFWI with 3 upsampling and dropout layers based on the Marmousi model: (a) inversion result; (b) inversion error map; (c) estimated standard deviation through Monte Carlo samplings with a dropout rate of 0.1; (d) velocity profiles with standard deviation ranges plotted in gray. Their locations are marked by black vertical lines in (a, b, c).}
    \label{fig:marmousi_UQ_3}
\end{figure}

\begin{figure}
    \centering
    \begin{subfigure}{0.48\textwidth}
        \centering
        \includegraphics[width=\textwidth]{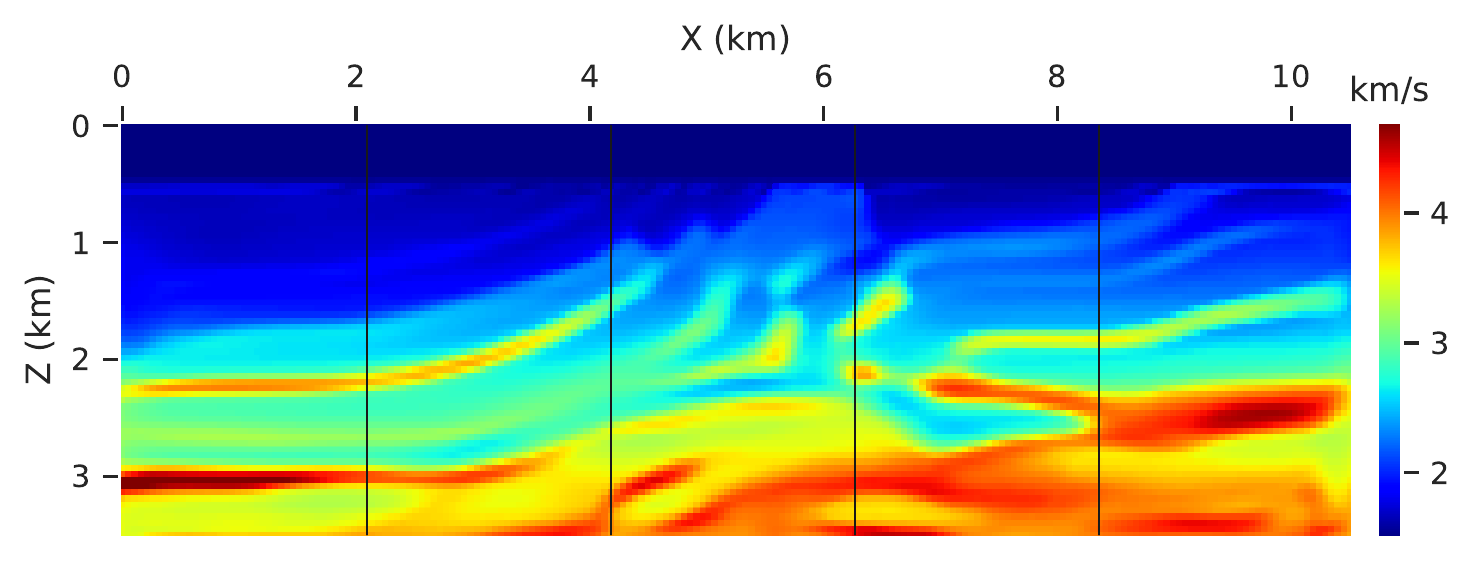}
        \caption{}
    \end{subfigure}
    \begin{subfigure}{0.48\textwidth}
        \centering
        \includegraphics[width=\textwidth]{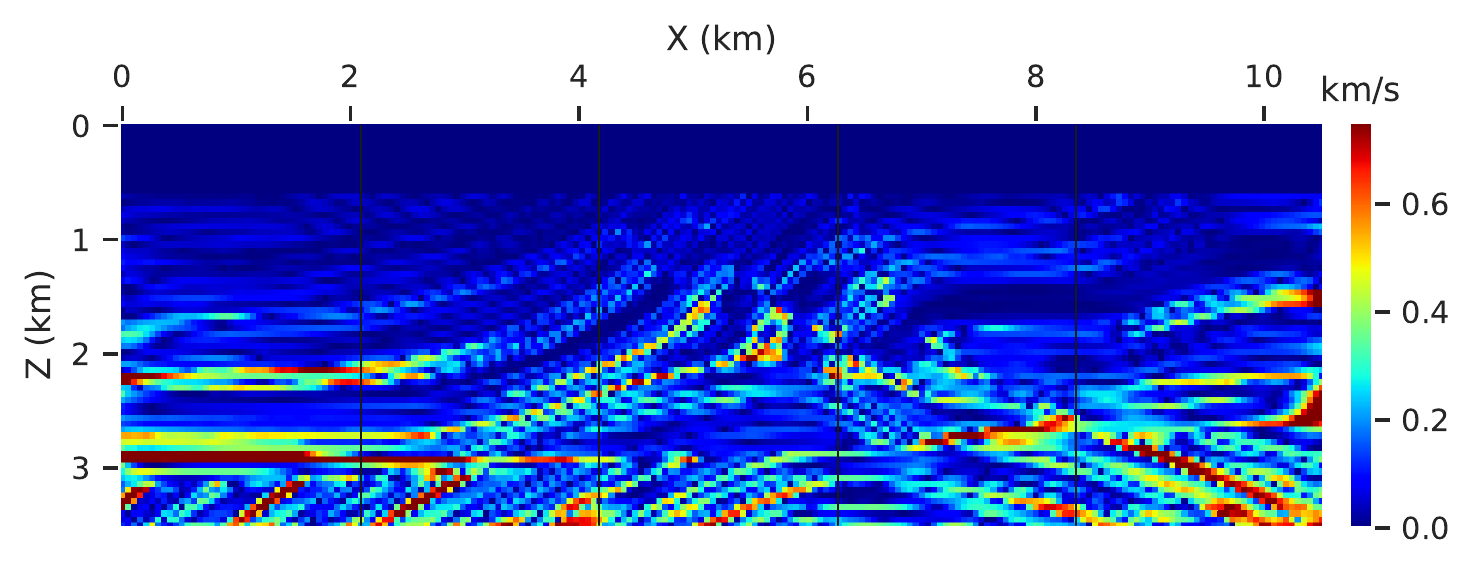}
        \caption{}
    \end{subfigure}
    \begin{subfigure}{0.48\textwidth}
        \centering
        \includegraphics[width=\textwidth]{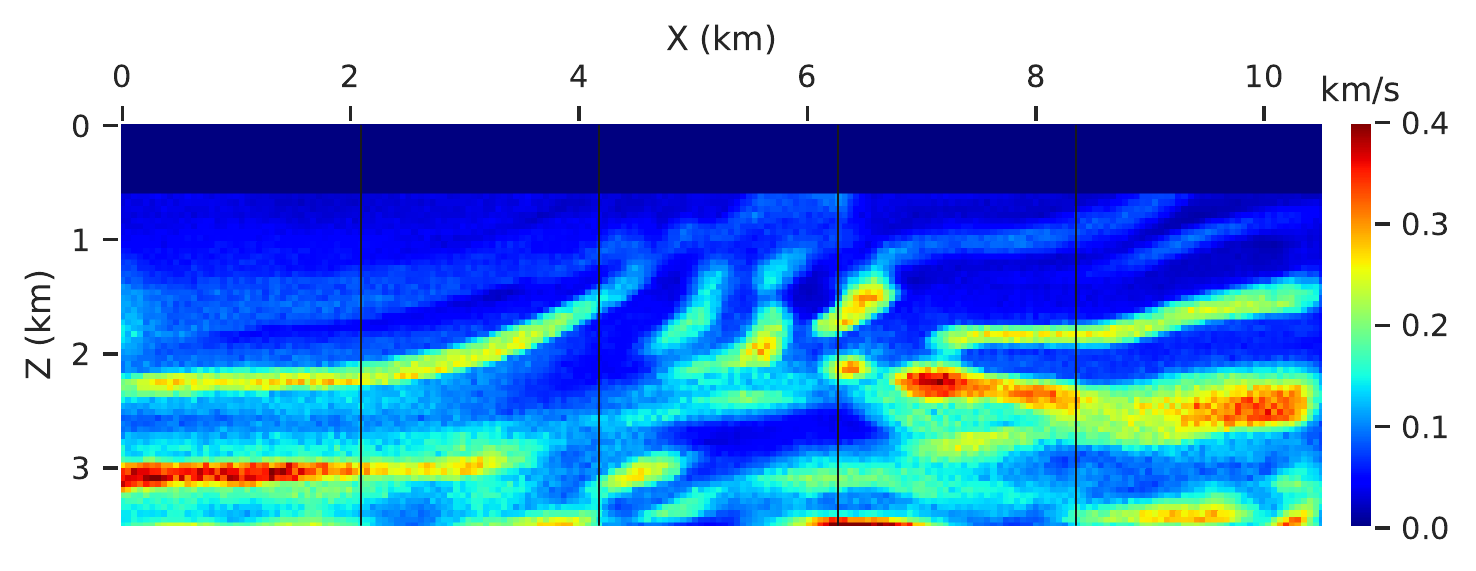}
        \caption{}
    \end{subfigure}
    \begin{subfigure}{0.48\textwidth}
        \centering
        \includegraphics[width=\textwidth]{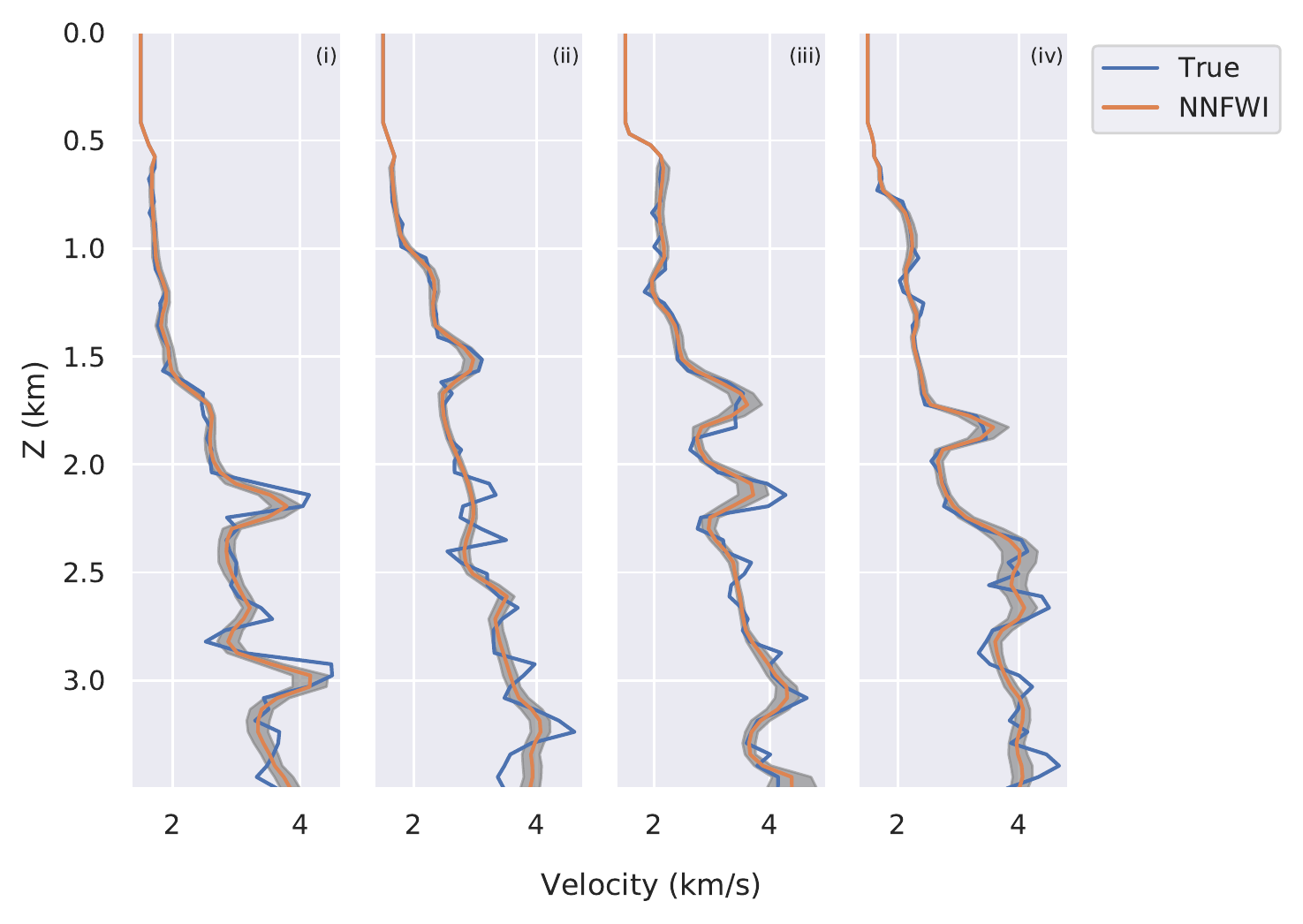}
        \caption{}
    \end{subfigure}
    \caption{Uncertainty quantification of NNFWI with a dropout rate of 0.2 based on the Marmousi model: (a) inversion result; (b) inversion error map; (c) estimated standard deviation through Monte Carlo samplings with a dropout rate of 0.2; (d) velocity profiles with standard deviation ranges plotted in gray. Their locations are marked by black vertical lines in (a, b, c).}
    \label{fig:marmousi_UQ_dp2}
\end{figure}

\begin{figure}
    \centering
    \begin{subfigure}{0.48\textwidth}
        \centering
        \includegraphics[width=\textwidth]{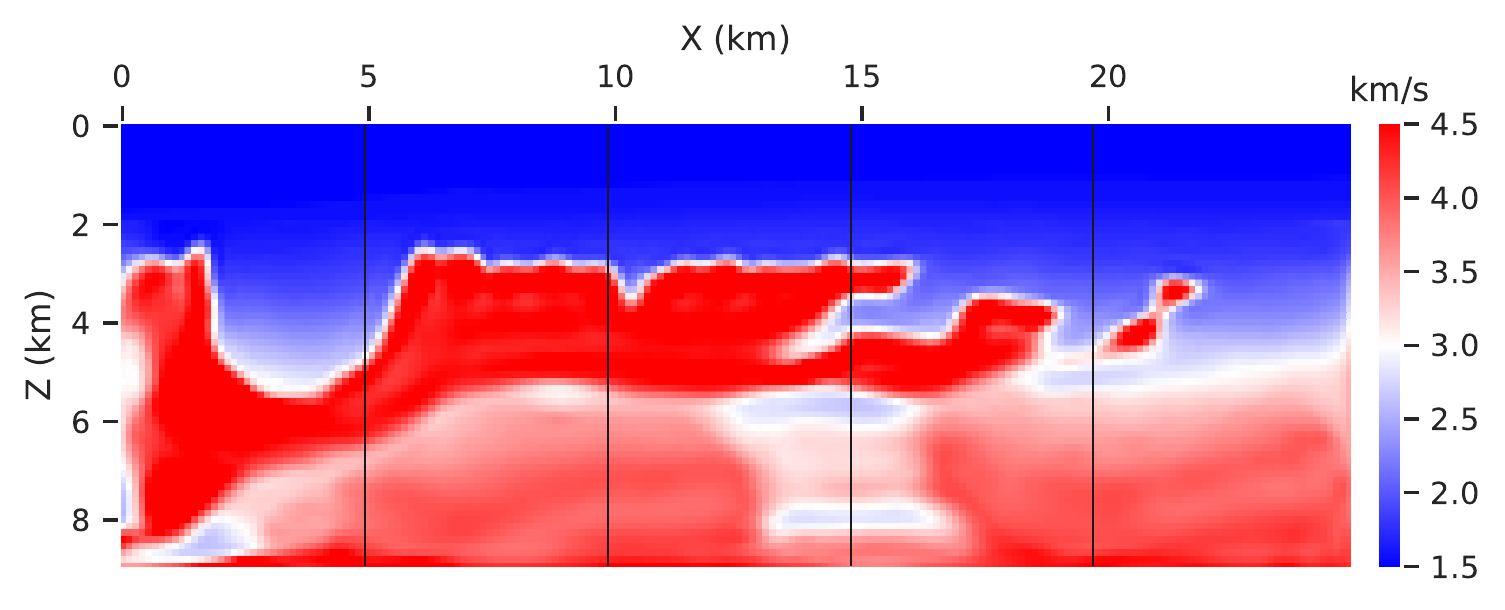}
        \caption{}
    \end{subfigure}
    \begin{subfigure}{0.48\textwidth}
        \centering
        \includegraphics[width=\textwidth]{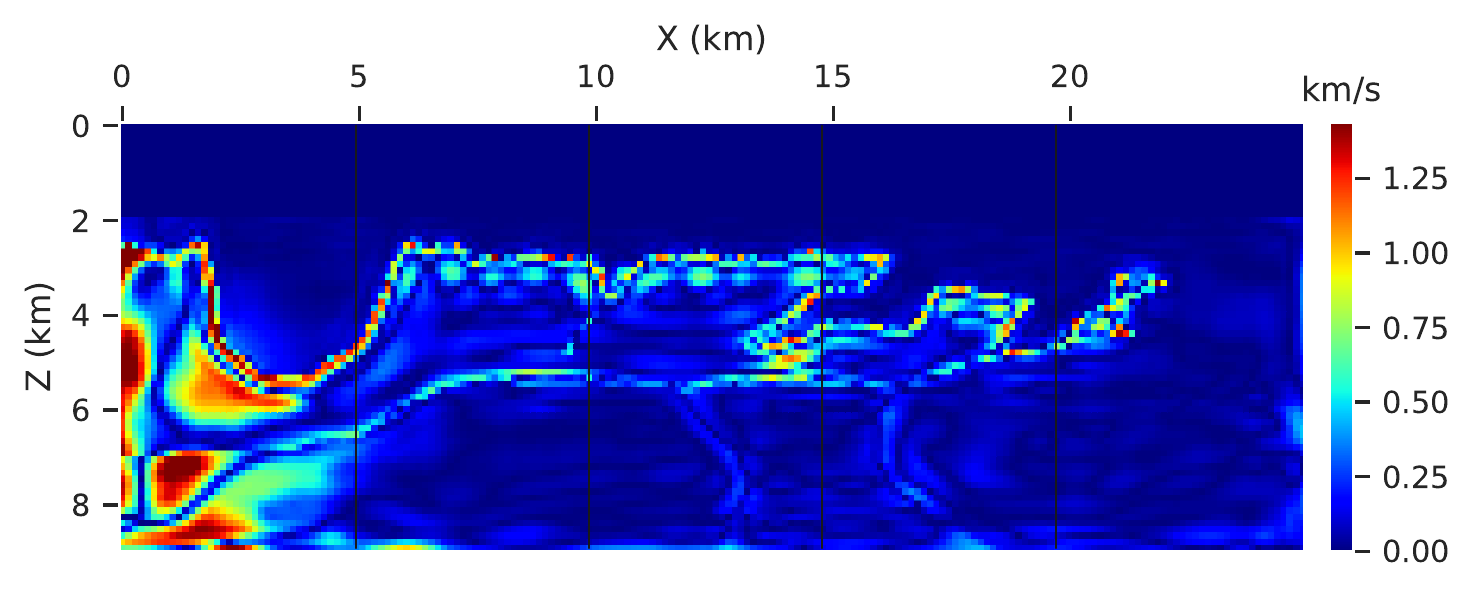}
        \caption{}
    \end{subfigure}
    \begin{subfigure}{0.48\textwidth}
        \centering
        \includegraphics[width=\textwidth]{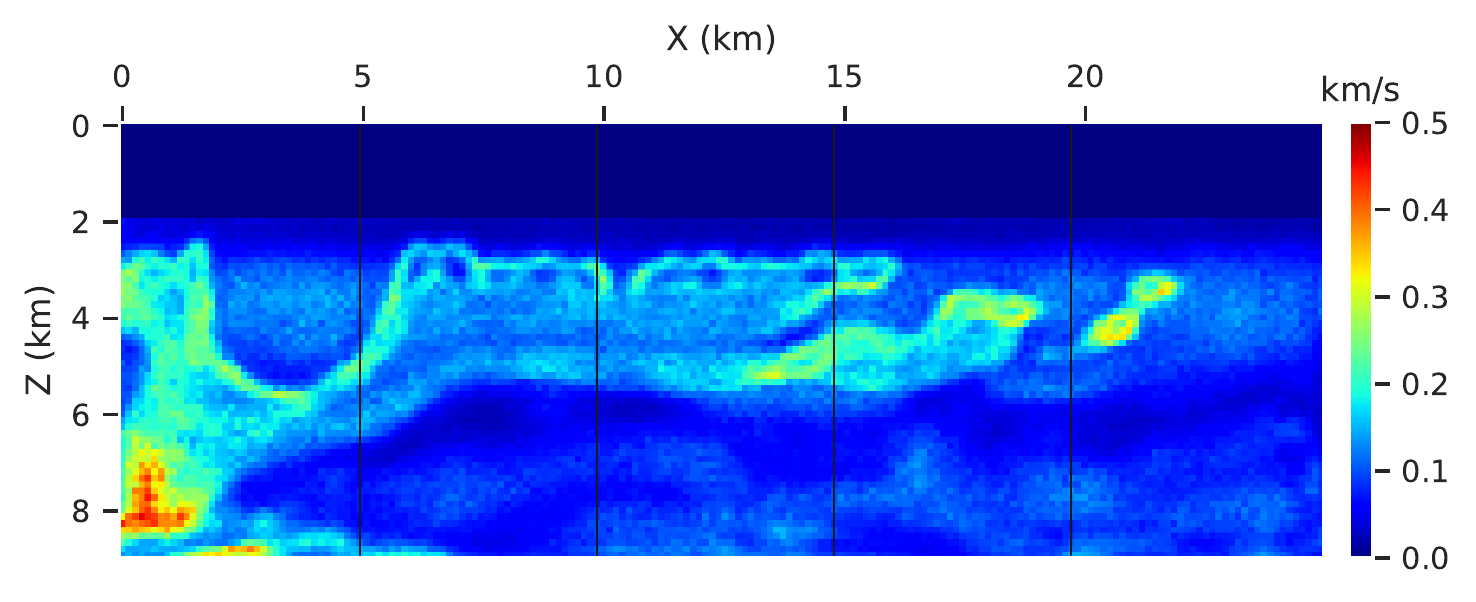}
        \caption{}
    \end{subfigure}
    \begin{subfigure}{0.48\textwidth}
        \centering
        \includegraphics[width=\textwidth]{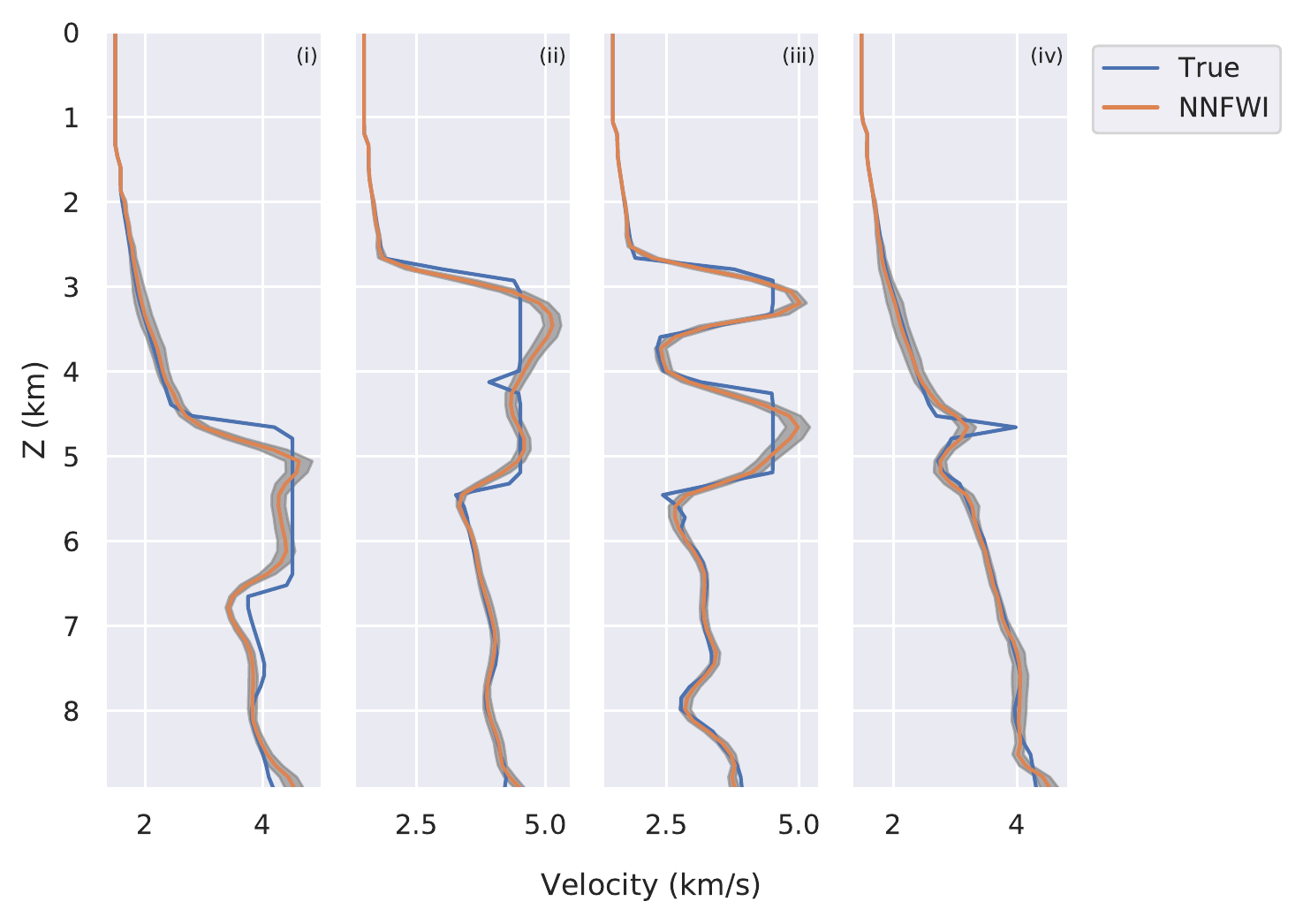}
        \caption{}
    \end{subfigure}
    \caption{Uncertainty quantification of NNFWI with 3 upsampling and dropout layers based on the 2004 BP model: (a) inversion result; (b) inversion error map; (c) estimated standard deviation through Monte Carlo samplings with a dropout rate of 0.1; (d) velocity profiles with standard deviation ranges plotted in gray. Their locations are marked by black vertical lines in (a, b, c).}
    \label{fig:BP_UQ_3}
\end{figure}

\begin{figure}
    \centering
    \begin{subfigure}{0.48\textwidth}
        \centering
        \includegraphics[width=\textwidth]{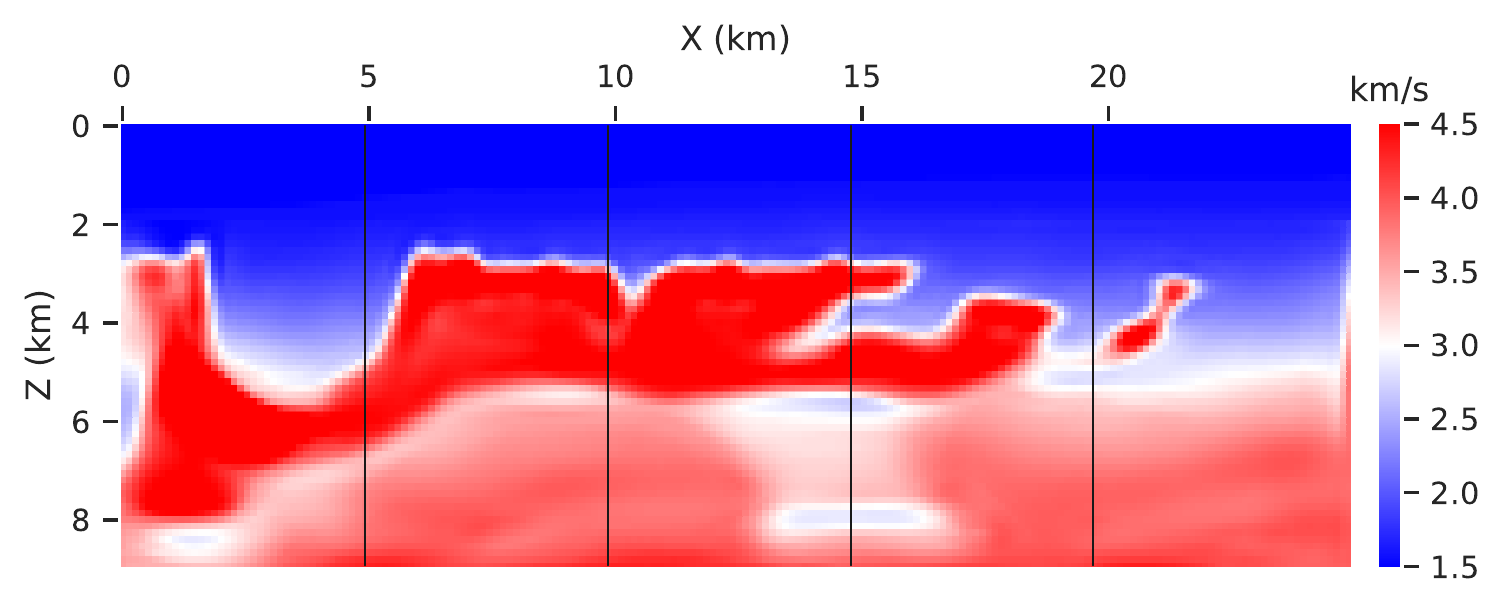}
        \caption{}
    \end{subfigure}
    \begin{subfigure}{0.48\textwidth}
        \centering
        \includegraphics[width=\textwidth]{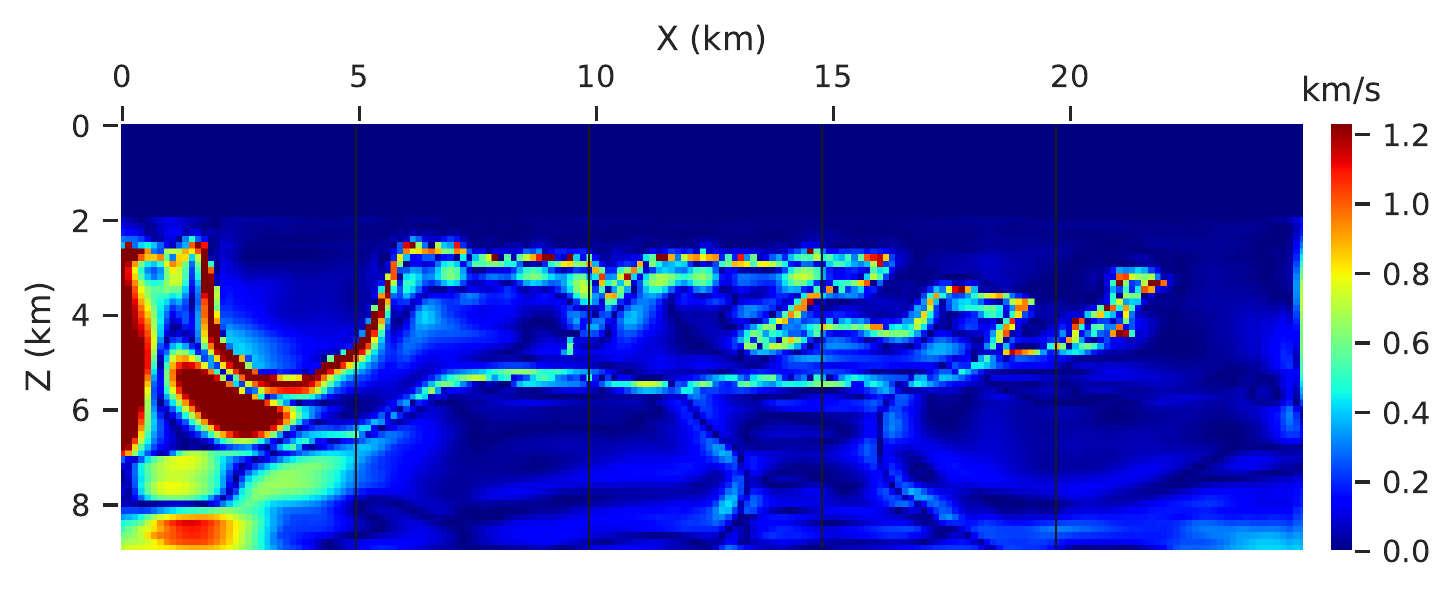}
        \caption{}
    \end{subfigure}
    \begin{subfigure}{0.48\textwidth}
        \centering
        \includegraphics[width=\textwidth]{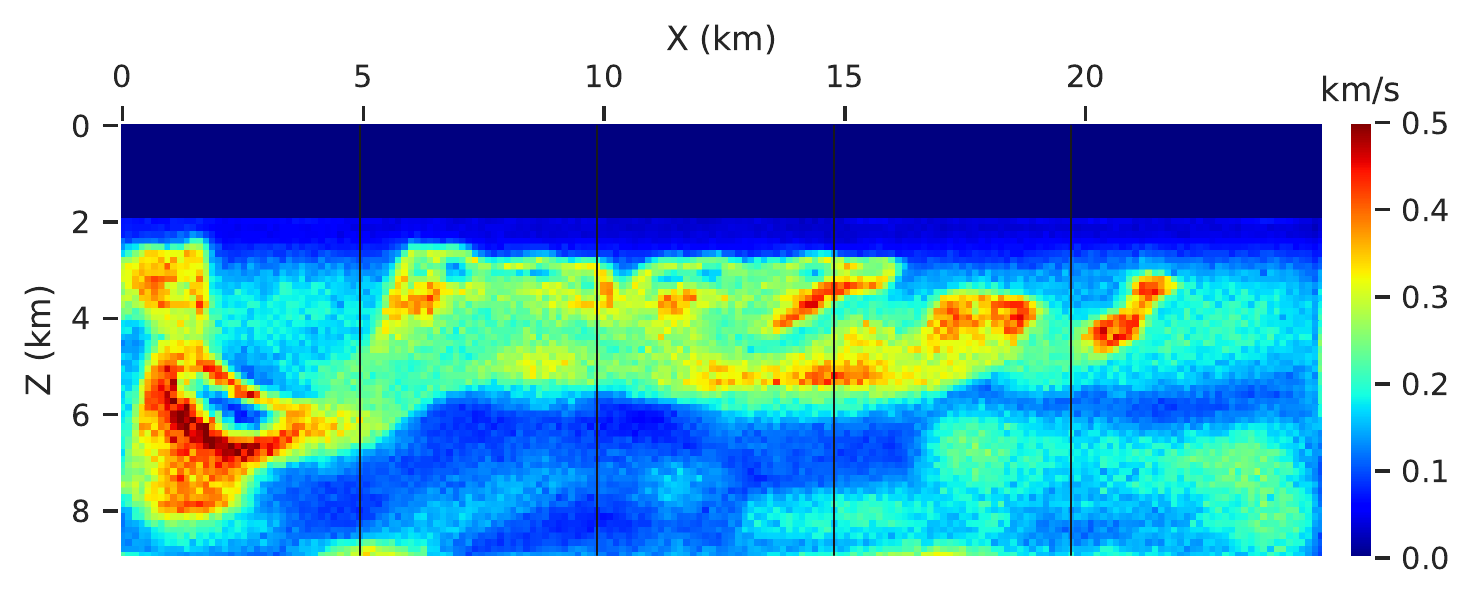}
        \caption{}
    \end{subfigure}
    \begin{subfigure}{0.48\textwidth}
        \centering
        \includegraphics[width=\textwidth]{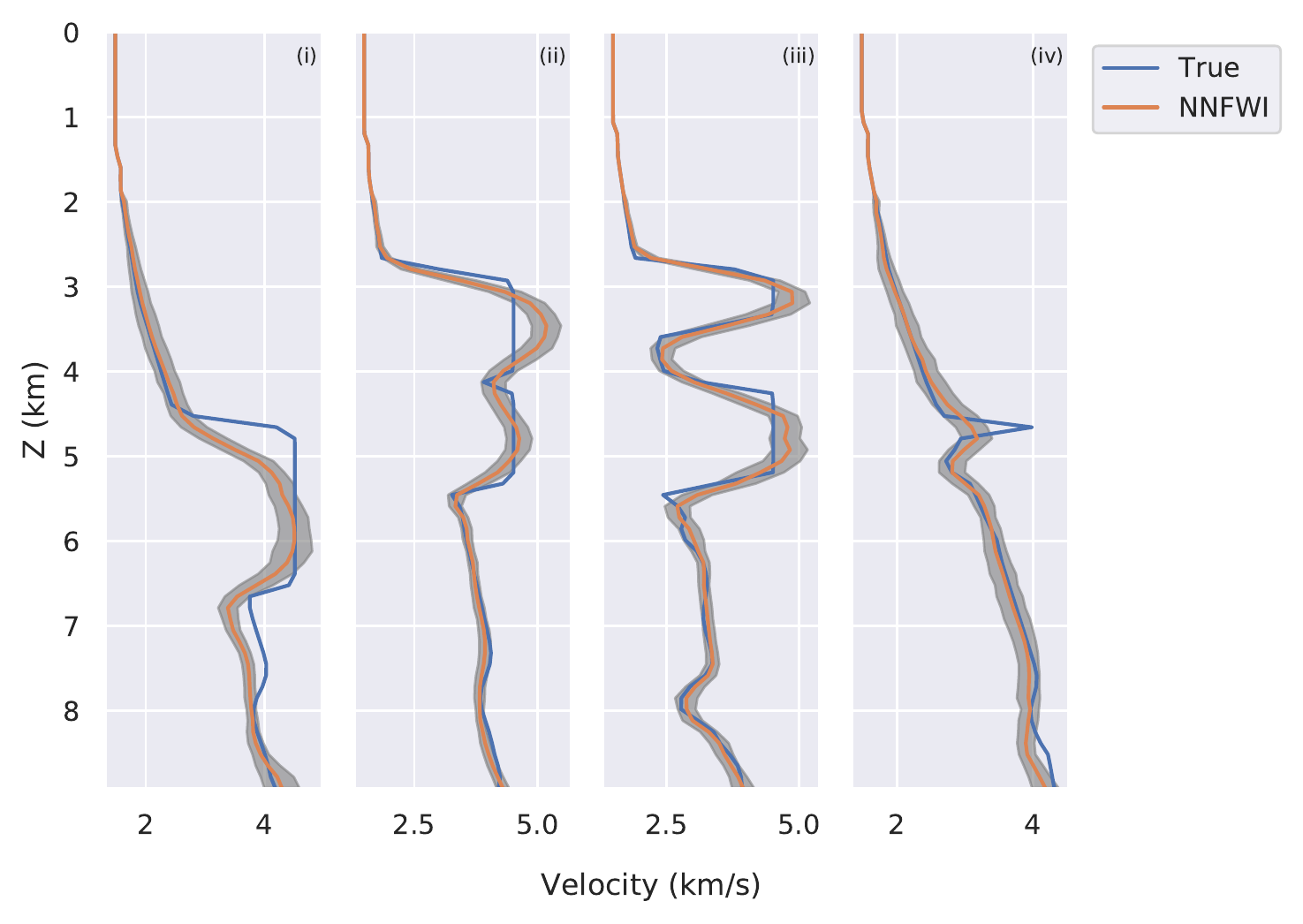}
        \caption{}
    \end{subfigure}
    \caption{Uncertainty quantification of NNFWI with a dropout rate of 0.2 based on the 2004 BP model: (a) inversion result; (b) inversion error map; (c) estimated standard deviation through Monte Carlo samplings with a dropout rate of 0.2; (d) velocity profiles with standard deviation ranges plotted in gray. Their locations are marked by black vertical lines in (a, b, c).}
    \label{fig:BP_UQ_dp2}
\end{figure}

\clearpage
\bibliography{reference.bib}

\begin{thebibliography}{}
\itemsep0pt

\bibitem[Abadi et~al., 2016]{abadi2016tensorflow}
Abadi, M., P. Barham, J. Chen, Z. Chen, A. Davis, J. Dean, M. Devin, S.
  Ghemawat, G. Irving, M. Isard, et~al.,  2016, Tensorflow: A system for
  large-scale machine learning: 12th $\{$USENIX$\}$ symposium on operating
  systems design and implementation ($\{$OSDI$\}$ 16), 265--283.

\bibitem[Adler et~al., 2021]{adler2021deep}
Adler, A., M. Araya-Polo, and T. Poggio,  2021, Deep learning for seismic
  inverse problems: Toward the acceleration of geophysical analysis workflows:
  IEEE Signal Processing Magazine, {\bfseries 38}, 89--119.

\bibitem[{Allen-Zhu} et~al., 2019]{allen-zhuConvergence2019}
{Allen-Zhu}, Z., Y. Li, and Z. Song,  2019, A {{Convergence Theory}} for {{Deep
  Learning}} via {{Over}}-{{Parameterization}}: arXiv:1811.03962 [cs, math,
  stat].

\bibitem[Arora et~al., 2018]{aroraOptimization2018}
Arora, S., N. Cohen, and E. Hazan,  2018, On the {{Optimization}} of {{Deep
  Networks}}: {{Implicit Acceleration}} by {{Overparameterization}}:
  arXiv:1802.06509 [cs].

\bibitem[Asnaashari et~al., 2013]{asnaashariRegularized2013}
Asnaashari, A., R. Brossier, S. Garambois, F. Audebert, P. Thore, and J.
  Virieux,  2013, Regularized seismic full waveform inversion with prior model
  information: GEOPHYSICS, {\bfseries 78}, R25--R36.

\bibitem[Barnier et~al., 2018]{barnier2018full}
Barnier, G., E. Biondi, and B. Biondi,  2018, Full waveform inversion by model
  extension, {\itshape in} SEG Technical Program Expanded Abstracts 2018:
  Society of Exploration Geophysicists,  1183--1187.

\bibitem[Barnier et~al., 2019]{barnier2019waveform}
Barnier, G., E. Biondi, and R. Clapp,  2019, Waveform inversion by model
  reduction using spline interpolation, {\itshape in} SEG Technical Program
  Expanded Abstracts 2019: Society of Exploration Geophysicists,  1400--1404.

\bibitem[Billette and {Brandsberg-Dahl}, 2005]{billette20042005}
Billette, F.~J., and S. {Brandsberg-Dahl},  2005, The 2004 {{BP Velocity
  Benchmark}}: 67th {{EAGE Conference}} \& {{Exhibition}}, {European
  Association of Geoscientists \& Engineers},~cp.

\bibitem[Biondi and Almomin, 2014]{biondi2014simultaneous}
Biondi, B., and A. Almomin,  2014, Simultaneous inversion of full data
  bandwidth by tomographic full-waveform inversion: Geophysics, {\bfseries 79},
  WA129--WA140.

\bibitem[Bozda{\u{g}} et~al., 2011]{bozdaug2011misfit}
Bozda{\u{g}}, E., J. Trampert, and J. Tromp,  2011, Misfit functions for full
  waveform inversion based on instantaneous phase and envelope measurements:
  Geophysical Journal International, {\bfseries 185}, 845--870.

\bibitem[Bunks et~al., 1995]{bunks1995multiscale}
Bunks, C., F.~M. Saleck, S. Zaleski, and G. Chavent,  1995, Multiscale seismic
  waveform inversion: Geophysics, {\bfseries 60}, 1457--1473.

\bibitem[Burstedde and Ghattas, 2009]{bursteddeAlgorithmic2009}
Burstedde, C., and O. Ghattas,  2009, Algorithmic strategies for full waveform
  inversion: {{1D}} experiments: GEOPHYSICS, {\bfseries 74}, WCC37--WCC46.

\bibitem[Cao and Gu, 2019]{caoGeneralization2019}
Cao, Y., and Q. Gu,  2019, Generalization {{Error Bounds}} of {{Gradient
  Descent}} for {{Learning Over}}-parameterized {{Deep ReLU Networks}}:
  arXiv:1902.01384 [cs, math, stat].

\bibitem[Esser et~al., 2018]{esser2018total}
Esser, E., L. Guasch, T. van Leeuwen, A.~Y. Aravkin, and F.~J. Herrmann,  2018,
  Total variation regularization strategies in full-waveform inversion: SIAM
  Journal on Imaging Sciences, {\bfseries 11}, 376--406.

\bibitem[Gal and Ghahramani, 2016]{gal2016dropout}
Gal, Y., and Z. Ghahramani,  2016, Dropout as a bayesian approximation:
  Representing model uncertainty in deep learning: international conference on
  machine learning, PMLR, 1050--1059.

\bibitem[Gebraad et~al., 2020]{gebraadBayesian2020}
Gebraad, L., C. Boehm, and A. Fichtner,  2020, Bayesian {{Elastic
  Full}}-{{Waveform Inversion Using Hamiltonian Monte Carlo}}: Journal of
  Geophysical Research: Solid Earth, {\bfseries 125}, e2019JB018428.

\bibitem[Goodfellow et~al., 2016]{goodfellow2016deep}
Goodfellow, I., Y. Bengio, and A. Courville,  2016, Deep learning: MIT Press.

\bibitem[Guitton, 2012]{guittonBlocky2012}
Guitton, A.,  2012, Blocky regularization schemes for {{Full}}-{{Waveform
  Inversion}}: Geophysical Prospecting, {\bfseries 60}, 870--884.

\bibitem[Guitton et~al., 2012]{guitton2012constrained}
Guitton, A., G. Ayeni, and E. D{\'\i}az,  2012, Constrained full-waveform
  inversion by model reparameterization: Geophysics, {\bfseries 77},
  R117--R127.

\bibitem[He et~al., 2016]{he2016deep}
He, K., X. Zhang, S. Ren, and J. Sun,  2016, Deep residual learning for image
  recognition: Proceedings of the IEEE conference on computer vision and
  pattern recognition, 770--778.

\bibitem[He and Wang, 2021]{he2021reparameterized}
He, Q., and Y. Wang,  2021, Reparameterized full-waveform inversion using deep
  neural networks: Geophysics, {\bfseries 86}, V1--V13.

\bibitem[Hore and Ziou, 2010]{hore2010image}
Hore, A., and D. Ziou,  2010, Image quality metrics: Psnr vs. ssim: 2010 20th
  international conference on pattern recognition, IEEE, 2366--2369.

\bibitem[Hu et~al., 2009]{huSimultaneous2009}
Hu, W., A. Abubakar, and T.~M. Habashy,  2009, Simultaneous multifrequency
  inversion of full-waveform seismic data: GEOPHYSICS, {\bfseries 74}, R1--R14.

\bibitem[Hu et~al., 2021]{hu2021progressive}
Hu, W., Y. Jin, X. Wu, and J. Chen,  2021, Progressive transfer learning for
  low-frequency data prediction in full waveform inversion: Geophysics,
  {\bfseries 86}, 1--82.

\bibitem[Kalita et~al., 2019]{kalita2019regularized}
Kalita, M., V. Kazei, Y. Choi, and T. Alkhalifah,  2019, Regularized
  full-waveform inversion with automated salt flooding: Geophysics, {\bfseries
  84}, R569--R582.

\bibitem[Kazei et~al., 2021]{kazei2021mapping}
Kazei, V., O. Ovcharenko, P. Plotnitskii, D. Peter, X. Zhang, and T.
  Alkhalifah,  2021, Mapping full seismic waveforms to vertical velocity
  profiles by deep learning: Geophysics, {\bfseries 86}, 1--50.

\bibitem[Kendall and Gal, 2017]{kendallWhat2017a}
Kendall, A., and Y. Gal,  2017, What {{Uncertainties Do We Need}} in {{Bayesian
  Deep Learning}} for {{Computer Vision}}?, {\itshape in} Advances in {{Neural
  Information Processing Systems}} 30: {Curran Associates, Inc.},  5574--5584.

\bibitem[Kingma and Ba, 2014]{kingma2014adam}
Kingma, D.~P., and J. Ba,  2014, Adam: A method for stochastic optimization:
  arXiv preprint arXiv:1412.6980.

\bibitem[Li and Harris, 2018]{liFull2018}
Li, D., and J.~M. Harris,  2018, Full waveform inversion with nonlocal
  similarity and model-derivative domain adaptive sparsity-promoting
  regularization: Geophysical Journal International, {\bfseries 215},
  1841--1864.

\bibitem[Li et~al., 2019]{li2019deep}
Li, S., B. Liu, Y. Ren, Y. Chen, S. Yang, Y. Wang, and P. Jiang,  2019,
  Deep-learning inversion of seismic data: arXiv preprint arXiv:1901.07733.

\bibitem[Li and Demanet, 2016]{liFullwaveform2016}
Li, Y.~E., and L. Demanet,  2016, Full-waveform inversion with extrapolated
  low-frequency data: GEOPHYSICS, {\bfseries 81}, R339--R348.

\bibitem[Luo and Schuster, 1991]{luoWave1991}
Luo, Y., and G.~T. Schuster,  1991, Wave‐equation traveltime inversion:
  GEOPHYSICS, {\bfseries 56}, 645--653.

\bibitem[Maas et~al., 2013]{maas2013rectifier}
Maas, A.~L., A.~Y. Hannun, A.~Y. Ng, et~al.,  2013, Rectifier nonlinearities
  improve neural network acoustic models: Proc. icml, Citeseer,~3.

\bibitem[Mosser et~al., 2020]{mosser2020stochastic}
Mosser, L., O. Dubrule, and M.~J. Blunt,  2020, Stochastic seismic waveform
  inversion using generative adversarial networks as a geological prior:
  Mathematical Geosciences, {\bfseries 52}, 53--79.

\bibitem[Ovcharenko et~al., 2019]{ovcharenko2019deep}
Ovcharenko, O., V. Kazei, M. Kalita, D. Peter, and T. Alkhalifah,  2019, Deep
  learning for low-frequency extrapolation from multioffset seismic data:
  Geophysics, {\bfseries 84}, R989--R1001.

\bibitem[Plessix, 2006]{plessix2006review}
Plessix, R.-E.,  2006, A review of the adjoint-state method for computing the
  gradient of a functional with geophysical applications: Geophysical Journal
  International, {\bfseries 167}, 495--503.

\bibitem[Richardson, 2018a]{richardson2018generative}
Richardson, A.,  2018a, Generative adversarial networks for model order
  reduction in seismic full-waveform inversion: arXiv preprint
  arXiv:1806.00828.

\bibitem[Richardson, 2018b]{richardson2018seismic}
--------, 2018b, Seismic full-waveform inversion using deep learning tools and
  techniques: arXiv preprint arXiv:1801.07232.

\bibitem[Srivastava et~al., 2014]{srivastava2014dropout}
Srivastava, N., G. Hinton, A. Krizhevsky, I. Sutskever, and R. Salakhutdinov,
  2014, Dropout: a simple way to prevent neural networks from overfitting: The
  journal of machine learning research, {\bfseries 15}, 1929--1958.

\bibitem[Sun and Demanet, 2020]{sunExtrapolated2020}
Sun, H., and L. Demanet,  2020, Extrapolated full-waveform inversion with deep
  learning: GEOPHYSICS, {\bfseries 85}, R275--R288.

\bibitem[Symes, 2008]{symesMigration2008}
Symes, W.~W.,  2008, Migration velocity analysis and waveform inversion:
  Geophysical Prospecting, {\bfseries 56}, 765--790.

\bibitem[Tarantola, 1984]{tarantola1984inversion}
Tarantola, A.,  1984, Inversion of seismic reflection data in the acoustic
  approximation: Geophysics, {\bfseries 49}, 1259--1266.

\bibitem[Tarantola, 2005]{tarantolaInverse2005}
--------, 2005, Inverse {{Problem Theory}} and {{Methods}} for {{Model
  Parameter Estimation}}: {Society for Industrial and Applied Mathematics}.
\newblock Other {{Titles}} in {{Applied Mathematics}}.

\bibitem[Ulyanov et~al., 2018]{ulyanovDeep2018}
Ulyanov, D., A. Vedaldi, and V. Lempitsky,  2018, Deep image prior:
  arXiv:1711.10925 [cs, stat].

\bibitem[Versteeg, 1994]{versteeg1994marmousi}
Versteeg, R.,  1994, The marmousi experience: Velocity model determination on a
  synthetic complex data set: The Leading Edge, {\bfseries 13}, 927--936.

\bibitem[Virieux and Operto, 2009]{virieux2009overview}
Virieux, J., and S. Operto,  2009, An overview of full-waveform inversion in
  exploration geophysics: Geophysics, {\bfseries 74}, WCC1--WCC26.

\bibitem[Wu et~al., 2014]{wu2014seismic}
Wu, R.-S., J. Luo, and B. Wu,  2014, Seismic envelope inversion and modulation
  signal model: Geophysics, {\bfseries 79}, WA13--WA24.

\bibitem[Wu et~al., 2018]{wuInversionet2018}
Wu, Y., Y. Lin, and Z. Zhou,  2018, Inversionet: {{Accurate}} and efficient
  seismic-waveform inversion with convolutional neural networks: {{SEG
  Technical Program Expanded Abstracts}} 2018, {Society of Exploration
  Geophysicists}, 2096--2100.

\bibitem[Wu and McMechan, 2019]{wuParametric2019}
Wu, Y., and G.~A. McMechan,  2019, Parametric convolutional neural
  network-domain full-waveform inversion: GEOPHYSICS, {\bfseries 84},
  R881--R896.

\bibitem[Wu and McMechan, 2020]{wu2020cnn}
--------, 2020, Cnn-boosted full-waveform inversion, {\itshape in} SEG
  Technical Program Expanded Abstracts 2020: Society of Exploration
  Geophysicists,  1526--1530.

\bibitem[Xu and Darve, 2020]{xu2020adcme}
Xu, K., and E. Darve,  2020, Adcme: Learning spatially-varying physical fields
  using deep neural networks: arXiv preprint arXiv:2011.11955.

\bibitem[Yang and Ma, 2019]{yangDeeplearning2019}
Yang, F., and J. Ma,  2019, Deep-learning inversion: {{A}} next-generation
  seismic velocity model building method: GEOPHYSICS, {\bfseries 84},
  R583--R599.

\bibitem[Zhang et~al., 2017]{zhangSparse2017}
Zhang, P., L. Han, Z. Xu, F. Zhang, and Y. Wei,  2017, Sparse blind
  deconvolution based low-frequency seismic data reconstruction for multiscale
  full waveform inversion: Journal of Applied Geophysics, {\bfseries 139},
  91--108.

\bibitem[Zhang et~al., 2018]{zhang2018correcting}
Zhang, Z., J. Mei, F. Lin, R. Huang, and P. Wang,  2018, Correcting for salt
  misinterpretation with full-waveform inversion, {\itshape in} SEG Technical
  Program Expanded Abstracts 2018: Society of Exploration Geophysicists,
  1143--1147.

\bibitem[Zhu et~al., 1997]{zhu1997algorithm}
Zhu, C., R.~H. Byrd, P. Lu, and J. Nocedal,  1997, Algorithm 778: L-bfgs-b:
  Fortran subroutines for large-scale bound-constrained optimization: ACM
  Transactions on Mathematical Software (TOMS), {\bfseries 23}, 550--560.

\bibitem[Zhu et~al., 2017]{zhuSparsepromoting2017}
Zhu, L., E. Liu, and J.~H. McClellan,  2017, Sparse-promoting full-waveform
  inversion based on online orthonormal dictionary learning: GEOPHYSICS,
  {\bfseries 82}, R87--R107.

\bibitem[Zhu et~al., 2020]{zhuGeneral2020}
Zhu, W., K. Xu, E. Darve, and G.~C. Beroza,  2020, A {{General Approach}} to
  {{Seismic Inversion}} with {{Automatic Differentiation}}: arXiv:2003.06027
  [physics].

\end{thebibliography}
\bibliographystyle{seg}

\end{document}